\documentclass[11pt]{ucthesis}
\def\dsp{\def\baselinestretch{2.0}\large\normalsize}
\dsp
\usepackage{amsmath,amsfonts,amssymb,epsfig,amsthm}
\begin{document}

\newtheorem{theorem}{Theorem}
\newtheorem{lemma}{Lemma}
\newtheorem{sublemma}{Sublemma}
\newtheorem{problem}{Problem}
\newtheorem{pproblem}{Promise Problem}
\newtheorem{claim}{Claim}
\newtheorem{corollary}{Corollary}
\newtheorem{definition}{Definition}
\newtheorem{observation}{Observation}
\newtheorem{assumption}{Assumption}
\newtheorem{algorithm}{Algorithm}
\newtheorem*{oldlemmalem:rec}{Lemma \ref{lem:rec}}
\newtheorem*{oldlemmalem:falloff}{Lemma \ref{lem:falloff}}
\newtheorem*{oldlemmalem:intcl}{Lemma \ref{lem:intcl}}
\newtheorem*{oldlemmalem:cdn}{Lemma \ref{lem:cdn}}
\newtheorem*{oldlemmalem:twid}{Lemma~\ref{lem:twid}}
\newtheorem*{oldlemmalem:sma}{Lemma~\ref{lem:sma}}

\newenvironment{solution}{{\bf Solution:  }}{\hfill\rule{2mm}{2mm}}
\newenvironment{example}{{\bf Example:  }}{\hfill\rule{2mm}{2mm}}
\newenvironment{comments}{{\bf Comments:  }}{\hfill\rule{2mm}{2mm}}
\newcommand{\bcdot}{{\bf\cdot}}
\newcommand{\ket}[1]{|#1\rangle}
\newcommand{\bra}[1]{\langle #1|}
\newcommand{\braket}[2]{\langle #1|#2\rangle}
\newcommand{\clint}[1]{\lfloor #1 \rceil}
\newcommand{\li}[1]{\lfloor #1 \rfloor}
\newcommand{\om}[2]{\omega_{#1}^{#2}}
\newcommand{\ignore}[1]{}
\newcommand{\balpha}{{\bf\alpha}}
\newcommand{\bbeta}{{\bf\beta}}
\newcommand{\bgamma}{{\bf\gamma}}
\newcommand{\bzeta}{{\bf\zeta}}
\newcommand{\cartj}[1]{[#1_1]\times [#1_2]\times\cdots \times [#1_j]}
\newcommand{\norm}[1]{\left|{#1}\right|}
\newcommand{\distD}{ {\mathcal D} }
\newcommand{\yes}{ {\mathcal Y} }
\newcommand{\no}{ {\mathcal N} }
\newcommand{\distP}{ {\mathcal P} }

\newcommand{\ftc}[4]{\sum_{#3<#2}#4_{#3}\om{#2}{#3#1}}
\newcommand{\zm}[1]{\bigoplus_{i<{#1}}Z_{m_i}}
\newcommand{\keta}{\ket{\alpha}}
\newcommand{\rketa}{\ket{\tilde{\alpha}}}
\newcommand{\frketa}{\ket{\hat{\tilde{\alpha}}}}
\newcommand{\ketha}[0]{\ket{\hat{\alpha}}}
\newcommand{\ha}[0]{\hat{\alpha}}

\newcommand{\inprod}[2]{\big\langle{#1},{#2}\big\rangle}
\newcommand{\ngslice}[1]{\ket{\phi'_{#1}}}
\newcommand{\ftv}[0]{\ket{\hat{v}}}
\newcommand{\fta}[0]{\ket{\hat{a}}}
\newcommand{\zton}[0]{\bigoplus_nZ_2}
\newcommand{\litm}[0]{^{_M}}
\newcommand{\litmp}[0]{^{_{M\prime}}}
\newcommand{\litn}[0]{^{_N}}
\newcommand{\kfv}[0]{\ket{\hat {v}}}
\newcommand{\kfmv}[0]{\ket{\hat {v}^{\litm}}}
\newcommand{\kfw}[0]{{\ket{\hat{w}}}}
\newcommand{\kfmw}[0]{{\ket{\hat{w}^{\litm}}}}
\newcommand{\afv}[0]{\ket{{\hat {v}^{\litm}}}^\prime}
\newcommand{\fmw}[0]{{\hat{w}^{\litm}}}
\newcommand{\kethj}[0]{{\ket{\hat{j}}}}
\newcommand{\kjm}[0]{{\ket{j^{\litm}}}}
\newcommand{\kim}[0]{{\ket{i^{\litm}}}}
\newcommand{\im}[0]{i^{\litm}}
\newcommand{\jm}[0]{j^{\litm}}
\newcommand{\kaj}[0]{\ket{j^{\litm}}^\prime}
\newcommand{\aj}[0]{j^{\litmp}}
\newcommand{\fv}[0]{\hat{v}}
\newcommand{\pri}[0]{i^\prime}
\newcommand{\prj}[0]{j^\prime}
\newcommand{\prk}[0]{k^\prime}

\newcommand{\ufqw}[0]{{\hat {w}}^q}

\newcommand{\Dv}[0]{{\distD}_{\kfv}}
\newcommand{\Dw}[0]{{\distD}_{\kfmw}}
\newcommand{\Dx}[0]{\distD_{\ket{x}}}
\newcommand{\uDw}[0]{d_{\kfmw}}
\newcommand{\bumpi}[0]{{\ket{b_{i}}}}
\newcommand{\bumpo}[0]{{\ket{b_{0}}}}
\newcommand{\shibumpo}[0]{\ket{b_{0}}^{\pri}}
\newcommand{\taili}[0]{{\ket{t_{i}}}}
\newcommand{\tailj}[0]{{\ket{t_{j}}}}
\newcommand{\kfal}{\ket{\hat{\alpha}}}
\newcommand{\fal}{\hat{\alpha}}
\newcommand{\kbeta}{\ket{\beta}}
\newcommand{\kgam}{\ket{\gamma}}
\newcommand{\fmkb}{\ket{\hat{\beta}^{\litm}}}
\newcommand{\cy}[1]{\langle{#1}\rangle}
\newcommand{\fh}[0]{f_H}
\newcommand{\hp}[0]{H^\prime}
\newcommand{\fk}[0]{f_{K}}
\newcommand{\fn}[0]{f_{\langle N\rangle}}
\newcommand{\bD}[0]{{\bf D}}
\newcommand{\iso}[0]{{\tilde{=}}}
\newcommand{\zpp}[0]{\bigoplus_{i<n} Z_{p_i}}
\newcommand{\perH}[0]{H^{\perp}}
\newcommand{\perK}[0]{K^{\perp}}
\newcommand{\distDf}[2]{{\mathcal D}^f_{F_{#1}{#2}}}
\newcommand{\dD}[2]{\distD_{F_{#1}{#2}}}
\newcommand{\fp}[0]{f_p}
\newcommand{\fq}[0]{f_q}
\newcommand{\twid}[2]{{#1}\sim_{\scriptstyle R(x,y)}{#2}}


\title{The Quantum Fourier Transform and
Extensions of the Abelian Hidden Subgroup Problem}
\author{Lisa Ruth Hales}
\degreeyear{Spring 2002}
\degree{Doctor of Philosophy}
\chair{Professor Umesh V. Vazirani}
\othermembers{Professor W. Hugh Woodin\\
Professor Christos H. Papadimitriou\\
Professor John W. Addison Jr.\\
Professor K. Birgitta Whaley}
\numberofmembers{5}
\prevdegrees{B.A. (University of California at Berkeley) 1991}
\field{Logic and the Methodology of Science}
\campus{Berkeley}

\maketitle
\begin{abstract}
The quantum Fourier transform (QFT) has emerged as the primary tool
in quantum algorithms which achieve exponential advantage over
classical computation and lies at the heart of the solution to
the abelian hidden subgroup problem, of which Shor's celebrated
factoring and discrete log algorithms are a special case.
We begin by addressing various computational issues surrounding the QFT
and give improved parallel circuits for both the QFT over a power
of $2$ and the QFT over an arbitrary cyclic group.  These circuits
are based on new insight into the relationship between the discrete Fourier
transform over different cyclic groups.  We then exploit this
insight to extend the class of hidden subgroup problems with
efficient quantum solutions.  First we relax the condition
that the underlying hidden subgroup function be 
distinct on distinct cosets
of the subgroup in question 
and show that this relaxation can be solved whenever $G$ is
a finitely-generated abelian group.  We then extend this reasoning
to the hidden cyclic subgroup problem over the reals, showing
how to efficiently generate the bits of the period of any
sufficiently piecewise-continuous function on $\Re$.   
Finally, we show that this problem of period-finding over $\Re$,
viewed as an oracle promise problem, is strictly harder than
its integral counterpart.  In particular, period-finding over
$\Re$ lies outside the complexity class $MA$, a class which contains
period-finding over the integers. 

\abstractsignature
\end{abstract}
\begin{frontmatter}
\setcounter{page}{1}
\begin{dedication}
\null\vfil
{\large
\begin{center}
To Samantha,\\\vspace{12pt}
a faithful friend.\\\vspace{12pt}
\end{center}}
\vfil\null
\end{dedication}

\tableofcontents
\listoffigures
\begin{acknowledgements}
I want to thank my advisor for his patience, his insight and
his sense of humor without which this thesis would never have been
completed.

I am grateful to Sean Hallgren, the co-author of many of the results 
in this thesis. I really enjoyed the countless hours spent together
in cafes over the years and hope to collaborate again.

I want to thank the many members of the Logic group who 
have given me support and encouragement during my years at Berkeley.
Professor John Addison deserves particular thanks for introducing
me to the complexities of Complexity theory and for his 
ever-present sense of humor.  I would certainly never 
have finished without the support and friendship of
the group secretary, Catalina Cordoba. Good luck in your
retirement!  And thanks to Richard Zach for making
the ``middle years'' of my graduate career more fun.

I also want to thank the faculty, staff, and students of my adopted 
department, Computer Science, for making me feel welcome.

Finally, I must thank my family.  First, my Mom and Dad
for their constant love and support without which none of this
would have been possible. Second, my husband for putting up with
my seemingly infinite thesis-writing and for his almost hourly
help with my computer.  My sister Kathy for doing the dishes,
taking out the dog, and mowing the lawn even though she had
her own thesis to write.  Charlotte for her
smiling face and hugs.  Eve for her
kicks and prods over the past few months which have served as
a continuous reminder of the urgency of my task.  And finally
Sal for helping to heal a broken heart.
\end{acknowledgements}

\end{frontmatter}
\chapter{Introduction}The primary tool underlying all quantum algorithms
which achieve exponential advantage over classical
computation is the quantum Fourier transform (QFT).
The fact that the QFT over exponentially large groups 
can be computed efficiently is at the heart
of the solution to the Abelian hidden subgroup
problem, of which Shor's celebrated factoring and discrete
logarithm algorithms \cite{SICOMP::Shor1997} are a special case. 
The aim of this dissertation is twofold.
First, we give improved quantum circuits for computing the QFT.
Second, we use the resulting insight into the structure of
the QFT to extend the class of hidden subgroup problems
with efficient quantum solutions.

In particular, after surveying existing techniques
computing the QFT over finite Abelian
groups, we give explicit parallel circuits for approximating
the QFT over a power of $2$, tightening
the results of \cite{cleve00fast}.
We then give improved parallel circuits 
for approximating the
QFT over an arbitrary cyclic
group, based on new insight
into the relationship between the discrete Fourier 
transforms over different cyclic groups.   
This insight also leads to a particularly elegant method
of ``Fourier sampling''
(\cite{SICOMP::BernsteinV1997},\cite{STOC::HalesH1999},\cite{Hoyer2000})
and simplifies the
presentation of the standard Abelian hidden subgroup
algorithm.

Second, we extend the class of
Abelian hidden subgroup promise problems
which have efficient quantum algorithms.
Given oracle access to a function $f$
defined on a group $G$ and constant on
cosets of some unknown subgroup $H\leq G$,
a solution to the hidden subgroup problem
is a list of generators for the subgroup $H$.
This problem can be solved efficiently
on a quantum computer whenever $G$ is
a finitely-generated Abelian group and
$f$ is distinct on distinct cosets(\cite{Jozsa}).

We first use our Fourier sampling procedure to
relax this distinctness requirement,
requiring only that the encoding of $H$ 
by $f$ be probabilistically unambiguous. 
This extends the
results of \cite{BonehL1995} and \cite{MosEke98} who relax
the distinctness condition only slightly.
Moreover, our result is tight -- we give a corresponding 
lower bound which shows that, in the absence of such
an unambiguous encoding, no polynomial-time
algorithm, classical or quantum, can recover 
the desired hidden subgroup.   

Finally, we give an efficient quantum algorithm
for the hidden cyclic subgroup problem over the reals $\Re$.
More specifically, given a sufficiently piecewise-continuous 
periodic function defined on $\Re$, we show how
to efficiently generate the bits of its period.
Again we must require that the encoding of the period
be probabilistically unambiguous.  This generalizes
a result of \cite{hallgren2002} which gives a quantum algorithm 
finding the period of 
a subclass of such functions and an important application,
namely an efficient quantum solution to Pell's equation.
Furthermore, we show that the hidden cyclic subgroup
problem over $\Re$ is harder than the analogous problem
over $Z$.  In particular we show that a decision version of the
problem over $\Re$
is outside of the complexity class $MA$, whereas any decision problem
which reduces to the problem over $Z$ lies inside of this class. 
\section{Outline}
The remainder of this Chapter is devoted to setting up our
quantum circuit model. Chapter~\ref{chap:qft_hsp} introduces
the QFT and its relation to the hidden subgroup problem while
Chapter~\ref{chap:compqft} surveys earlier 
techniques for implementing the QFT.
In Chapter~\ref{chap:pcs} we give new parallel circuits for computing the
QFT over a power of $2$.  Chapter~\ref{chap:fta} contains both our algorithm
for computing the QFT over an arbitrary cyclic group and also
the associated Fourier sampling procedures.  The technical results
leading to these algorithms can be found in Chapter~\ref{chap:ftt}.

We then turn to extensions of the hidden subgroup algorithm.
Chapter~\ref{chap:nmulti} extends the hidden subgroup algorithm over
finitely generated Abelian groups to the case
where the given function is not distinct on distinct cosets.
The associated lower bound is found in Section~\ref{lowbou}.  
Chapter~\ref{chap:overr}
is devoted to period-finding over the reals and
and Chapter~\ref{chap:or} to the proof that this problem is outside of $MA$.

\section{Notation}
We shall be primarily concerned with the
vector space $V_n$ over the field of complex
numbers $C$ consisting of formal linear
combinations of bit-strings $k\in\{0,1\}^n$.
We use the Dirac ``ket'' notation, $\ket{\cdot}$, for vectors in this space
and reserve $\ket{i}$, $\ket{j}$ and $\ket{k}$ for the
basis vectors corresponding to the bit-strings $i,j,k\in\{0,1\}^n$.
$\ket{v}$, $\ket{w}$, and $\ket{u}$ denote arbitrary vectors
in this space with 
$$\ket{v}=\sum_{j\in\{0,1\}^n}v_j\ket{j}.$$
$V_n$ is a Hilbert Space with {\bf inner product}
$$\inprod{\ket{v}}{\ket{w}}=\sum_{i\in\{0,1\}^n}v_i\overline{w_i},$$
where $\overline{w_i}$ is the complex conjugate of
$w_i$.  By using the standard ``bra'' notation, $\bra{\cdot}$, to denote
the dual vector, i.e. $\bra{v}$
is the linear operator from $V_n$ to $C$ defined by
$$\bra{v}\ket{w}=\inprod{\ket{v}}{\ket{w}},$$
the vertical bars of the adjacent ``bra'' and ``ket'' in
the inner product of $\ket{v}$ and $\ket{w}$ can be conveniently
merged and written as $\braket{v}{w}$. 
We let $\|\cdot\|$ denote the {\bf norm} associated with this inner product,
$$\|\ket{v}\|=\sqrt{\braket{v}{v}}.$$
Note that the vectors $\ket{k}$ form an
orthonormal basis for $V_n$ under this inner product.
We will also use $\|\cdot\|$ to denote the associated
{\bf operator norm}.  In particular if $U$ is a (linear) operator on $V_n$ then
$$\|U\|=max_{\ket{v}\in V}\frac{\|U\ket{v}\|}{\|\ket{v}\|}.$$

For any linear operator $U$ on $V_n$ there is a unique linear
operator $U^\dagger$ on $V_n$
satisfying
$$\braket{Uv}{w}=\braket{v}{U^\dagger w}$$
for all vectors $\ket{v}$ and
$\ket{w}$.
$U^\dagger$ is called the {\bf Hermitian adjoint} of $U$.
Its matrix representation is the conjugate transpose
of the matrix representing $U$, in other words if
$U$ has matrix representation
$(u_{ij})$ then the matrix representation of $U^\dagger$
has $ij$th entry $\overline{u_{ji}}$.
An operator $U$ is called {\bf unitary} if 
its Hermitian adjoint is also its inverse, that is, if $UU^\dagger=I$.
This is equivalent to the condition that the
vectors $U\ket{k}$ form an orthonormal basis for $V$.  

An important construction underlying the quantum mechanics
of multiparticle systems is the
{\bf tensor product} of vector spaces.
Given any pair of bases for the vector spaces $V$ and $W$,
their Cartesian product forms a basis for the tensor product
of $V$ and $W$, denoted $V\otimes W$.
That is, if $\ket{j}$ and $\ket{k}$ are elements of $V$ and $W$'s
respective bases then $\ket{j}\otimes\ket{k}$ is an element of
the resulting basis for $V\otimes W$.  $V\otimes W$ then consists of all
linear combinations of these basis elements 
modulo the following equivalences:

\begin{enumerate}\item For any scalar $c\in C$ and elements
$\ket{v}\in V$ and $\ket{w}\in W$,
$$c\left(\ket{v}\otimes\ket{w}\right)=\left(c\ket{v}\right)\otimes\ket{w}=
\ket{v}\otimes\left(c\ket{w}\right).$$
\item  For any $\ket{v}$ and $\ket{v^\prime}$ in $V$ and $\ket{w}$ and
$\ket{w^\prime}$ in $W$ the following hold
$$\left(\ket{v}+\ket{v^\prime}\right)\otimes\ket{w}=\ket{v}\otimes\ket{w}+
\ket{v^\prime}\otimes\ket{w},$$
and
$$\ket{v}\otimes\left(\ket{w}+\ket{w^\prime}\right)=
\ket{v}\otimes\ket{w}+
\ket{v}\otimes\ket{w^\prime}.$$
\end{enumerate}
These relations can also be used to give a basis independent
construction of $V\otimes W$ -- it is the free product of
$V$ and $W$ modulo these equivalences. 
It is not hard to see that these equivalence relations do not collapse
any elements of the basis for $V\otimes W$ described above and thus 
$\dim\left(V\otimes W\right)=\dim\left(V\right)\dim\left(W\right)$.
It is worth noting that, while for any pair of vectors 
$\ket{v}\in V$ and $\ket{w}\in W$ the vector 
$\ket{v}\otimes\ket{w}$ is in $V\otimes W$, vectors of this form
comprise only a tiny fraction of the tensor space.  In particular,
their description only has dimension $\dim(V)+\dim(W)$.   

The tensor product $V\otimes W$ inherits a natural
inner product structure from $V$ and $W$ by taking as an orthonormal
basis any Cartesian product of orthonormal bases of $V$ and $W$.  
Also, for any two linear operators 
$A$ and $B$ on $V$ and $W$ respectively,
we can define their tensor product $A\otimes B$ 
which is the linear operator
on $V\otimes W$ satisfying
$$A\otimes B\left(\ket{j}\otimes\ket{k}\right)
=A\ket{j}\otimes B\ket{k},$$
for basis elements $\ket{j}\in V$ and $\ket{k}\in W$. 
More generally, any bilinear map on the cartesian product $V\times W$
induces a linear transformation of the tensor product
$V\otimes W$ -- a category
theoretic definition of the tensor product can be formulated
in these terms.  

\begin{example}
As a concrete example of the tensor product, recall the vector spaces
$V_n$ defined previously.  The tensor product of any two spaces
$V_n$ and $V_m$
has an orthonormal basis consisting of elements of the 
form $\ket{j}\otimes \ket{k}$
where $j$ and $k$ are bitstrings of length $n$ and $m$ respectively.
The resulting space 
$V_n\otimes V_m$ is clearly isomorphic to $V_{n+m}$ by extending the 
obvious map of basis elements
$$\ket{j}\otimes \ket{k}\longrightarrow \ket{jk}$$
Notice that this map is also preserves the corresponding
inner product.
\end{example}
\section{Qubits}
A {\bf qubit} is the abstraction of a two-level quantum particle,
in the same sense that a bit is the abstraction
of a classical storage device which can be in one 
of two positions, $0$ or $1$.  While such a classical storage device
is always either in position $0$ or in position $1$,  
quantum particles can exist in a complex combination
or ``superposition'' of levels and are described by 
a unit vector in $V_{1}$, that is, a vector 
\begin{equation}\label{qubit}
\alpha_0\ket{0}+\alpha_1\ket{1}\end{equation}
where the $\alpha_i$ are complex numbers
satisfying $\norm{\alpha_0}^2+\norm{\alpha_1}^2=1$.

A {\bf measurement} is the abstraction of a physical 
procedure which obtains classical 
information about the state of the quantum particle.  A
measurement is represented mathematically as the projection 
of the state vector onto a pair of orthogonal subspaces.  
For instance, a measurement of
the state (\ref{qubit}) in the standard basis
projects the state vector onto the
subspaces generated by $\ket{0}$ and $\ket{1}$
respectively, yielding the
state $\ket{0}$ with probability $\norm{\alpha_0}^2$ and
$\ket{1}$ with probability $\norm{\alpha_1}^2$.

Quantum computation entails the manipulation 
multi-particle quantum systems.  A system of
$n$ qubits is described by a unit vector in $V_n$,
the tensor product of the individual
vector spaces $V_1$ inhabited by each qubit.  The state vector
itself, however, need not be a product of vectors in these
component spaces -- recall that such product vectors form but a tiny fraction
of the entire tensored space.  A state vector
which cannot be decomposed as such a product is 
{\bf entangled}.  Entangled states play a critical role
both in quantum computation and quantum information theory.

\section{Circuits:  Classical vs. Quantum}
Various models for quantum computation have been proposed.  
For our purposes it will
be most convenient to work in the quantum circuit
model.
Before we specify the particulars of our model
we review some of the features peculiar to quantum computation
by contrasting a particular classical probabilistic circuit
model with its quantum counterpart.

A classical probabilistic circuit takes as input
a string of bits, runs them through a sequence
of one and two-bit probabilistic gates, and outputs a
string according to the probability
distribution induced by the array of gates. 
For our purposes it will be convenient to 
assume these gates have equal length input and output. 
In particular, we take the deterministic
$NOT$ and $FAND$ (fan-out and) together with a 
probabilistic $NOT_{1/2}$ gate
as our basis.  The $NOT_{1/2}$ gate acts like the
deterministic $NOT$ gate with 
probability $1/2$ and with probability $1/2$ 
allow the bits to pass through unaffected. We can represent these
gates by their transition matrices,
$$\begin{array}{ccc}
NOT=\left(\begin{array}{cc}0&1\\
1&0\end{array}\right),&
FAND=\left(\begin{array}{cccc}1&0& 0& 0\\
        1&0& 0& 0\\ 1& 0& 0& 0\\ 0& 0& 0& 1\end{array}\right),&
NOT_{1/2}=\left(\begin{array}{cc}1/2&1/2\\
1/2&1/2\end{array}\right),
\end{array}$$
where the rows and columns are indexed by 
bit strings in lexicographic order and the matrix entries represent the 
transition probabilities induced by the gate.
The state of the circuit at any stage $s$ 
can be described
by a probability distribution on $n$-bit strings, or equivalently
as a vector in $V_n$
$$\sum_{i\in\{0,1\}^n}p_{i,s}\ket{i},$$
where $p_{i,s}$ denotes the probability that
the bits are in the configuration $\ket{i}$ at stage $s$.
Thus the $p_{i,s}$ are nonnegative reals satisfying
$$\sum_{i\in\{0,1\}^n}p_{i,s}=1.$$
The probabilities $p_{i,s}$ can be gotten from
the $p_{i,s-1}$ by the formula
$$p_{i,s}=\sum_{j\in \{0,1\}^n}p_{j,s-1}t_{j,i,s-1},$$
where $t_{j,i,s-1}$ is the probability that the string $\ket{j}$
transitions to the string $\ket{i}$ when the gate of
stage $s-1$ is applied.  Again, the $t_{j,i,s}$'s are 
nonnegative reals satisfying 
$$\sum_{i\in\{0,1\}^n}t_{j,i,s}=1$$
for all $j,s$.  The matrix of values $(t_{j,i,s})$ for a fixed $s$
is a tensor product of the identity matrix and 
the transition matrix of the gate (see above)
applied at stage $s$.
A classical probabilistic circuit
with final state
$$\sum_{i\in \{0,1\}^n}p_{i,s_f}\ket{i}$$
outputs the string $i$ with probability 
$p_{i,s_f}$ at the conclusion of the algorithm.

In analogy to the classical case, a
quantum circuit takes as input a 
string of qubits
and runs them through a 
sequence of one and two-bit quantum gates.
In this case the state of the machine
at any stage $s$ is unit vector in $V_n$,
$$\ket{\balpha_s}=\sum_{i\in\{0,1\}^n}\alpha_{i,s}\ket{i}$$
where the $\alpha_i$ are complex numbers
satisfying 
$$\sum_{i\in\{0,1\}^n}|\alpha_i|^2=1.$$
As in the classical case we assume that
the input is a determinate bitstring, i.e.
a quantum state of the form $\ket{i}$
for some $i\in\{0,1\}^n$.
As before, the amplitude $\alpha_{i,s}$ of a state $\ket{i}$
at stage $s$ can be gotten
from the $\alpha_{i,s-1}$ by the formula 
$$\alpha_{i,s}=\sum_{j\in \{0,1\}^n}\alpha_{j,s-1}\tau_{j,i,s-1},$$
where $\tau_{j,i,s-1}$ is the amplitude with which the state $\ket{j}$
transitions to the state $\ket{i}$ when the gate of
stage $s-1$ is applied.  Again the the matrix of values $(\tau_{j,i,s})$
is a tensor product of the identity matrix with 
the transition matrix of the gate
applied at stage $s$, but in the quantum model
the $\tau_{j,i,s}$'s are not positive real probabilities, but
complex numbers whose amplitudes squared satisfy 
$$\sum_{i\in\{0,1\}^n}|\tau_{j,i,s}|^2=1$$
for all $j,s$.  

In order to obtain classical information
from the final quantum state $\ket{\balpha_{s_f}}$
output by the array,
a measurement in the standard basis
is performed at the conclusion of the
algorithm.  The probability of measuring a
particular string $\ket{i}$ is given by
$$\left|\alpha_{i,s_f}\right|^2.$$

We now isolate the aspects of the quantum 
model which distinguish it from its classical
counterpart.  First, we focus on the class of
allowable gates.  The principles
of quantum mechanics require that the evolution
of a quantum state be reversible -- no information
can be gained or lost.  The classical $FAND$ gate,
for example, violates this principle since it maps both the strings
00 and 01 to the string
00. This reversibility requirement restricts the
class of allowable gates to those whose
transition matrices are unitary and raises
the question of whether a quantum
device is capable of even simulating a classical 
probabilistic circuit,
much less moving beyond it.  Such a simulation is possible
if we allow the circuit to maintain a copy of its
input throughout the computation.
That is, if there is a classical probabilistic
circuit mapping
$$\ket{i}\longrightarrow\ket{j}\mbox{ with probability }p_{ij}$$
then there is a quantum circuit of polynomially-related size 
mapping
$$\ket{i}\ket{0}\longrightarrow\alpha_{ij}\ket{i}\ket{j}\mbox{ where }
\left|\alpha_{ij}\right|^2=p_{ij}.$$
The following three gates, known as the Hadamard, the 
controlled-not, and the $1/8$-rotation respectively,  
together suffice for such a simulation. 
$$\begin{array}{ccc}
H=\left(\begin{array}{cc}\frac{1}{\sqrt{2}} &\frac{1}{\sqrt{2}} \\
\frac{1}{\sqrt{2}} &-\frac{1}{\sqrt{2}}\end{array}\right)&

CNOT=\left(\begin{array}{cccc}1&0& 0& 0\\
        0 & 1 & 0& 0\\ 0& 0& 0& 1\\ 0& 0& 1& 0\end{array}\right)&

R_{3}=\left(\begin{array}{cc}1 & 0\\
0 &e^{2\pi i/8}\end{array}\right)\end{array}.$$
Moreover, they are 
universal for quantum computation, that is, they can be used
to approximate any unitary transformation on $n$ qubits with arbitrary
precision.
\begin{theorem}\label{thm:approxun}
Any unitary transformation on $n$ qubits can be approximated
to within $\epsilon$ by a quantum circuit of size 
$O\left(n^24^n\log^c\left(n^24^n/\epsilon\right)\right)$ over the gates 
$\{H,CNOT,R_3\}$.
\end{theorem}
Theorem \ref{thm:approxun} was proved independently by Solovay
and Kitaev.  See \cite{QuantumBook} for a nice proof and history.

While quantum circuits can efficiently simulate
their classical probabilistic counterpart, the 
converse appears to be false.  What are some of the difficulties
inherent in such a simulation?
One fundamental difference between the quantum and classical
models is that in the quantum setting the 
transition function $\tau$ is complex-valued.
This expresses the phenomenon of Quantum Interference --
nontrivial computational paths can cancel
each other out and disappear -- a property which lies
at the heart of the apparent exponential power
of the quantum model over its classical counterpart.  
Another difference is 
that the norms of the amplitudes of the state vector and the
transition function are 
only quadratically related to their associated probabilities.  
This property
is exploited by the data-base search algorithm of Grover 
\cite{STOC::Grover1996}
and its extensions
which achieve a corresponding quadratic speedup over 
probabilistic classical computation. 

So far the only classical simulations of quantum circuits
involve keeping an explicit record of the exponentially many amplitudes
associated with each step of the computation.  Such a brute-force
simulation can be accomplished using polynomial-space 
(but exponential-time) on a classical
Turing machine and, more specifically, inhabits the 
complexity class 
$P^{\#}\subseteq PSPACE$ \cite{SICOMP::BernsteinV1997}.  This indicates
that proving outright that quantum circuits cannot be efficiently
simulated by classical computation is very difficult -- such a proof
would imply that $P\neq PSPACE$, a long-standing open question in complexity
theory.  On the other hand, Shor's quantum algorithms for factoring
and discrete logarithms, well-studied problems for which there
is no efficient classical solution, 
together with various oracle results,
provide indirect evidence that no such simulation
exists.
 
Finally, we note that classical
devices have been proposed that purportedly solve $NP$-complete
problems in polynomial-time \cite{Vergis-et-al-86}.  
In each case it was shown that
the device in question required either exponential precision 
or energy and that
its apparent power was hidden in one of these untenable physical 
assumptions.  It is critical to point out that this is not
the case in our quantum circuit model.  In particular,
we need not implement our basic gates exactly, nor even with
exponential precision, to achieve the apparent power over classical
circuits.  It suffices to be able to approximate these gates to
within an arbitrary inverse polynomial, that is, to implement a 
unitary transformation $U$ satisfying
$$\|U-G\|<1/p(n),$$ 
where $G$ is the unitary transformation induced
by the desired gate.  This follows from fundamental work of 
\cite{SICOMP::BernsteinV1997}
showing that the errors incurred by such gate approximations are additive.
Thus for any given polynomial-size circuit each of these errors 
need only be less than a (larger) inverse polynomial in order 
for the distribution output by the approximation to be close 
to that of the exact circuit.

\section{A Quantum Circuit Model} 
\begin{figure}[hbt]
  \begin{center} \psfig{file=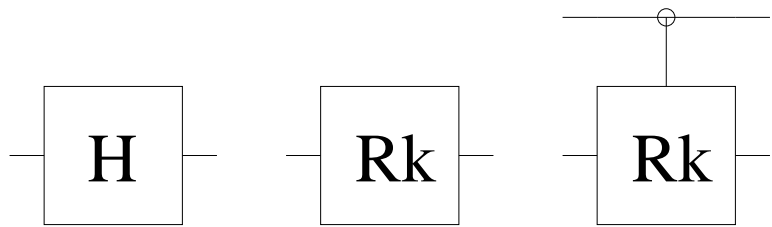,width=4in} \end{center}
  \caption{Quantum Gates:  Hadamard, Rotation, Controlled Rotation.}
  \label{fig:gates}
\end{figure}

We now turn to the particular quantum circuit model used in this thesis.
Our circuits will use the following slightly redundant
set of gates, in keeping with \cite{cleve00fast}.  
The Hadamard,
\begin{equation}\label{eq:had}
H=\left(\begin{array}{cc}\frac{1}{\sqrt{2}} &\frac{1}{\sqrt{2}} \\
\frac{1}{\sqrt{2}} &-\frac{1}{\sqrt{2}}\end{array}\right).
\end{equation}
The single qubit rotation gates,
\begin{equation}
R_{k}=\left(\begin{array}{cc}1 & 0\\
0 &e^{2\pi i/2^k}\end{array}\right).
\end{equation}
And, finally, the $2$-qubit controlled rotation 
gates which perform the rotation $R_k$
if and only if the control bit is a $1$. These
three types of gates are shown in Figure (\ref{fig:gates}).
For simplicity we assume that we are also able to run these gates
in reverse.  Multiple gates may be performed in parallel (i.e. to distinct
sets of bits) at any given stage, allowing for 
both size (total number of gates)
and depth (number of stages) analyses of our
algorithms. 
In our discussion of the hidden subgroup Problem
we will require
quantum circuits which have oracle access to some function
$f$. For our purposes we can assume without loss of generality 
that our input includes
a special string of clean (all zero) qubits which are left untouched
throughout the algorithm except for a single function
call.   At this point the oracle is invoked and 
the result is copied into the string of clean bits.
A more general model would have to allow for multiple calls to the oracle
and for manipulations of the resulting strings, but this restricted
version suffices for our purposes.
\subsection{Arithmetic Quantum Circuits}\label{ssect:arith}
It will be useful for us in presenting
our results to build a small repetoire of important subcircuits,
in particular, quantum circuits for basic arithmetic
operations.  The following two lemmas
allow us to translate classical results about
arithmetic circuits to the quantum setting.

\begin{lemma}\label{lem:unreverse}
Suppose the map 
$$\ket{x}\longrightarrow\ket{f(x)}$$
is computable by classical deterministic circuits of
size $s(n)$ and depth $d(n)$.  Then the map 
$$\ket{x}\ket{0}\longrightarrow\ket{x}\ket{f(x)}$$
is computable by quantum circuits of size and depth
$O\left(s(n)\right)$
and $O\left(d(n)\right)$ respectively. 
\end{lemma}

\begin{lemma}\label{lem:reverse}
Suppose that $f$ is 1-1 and the maps
$$\begin{array}{lcr}
\ket{x}\longrightarrow\ket{f(x)}&\mbox{and}&\ket{x}\longrightarrow\ket{f^{-1}(x)}
\end{array}$$
are computable by classical deterministic circuits of
size $s_1(n)$ and $s_2(n)$ and depth $d_1(n)$ and $d_2(n)$ respectively.  
Then the map 
$$\ket{x}\longrightarrow\ket{f(x)}$$
is computable by quantum circuits of size $O\left(s_1(n)+s_2(n)\right)$
and depth $O(d_1(n)+d_2(n))$.
\end{lemma}
These lemmas were originally proved in the context of
classical reversible computation \cite{Bennett} 
-- there is nothing inherently
``quantum'' about their proofs, which we proceed to
sketch.  

\begin{figure}[hbt]
  \begin{center} \psfig{file=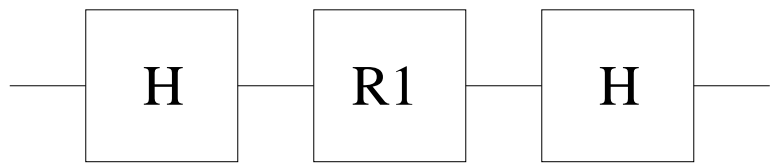,width=4in} \end{center}
  \caption{Quantum Not.}
  \label{fig:not}
\end{figure}

\begin{figure}[hbt]
  \begin{center} \psfig{file=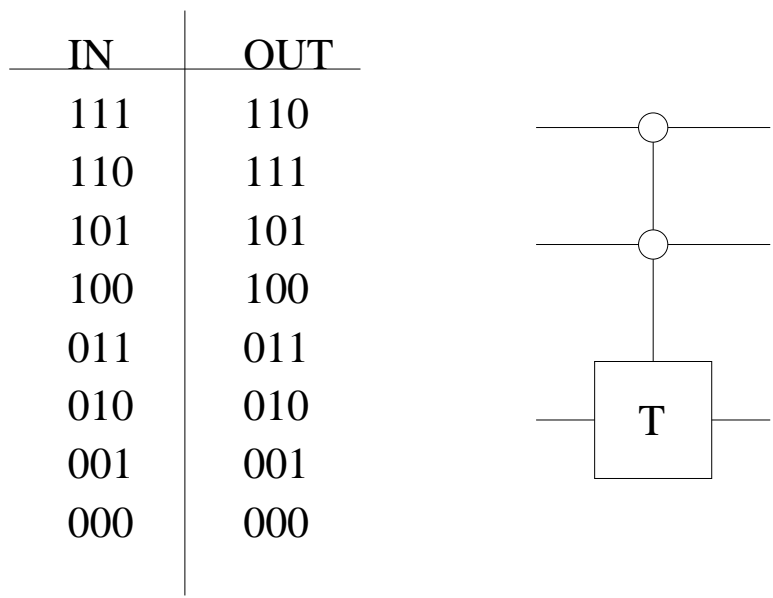,width=4in} \end{center}
  \caption{Toffoli Gate.}
  \label{fig:toff}
\end{figure}

The first step in 
constructing a quantum (or classical reversible) circuit 
from a deterministic one is developing subcircuits
which can simulate a universal set of classical boolean gates, such as
NOT and AND.  Simulating the classical NOT gate is easy since
it is already reversible (Figure \ref{fig:not}).    
Simulating the classical AND gate proves trickier.
The three-qubit transformation pictured with
its truth table in Figure \ref{fig:toff} can be accomplished
in constant size and depth by a surprisingly complicated configuration of
our basic gates (See, for example, \cite{QuantumBook}, page 182).
 
This gate is sometimes
referred to as the controlled-controlled not, since it
performs a controlled-not on the last two qubits if and
only if the first is a $1$, but more often it is called
the Toffolli gate in reference to \cite{Toffoli-80a}.  
It is easy to see
from its truth table representation that, if the third 
qubit is set to $\ket{0}$, the $AND$ of the first two input
qubits is recorded in the third output.

With these two subcircuits now in hand, suppose we are
given a classical circuit computing our function $f$.  We
replace each NOT gate by subcircuit (\ref{fig:not})
and each AND by subcircuit (\ref{fig:toff}) supplemented
with a clean qubit in its third register.
This necessitates a supply of at most $s(n)$
clean qubits and yields a quantum circuit mapping
$$\ket{x}\ket{0}\longrightarrow\ket{x}\ket{junk_x}\ket{f(x)},$$
where $junk_x$ are the junk-bits output by the first two registers
of each Toffolli.
The size and depth of this portion of the circuit are 
related to the classical circuit by constants
deriving from the size and depth of 
the subcircuits (\ref{fig:not}) and (\ref{fig:toff}). 
 
\begin{figure}[hbt]
  \begin{center} \psfig{file=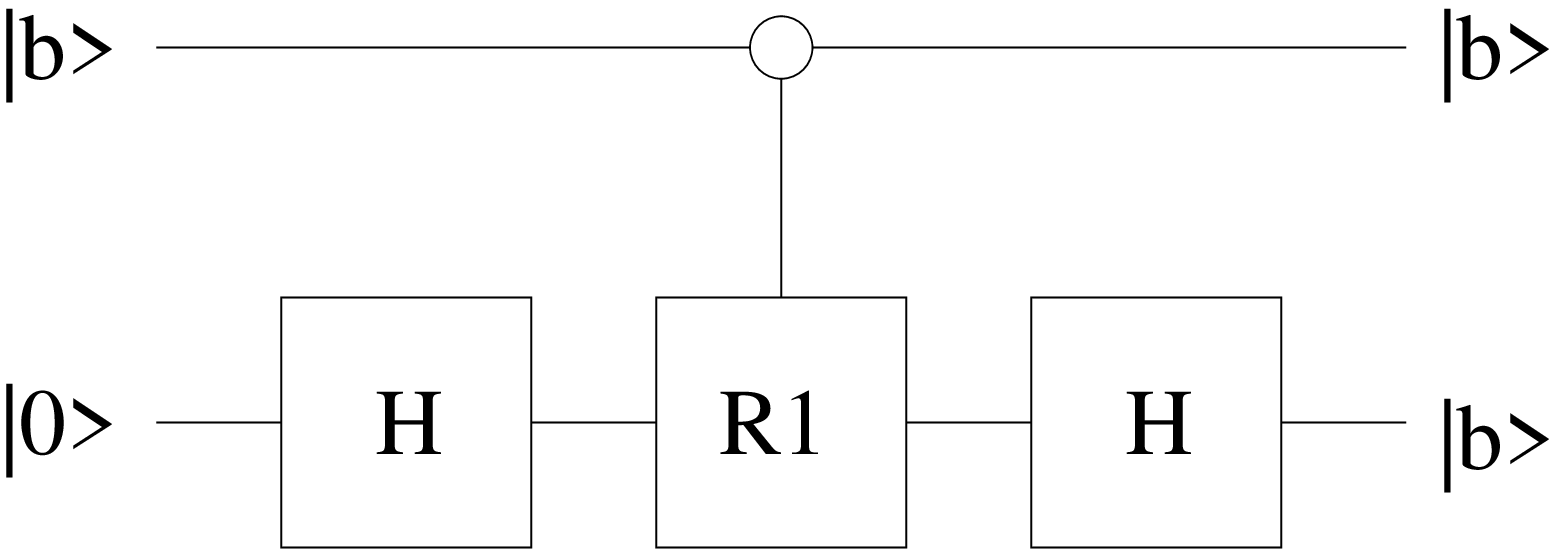,width=2in}\end{center} 
  \begin{center}\psfig{file=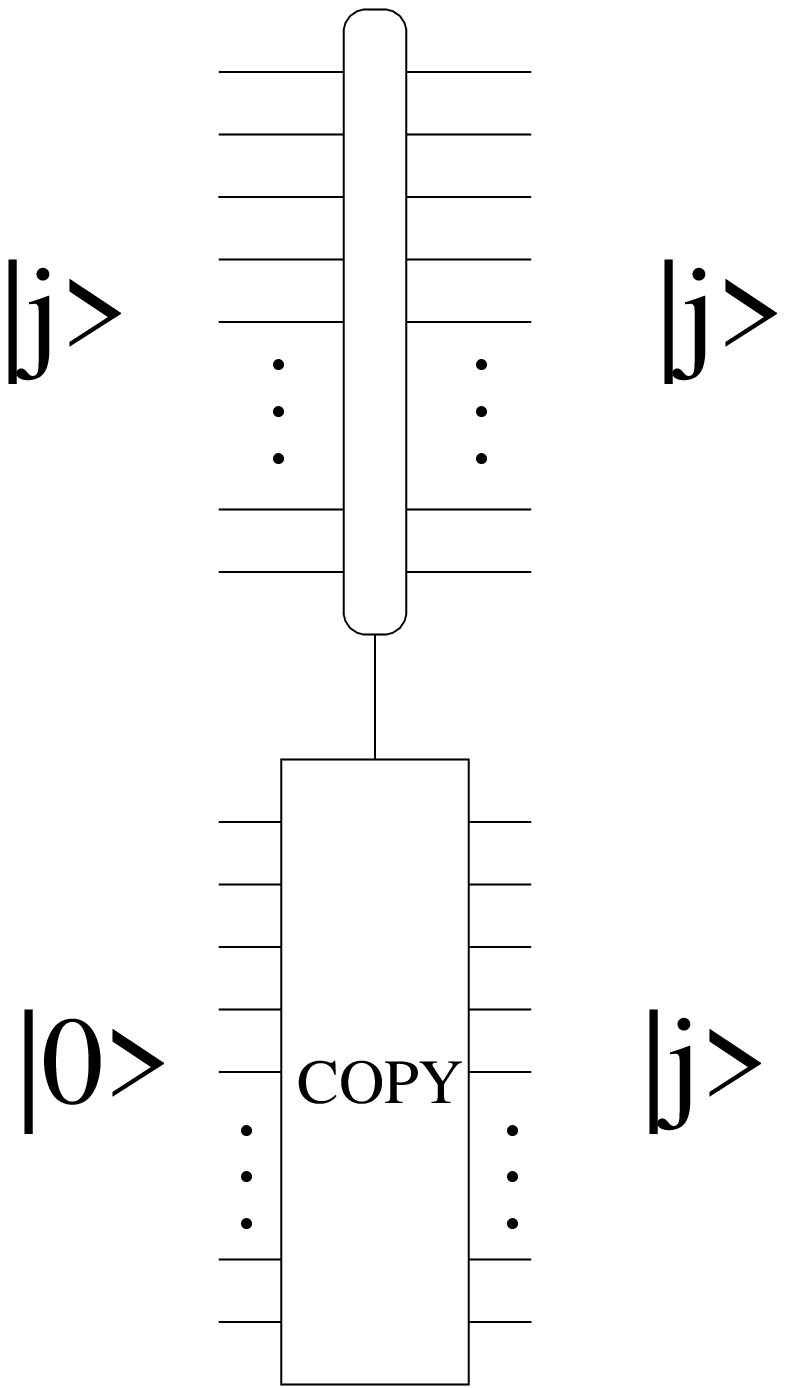,width=1.5in}\end{center}
  \caption{Controlled Not and Quantum Copy.}
  \label{fig:cnot}
\end{figure}

$f(x)$ is then copied into a remaining set of clean bits,
$$\ket{x}\ket{junk_x}\ket{f(x)}\ket{0}
\longrightarrow\ket{x}\ket{junk_x}\ket{f(x)}\ket{f(x)}.$$
A single bit can be copied into a clean qubit via the
controlled-not subcircuit of Figure \ref{fig:cnot},
and we let COPY denote the 
important subcircuit of size 
$3n$ and depth $3$ consisting of $n$ of these
controlled-nots in parallel.

Finally, the initial computation is run in reverse, yielding the 
desired map
$$\ket{x}\ket{0}\longrightarrow\ket{x}\ket{0}\ket{f(x)}.$$
The size and depth bounds follow easily.

We now turn to the second lemma.  We have already shown how to
construct circuits performing both 
$$\ket{x}\ket{0}\longrightarrow\ket{x}\ket{f(x)}$$
and 
$$\ket{f(x)}\ket{0}\longrightarrow\ket{f(x)}\ket{x}.$$
These circuits need merely be composed -- the
second in reverse -- to obtain a circuit computing the 
desired
$$\ket{x}\ket{0}\longrightarrow\ket{0}\ket{f(x)}.$$

One drawback of this simple construction is the 
inflation of the number of qubits needed
to accomplish the computation in question.
In particular, since each Toffolli gate
requires at least one clean qubit, the 
number of qubits required is proportional
to the size of the classical circuits involved.
It is possible to improve this space
bound at the expense of the other parameters \cite{Bennett-89},
but we shall be primarily concerned with minimizing
the overall size and depth of our circuits.  See
also \cite{Vedral-et-al-95} for quantum arithmetic 
circuits constructed with an
emphasis on minimizing this space requirement.

\begin{figure}[hbt]
  \begin{center} \psfig{file=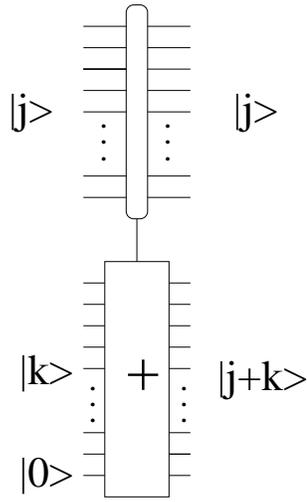,width=1.5in}
  \end{center}
  \caption{Quantum Addition}
  \label{fig:add}
\end{figure}

Addition and subtraction can both be accomplished
by classical circuits of size and depth $O(n)$ and $O(\log n)$
respectively.  By Lemma \ref{lem:reverse}, then, the quantum
addition circuit pictured in \ref{fig:add} also has size $O(n)$ 
and depth $O(\log n)$.
We shall also use a modular addition operation, denoted $+_N$, which
maps $\ket{j}\ket{k}$ to $\ket{j}\ket{j+k\bmod N}$ for
$j,k<N$.  It is easy to see that this bijection
can be accomplished with asymptotic
size and depth identical to regular addition.  Finally,
we will have occasion to run each of these circuits backwards,
performing $-$ and $-_N$ respectively.

\begin{figure}[hbt]
  \begin{center} \psfig{file=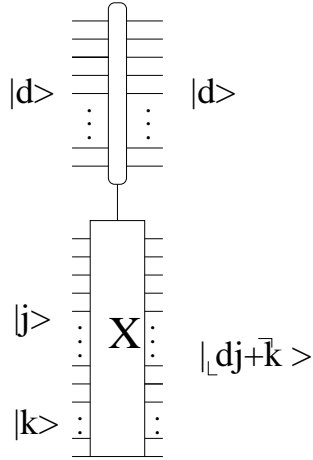,width=1.5in}
  \end{center}
  \caption{Quantum Multiplication with Remainder}
  \label{fig:mult_div}
\end{figure}

We shall also require two types of multiplication circuits.
The first, pictured in Figure \ref{fig:mult_div},
takes as input an $n$-bit decimal $d>1$, an $n$-bit integer $j$, and
an integer $k$ with $0\leq k<d$.  It outputs the nearest to $dj+k$
with ties broken by some consistent convention.  
The requirement $d>1$ ensures that this
map is a bijection and thus Lemma \ref{lem:reverse} can be invoked.  
We let $\div$ denote this multiplication circuit run in reverse.
Analyzing the circuit's size and depth
is more complicated than in the
case of addition.  
Currently the best classical circuits for multiplication 
have size $O(n\log n\log\log n)$ and depth $O(\log n)$.
But in order to apply Lemma \ref{lem:reverse} we must also
perform division, a notoriously stubborn operation
to parallelize.  There are classical division circuits
of size $O(n\log n\log\log n)$ 
which have depth $O(\log n\log\log n)$\cite{Reif1990}.
However, the smallest $O(\log n)$-depth classical division  
circuits have size $O(n^{1+\epsilon})$\cite{Hastad}.
Thus the quantum multiplication 
circuit (\ref{fig:mult_div}) can {\it either} be performed in simultaneous
size and depth $O(n\log n\log\log n)$ and $O(\log n\log\log n)$
{\it or} $O(n^{1+\epsilon})$ and $O(\log n)$.

\begin{figure}[hbt]
  \begin{center} \psfig{file=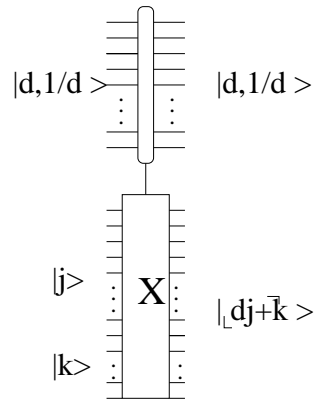,width=1.5in}
  \end{center}
  \caption{Quantum Multiplication with Inverse}
  \label{fig:multmult}
\end{figure}

If an $n$-bit approximation to $1/d$ is available
-- in Algorithm \ref{alg:approxft} we can prepare this inverse classically --
multiplication by $1/d$ can be substituted for division.
The multiplication circuit pictured in Figure \ref{fig:multmult} can thus be 
performed in simultaneous size and depth 
$O(n\log n\log\log n)$ and $O(\log n)$ respectively.

The classical multiplication and 
division techniques which achieve these various sub-quadratic
circuit sizes all make use of the discrete Fourier transform.
This raises the interesting question of whether there is some inherently 
quantum method which improves upon these techniques,
perhaps by using the QFT.
As noted in \cite{SICOMP::Shor1997} it could allow quantum
{\it de}coding of RSA encryption to run asymptotically 
faster than the corresponding classical
{\it en}coding.

\label{chap:prelim}
\chapter{Quantum Fourier Transforms and The Hidden Subgroup Problem}
\label{chap:qft_hsp}
\section{The Discrete Fourier Transform}

Let $G$ be a finite abelian group and let $V$ be the
vector space over the complex numbers consisting of 
formal linear combinations of elements of $G$,
$$\ket{v}=\sum_{g\in G}v_g\ket{g}.$$
We use $*$ to denote the group convolution operation 
induced by $G$ on $V$, that is, the operation
$$\ket{v}*\ket{w}=\sum_{g\in G}\left(\sum_{hk=g}v_hw_k\right)\ket{g}.$$
Notice that the $\ket{g}\in G$ form a group under this
operation which is trivially isomorphic
to $G$ itself.  

The discrete Fourier transform, or DFT, is a symmetric
unitary transformation $F$ of $V$ satisfying
\begin{equation}\label{FTG}
cF\left(\ket{g}*\ket{h}\right)=F\ket{g}\bcdot
F\ket{h}
\end{equation}
for all $g,h\in G$,
where $\bcdot$ denotes pointwise vector multiplication
and $c$ is the normalization factor $1/\sqrt{|G|}$.
The DFT thus exhibits a group isomorphism between the
$\ket{g}\in G$ under group convolution
and the $\ket{F(g)}\in G$ under pointwise
multiplication.
This characterization of the DFT is sufficient for the applications
discussed in this thesis.  For the definition of the 
DFT in terms of group representations, still in
the setting of quantum computation, see \cite{QuantumBook}.

\subsubsection{Cyclic G}

If $G=Z_N$, the cyclic group on $N$ elements, then 
the transformation
$$\ket{j}\longrightarrow \ket{k}\mbox{ with amplitude }
            \frac{1}{\sqrt{N}}\om{N}{jk}.$$
where $\omega_N=e^{\frac{2\pi i}{N}}$
satisfies (\ref{FTG}).
In fact, $\omega_N$ could be replaced
with any primitive $N$th root of unity and the resulting map
would still satisfy this condition.  It is not hard to show that these
are the only such transformations and thus
our characterization yields a unique
DFT up to isomorphism of the underlying cyclic group. 

\begin{example}\label{ex:ztwo}
A simple example is the DFT over $Z_2$,
which is the map sending 
$$\ket{0}\longrightarrow
\frac{1}{\sqrt{2}}\om{2}{0\cdot 0}\ket{0}+
\frac{1}{\sqrt{2}}\om{2}{0\cdot 1}\ket{1}=
\frac{1}{\sqrt{2}}\ket{0}+\frac{1}{\sqrt{2}}\ket{1}.$$
and
$$\ket{0}\longrightarrow
\frac{1}{\sqrt{2}}\om{2}{1\cdot 0}\ket{0}+
\frac{1}{\sqrt{2}}\om{2}{1\cdot 1}\ket{1}=
\frac{1}{\sqrt{2}}\ket{0}-\frac{1}{\sqrt{2}}\ket{1}.$$
The matrix representation of this map is thus
\begin{equation}
\left(\begin{array}{cc}\frac{1}{\sqrt{2}} &\frac{1}{\sqrt{2}} \\
\frac{1}{\sqrt{2}} &-\frac{1}{\sqrt{2}}\end{array}\right).
\end{equation}
\end{example}

\subsubsection{Finite Abelian G}
By the Fundamental Theorem on finite Abelian groups, 
any such $G$ can be decomposed as a direct
product of cyclic subgroups.  Its DFT is 
the tensor product of the DFT's corresponding
to each cyclic subgroup in this decomposition.
Again it is possible to show that there is
a unique map satisfying (\ref{FTG}) up to isomorphism of the
underlying group.  Before giving a description of these
maps we present a simple example.

\begin{example}
The simplest example of this tensor product construction
is the DFT over $\zton =(Z_2)^n$, sending
$$\ket{j}\longrightarrow\ket{k}$$
with amplitude
$$\prod_{i<n}\frac{1}{\sqrt 2}\om{2}{j_ik_i}=
\frac{1}{\sqrt{2^n}}\om{}{j\cdot_2 k},$$
where $\cdot_2$ denotes the mod $2$ dot product.
\end{example}

More generally if $G=\zpp$ is an arbitrary finite
abelian group given by its decomposition
as a direct product of cyclic groups
we can describe the DFT over $G$ in a uniform manner.
We first define the mod $G$ dot product,
$\cdot_G$ as follows.
\begin{definition}
Suppose $G=\zpp$.  Let $P=|G|=p_1p_2\dotsb p_n$ and $P_i=P/p_i$
Then $\cdot_G$ is the
binary operation on $G$ given by
$$g\cdot_G h =\frac{1}{P}\left[\left(\sum_{0<j\leq n}P_jg_jh_j\right)\bmod P\right],$$
where $g$ and $h$ equal $(g_1, g_2,\dotsc,g_n)$
and $(h_1, h_2,\dotsc, h_n)$ respectively.
\end{definition}
The value of this definition is that the DFT over $G$
can now be simply described as 
sending 
$$\ket{g}\longrightarrow\ket{h}\mbox{ with amplitude }
\frac{1}{\sqrt {|G|}}\om{}{g\cdot_G h}.$$

\subsubsection{DFT vs. QFT}
The classical task of computing the DFT over $Z_N$ of an
explicitly given vector of complex numbers $v=(v_1,v_2,\dots,v_N)$,
a task which naively appears to require $O(N^2)$ arithmetic operations,
can actually be accomplished in $O(N\log N)$ arithmetic operations,
by techniques referred to as the fast Fourier transform, or FFT
(See Sections \ref{ssect:FFT} and \ref{sect:qchirp}).
This nontrivial algorithm, together with the
fact that the DFT maps $*$, i.e. group convolution, to $\cdot$,
pointwise multiplication, is exploited by the many 
classical applications of the DFT, 
such as Fast Polynomial and Integer Multiplication.

In contrast to this classical computational task, the
quantum Fourier transform or QFT refers to the implementation
of the discrete Fourier transform
on the underlying quantum state space. In other words, the input is
not an explicit vector
of complex values, but a quantum state 
$$\keta=\sum_{i<N}\alpha_i\ket{i}$$
whose {\it amplitudes} 
represent the vector to be transformed.
The output of the QFT over $Z_{N}$ is
the quantum state
$$\ketha=\sum_{j<N}\hat\alpha_j\ket{j},$$
where 
$$\ha_j=\frac{1}{\sqrt{N}}\sum_{i<N}\alpha_i\om{N}{ij}.$$

\begin{example}
It is not hard to see that one of our basic quantum gates, namely
the Hadamard (See Equation \ref{eq:had} and Example \ref{ex:ztwo})
is precisely the 
QFT over $Z_2$.  Moreover, applying $n$ Hadamards
independently to each qubit as pictured in Figure
\ref{fig:zton} 
\begin{figure}[hbt]
  \begin{center} \psfig{file=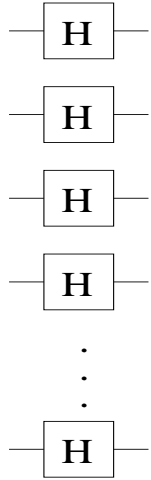,width=.75in,height=2.5in} \end{center}
  \caption{QFT over $(Z_2)^n$.}
  \label{fig:zton}
\end{figure}
accomplishes the 
QFT over $(Z_2)^n$, which we will denote $(F_2)^n$.
This is the first and simplest example of a polynomial
sized quantum circuit implementing the QFT
over an exponential-sized group.
\end{example}

The fact that the QFT over exponentially large groups
can be efficiently implemented is the basis for 
for all quantum algorithms achieving exponential
advantage over classical computation.  However,
it is important to notice that, in and of itself,
the ability to perform the QFT over an exponentially
large group does not represent an exponential speedup of any 
classical task, such as DFT computation. 
This contrast
between the classical DFT and the QFT has been likened to,
on the one hand, producing a list of all the probabilities of 
points in the sample space of some distribution (the classical DFT)
and, on the other hand, producing a method for sampling
from that distribution (the QFT).  
We now turn to a situation where the ability to compute
the quantum Fourier transform over an exponentially large
group {\it does} give quantum computation advantage over classical,
an astounding exponential advantage to be precise. 
\section{Simon's Algorithm}\label{sect:simon}
Simon \cite{SICOMP::Simon1997:1474} gave a polynomial-time 
quantum algorithm for the following
promise problem.

GIVEN: A function $f$ 
    defined on $(Z_2)^n$ which is 2-1 and satisfies $f(x)=f(x\oplus b)$
    for all $x$ and some fixed $b$.

FIND: $b$.

This is our first example of a hidden
subgroup problem.  The given function $f$ is defined on the
group $(Z_2)^n$ and constant on cosets of the unknown
subgroup $\{0,b\}$.
The goal is to reconstruct this subgroup.
The following 
quantum procedure is commonly referred to as Simon's algorithm.
It exploits a certain coset invariance property of
the QFT -- regardless of which coset $x$ of $\{0,b\}$ is input
to the QFT at Step \ref{QFT}, the output distributions are identical
The information about the particular coset is concentrated in the
complex phases of the final superposition, while its distribution
encodes just the underlying subgroup $\{0,b\}$.
\begin{algorithm}{Simon's Algorithm\footnote{In this and all later
quantum procedures we shall feel free to suppress global normalization
factors in order to preserve readability.}}
\label{alg:simon}

\begin{enumerate}
    
  \item\label{input} We prepare the input to the Fourier transform as
    follows:
    $$\ket{0}\ket{0}
\longrightarrow\sum_{x<2^n}\ket{x}\ket{0}
    \longrightarrow\sum_{x<2^n}\ket{x}\ket{f(x)}
    =\sum_{a\in R_f}\left(\ket{x}+\ket{x\oplus b}\right)\ket{a}$$
 where $R_f$ denotes the range of $f$.

  \item\label{QFT}Quantum Fourier Transform over $\zton$:
\begin{eqnarray*} 
\left(\ket{x}+\ket{x\oplus b}\right)\ket{a}
&\longrightarrow&\sum_{y<2^n}
            \left((-1)^{x\cdot y}+(-1)^{(x\oplus b)\cdot y}\right)
                \ket{y}\ket{a}\\
&=& \sum_{y, y\cdot_2 b=0}(-1)^{x\cdot y}\ket{y}\ket{a}, 
\end{eqnarray*}
 
where $\cdot_2$ denotes the mod $2$ dot product.
  \item Measure the first register.
  \end{enumerate}
Repeat this quantum subroutine $O(n)$ times and obtaining $\{y_i\}$.
Solve(classically) the system of equations $y_i\cdot_2 z=0$.  Output this
solution
\end{algorithm}
Our quantum subroutine outputs a $y_i$
uniformly at random from the set of $y$ such that
$y\cdot_2 b=0$.  It is not hard to show that after $O(n)$ 
repetitions of the subroutine the resulting system of linear equations
will have a unique solution with high probability and the
correctness of the algorithm follows.

It is possible to show that any classical probabilistic
algorithm for this problem has query complexity 
$\Omega(2^{\frac{n}{2}})$ \cite{SICOMP::Simon1997:1474},
thus this is an example of a (promise) problem
where quantum computation achieves exponential 
advantage over classical computation.  In fact,
we show in Chapter~\ref{chap:or}, Section~\ref{sect:ma} 
that, even in the presence of non-determinism,
any classical probabilistic method of
distinguishing functions that are $1-1$
on $(Z_2)^n$ from the $2-1$ functions described above
requires a similar exponential number of queries.
In contrast, this can be accomplished in
polynomial-time on a quantum computer by a slight
modification of Algorithm \ref{alg:simon}.

\section{Generalizing Simon's Algorithm: The Abelian Hidden Subgroup Problem}
\label{sect:gsa}
Algorithm \ref{alg:simon} is the prototype for all later hidden subgroup
algorithms, including Shor's 
celebrated algorithms factoring
and discrete logarithm\cite{SICOMP::Shor1997}.  
We reinterpret Algorithm \ref{alg:simon} in terms
of ``Fourier sampling'' over $G=\zton$, then show how
this procedure generalizes to an arbitrary finite Abelian group.
Our approach is similar to that of \cite{Jozsa}.
\begin{algorithm}\label{alg:hs}
\begin{enumerate}
\item\label{it:eqsup} Prepare $$\keta=\sum_{x\in G}\ket{x}\ket{\fh(x)}$$ where $f_H$
is constant and distinct on the cosets of $H\leq G$
\item\label{it:fs} Sample from $\dD{G}{\keta}$, the distribution
gotten by  measuring the first register of
$$F_{G}\keta=\sum_{x\in G}F_{G}\ket{x}\ket{\fh(x)},$$
where $F_G$ denotes the QFT over $G$.\end{enumerate}
Repeat this quantum subroutine 
$O(n^2)$ times where $n=\log |G|$ 
obtaining samples $\{y_i\}$.
Solve(classically) the system of equations $y_i\cdot_G z=0$.  Output this
solution
\end{algorithm}
We can describe the distribution sampled by the quantum procedure 
using the following definition:
\begin{definition}Let $G$ be a finite abelian group.  
For any subgroup $H\leq G$
let $\perH\leq G$ be the subgroup consisting of the elements
$g\in G$ satisfying
$$g\cdot_G h=0$$
for all $h\in H$.
\end{definition}
In the particular case $G=\zton$ and $H=\{0,b\}$, i.e. Simon's 
algorithm, we have already seen that the distribution
$\dD{G}{\keta}$ is supported uniformly on the subgroup $\perH$.
We now show that for {\it any} finite abelian $G$
and $H\leq G$, $\dD{G}{\keta}$ is uniformly
supported on $\perH$. 

Recall that 
$$\keta=\sum_{x\in G}\ket{x}\ket{\fh(x)}.$$
For any $g\in G$ let
$$\ket{g*\alpha}=\ket{g}*\keta=\frac{1}{\sqrt{|G|}}
\sum_{x\in G}\ket{g+x}\ket{\fh(x)},$$
be the convolution of $\ket{g}$ and the first register of $\ket{\alpha}$.
Then for any $h\in H$, 
\begin{equation}
cF_G\keta=cF_G\ket{h*\alpha}
=F_G\ket{h}\bcdot F_G\keta,
\end{equation}
for $c=\frac{1}{\sqrt{|G|}}$.  The first equality follows from the fact that,
since $\fh$ is constant on cosets of $H$, $\keta=\ket{h*\alpha}$,
and the second 
from our definition of the Fourier transform.
Since the amplitude of $F_G\ket{h}$
at $\ket{x}$ is $\frac{1}{\sqrt{|G|}}\om{}{x\cdot_G h}$ this equality can
hold only if $F_G\keta$
is supported on $\ket{x}$ satisfying $\om{}{x\cdot_G h}=1$
for all $h\in H$. This is precisely $\perH$ as claimed. 

Showing that the distribution is uniform on $\perH$ 
requires more detail.  Fix any state $\ket{y}\ket{a}$
with $y\in \perH$ and $a$ in the range of $\fh$.
The amplitude at this point
is determined by the Fourier transform of the superposition
$$\frac{1}{\sqrt{|G|}}
\sum_{f(x)=a}\ket{x}\ket{a}=
\frac{1}{\sqrt{|G|}}\sum_{h\in H}\ket{k+h}\ket{a}$$
for some fixed $k$, since $\fh$ is both constant and distinct
on the cosets of $H$.  The resulting amplitude at
$\ket{y}\ket{a}$ is thus
$$\frac{1}{|G|}\sum_{h\in H}\om{}{(k+h)\cdot_G y}
=\frac{1}{|G|}\om{}{k\cdot_G y}
\left(\sum_{h\in H}\om{}{h\cdot_G y}\right)=\om{}{k\cdot_G y}\frac{|H|}{|G|}$$
where the last equality follows from the assumption that $y\in \perH$.
Clearly the norm of this amplitude is independent of $y$ (and $a$),
whose influence is only seen in the complex phase $\om{}{k\cdot_G y}$,
and the probabilities arising from the squares
of these norms are thus uniformly distributed over $\perH$.

How many samples are required in order to generate $\perH$
and thus solve for the generators of $H$?  In the special case
of $G=\zton$ and $H=\{0,b\}$ a simple argument shows that 
$O(n)$ samples suffice:
For any $n$-bit $y\neq b$ the probability that a random
element of $x\in \{0,b\}^\perp$ satisfies $y\cdot_2 x=0$
is at most $1/2$.  Since there are $2^n-1$ such $y$
the probability that $b$ is not uniquely determined falls off as
$2^n(1/2)^t$ where $t$ is the number of samples.

In general we need to ensure that our samples
samples are not contained in any proper subgroup of $\perH$.
This is achieved after $O(n^2)$ samples where $n=\log |G|$.  
In particular, there are
at most $|G|^n<2^{n^2}$ such subgroups -- each is determined
by a set of at most $n$ generators chosen from $G$.
Moreover, each has size at most half of $\perH$.
Thus the probability that there exists one such
subgroup containing all $t$ samples decreases as
$$2^{n^2}\left(\frac{1}{2}\right)^t.$$

In theory, then, Algorithm \ref{alg:hs} 
solves the hidden subgroup problem
for any finite abelian group $G$.  But so far we have only seen
circuits implementing this procedure in the special
case $G=\zton$ (Figure \ref{fig:zton}).  
Step \ref{it:eqsup} requires the preparation of an equal superposition
over the group G and an evaluation of the function $\fh$.
This is easily accomplished. 
Generalizing Step \ref{it:fs}
hinges upon extending the class of groups with efficient QFT's.
More specifically, the class of cyclic groups with efficient
QFT's must be extended since these can be tensored together to produce the QFT
over an arbitrary finite abelian group.  We turn to this topic in the
next Chapter.

\chapter{Computing the Quantum Fourier Transform}\label{chap:compqft}
\section{The QFT over $Z_N$, $N$ Smooth}\label{sect:smooth}
Two separate methods emerged
for extending the class of cyclic groups with efficient QFT's.
The first, developed by Shor \cite{Shor94} and subsequently Cleve 
\cite{Cleve94},
was based on the recognition
that the component QFT's over $Z_2$ in the circuit for the 
QFT over $\zton$
could be replaced by QFT's over any sufficiently small
cyclic group.  In particular,
since any $n$-bit unitary operation can be approximated 
by exponential-size quantum
circuits via Theorem \ref{thm:approxun},
the QFT over $Z_{m}$ can always be approximated by a circuit
of size $O(m^2)$.   It follows that the QFT over
any group of the form $\zm{n}$
where the $m_i=O(n^k)$ can be efficiently computed.
\begin{figure}[hbt]
  \begin{center} \psfig{file=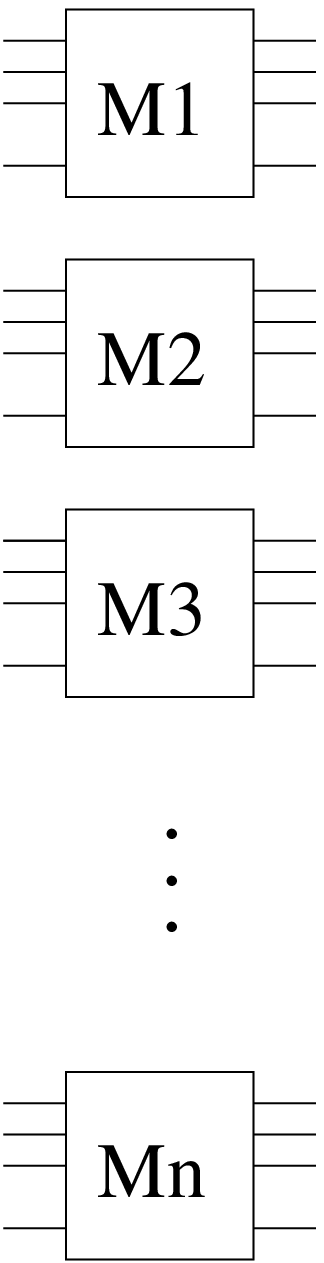,width=.75in,height=2.5in} \end{center}
  \caption{QFT over $\zm{n}$.}
  \label{fig:zmi}
\end{figure}
More importantly, this insight allows us to efficiently compute
the QFT over a special class of exponentially large cyclic groups.  
In particular, suppose that $N=m_1m_2...m_k$ where the $m_i$
are pairwise relatively prime.  By the Chinese remainder theorem
we have $Z_N\cong\zm{k}$
via the isomorphism

\begin{equation}\label{eq:iso}
a \bmod N\longrightarrow (a\bmod m_1,a\bmod m_2,\dotsc,a \bmod m_k).
\end{equation}
This isomorphism is easy to compute, and can be inverted as well
using the formula
\begin{equation}\label{eq:inv}
a \bmod N=\sum_{i<k}(a \bmod m_i)N_i\left(N_i^{-1}\bmod m_i\right),
\end{equation}
where $N_i=N/m_i$.

Thus if we are given an $N$ which factors into pairwise relatively
prime $m_i$ satisfying $m_i=O(\log^k N)$, we can compute the QFT mod $Z_N$
by first computing \ref{eq:iso}, then performing the QFT over 
$\zm{k}$, and then inverting by 
\ref{eq:inv}.  Obviously we need to know
the factorization $m_1m_2\cdots m_k$ but
these can easily be computed classically since they are so small.
This reasoning thus extended the class of groups whose
QFT could be efficiently implemented to include
cyclic groups $Z_N$ for $N$ smooth, i.e. with
prime power factors all of size $O(n=\log N)$\cite{Shor94},
and, more generally, for any $N$ with prime power factors
equal to $O(n^k)$\cite{Cleve94}.

\section{The QFT over $Z_{2^n}$}
The second method was developed independently
by Coppersmith \cite{Coppersmith1994} and Deutsch.
By exploiting the same recursive structure
of the DFT over $Z_{2^n}$
which leads to the classical FFT, the QFT
over $Z_{2^n}$ can be computed exactly by a quantum 
circuit of size and depth $n^2$.  As in the classical
setting the method can be
generalized with only minor changes 
to any $N=c^m$ where $c$ is a constant.  

The recursive structure of the DFT over $Z_{2^n}$
is encapsulated by the following product representation.
Let $j=j_1j_2\dotsb j_n$ is
be the bit representation of $j$
where $j_1$ is the most significant.  
Then the DFT over $Z_{2^n}$ can be written as
\begin{equation}\label{prod}
\ket{j}\longrightarrow\frac{
        \left(\ket{0}+\omega^{.j_n}\ket{1}\right)
        \left(\ket{0}+\omega^{.j_{n-1}j_n}\ket{1}\right)
        \dotsb
        \left(\ket{0}+\omega^{.j_1j_2\dotsb j_n}\ket{1}\right)}
{2^{n/2}}.
\end{equation}
We digress briefly to show how to derive the
classical FFT from this expression. 
\subsection{The Classical FFT}\label{ssect:FFT}
The classical FFT algorithm was first proposed
in \cite{cooltuk} but its motivation goes back to Gauss.
The product representation of Equation \ref{prod} lends itself
most nicely to the decimation-in-frequency, as opposed
to decimation-in-time, version
of the classical FFT.  Suppose the input
to the DFT is the vector $\ket{v}$.  
Let $\ket{w}$ and $\ket{z}$ be the length
$2^{n-1}$ vectors with amplitudes

$$w_{j}=v_{0j}+v_{1j}$$

and

$$z_{j}=\omega^{.0j}
        v_{0j}+\omega^{.1j}
        v_{1j}.$$

Then the amplitude of the product expression (\ref{prod}) 
at an 
even integer $\ket{k}=\ket{k_1k_2\dotsb k_n}$
is just the amplitude at 
$\ket{k_1k_2\dotsb k_{n-1}}$ of the
DFT over $2^{n-1}$ of $\ket{w}.$ 
Furthermore, the amplitude at an
odd $\ket{k}$
is just the amplitude at $\ket{k_1k_2\dotsb k_{n-1}}$
of the DFT over $2^{n-1}$ of $\ket{z}$.

The DFT over $2^n$ can then be gotten
by performing these 2 related DFT's over 
$2^{n-1}$.  Computing the vectors $\ket{w}$ and 
$\ket{z}$ from $\ket{v}$ 
requires $O(N)$ arithmetic operations.
Thus we get the recurrence relation 

$$T(N)=2T(N/2)+O(N)$$
for the arithmetic complexity of the classical
DFT over $N=2^n$, leading to
the well-known bound of $O(N\log N)$.

Recall our definition of the Fourier transform
as a map taking convolution to pointwise
multiplication and back again.  Computing the 
convolution of two vectors $\ket{v}$ and $\ket{w}$ 
in a brute force manner requires $2N$
arithmetic operations for each amplitude
$$\sum_{i<N}v_iw_{l-i}$$
and thus $2N^2$ for the vector as a whole.
If we instead perform an FFT, pointwise multiply,
then invert the FFT, only $O(N\log N)$ arithmetic
operations are involved. This rather
startling fact is the basis for the
well known fast polynomial and integer multiplication
algorithms \cite{schstr}.
\subsection{The QFT over $Z_{2^n}$}\label{ssect:qftton}
The product representation (\ref{prod}) leads even more directly to the
following quantum gate array which computes the QFT mod $Z_{2^n}$.
exactly.
\begin{figure}[hbt]
  \begin{center} \psfig{file=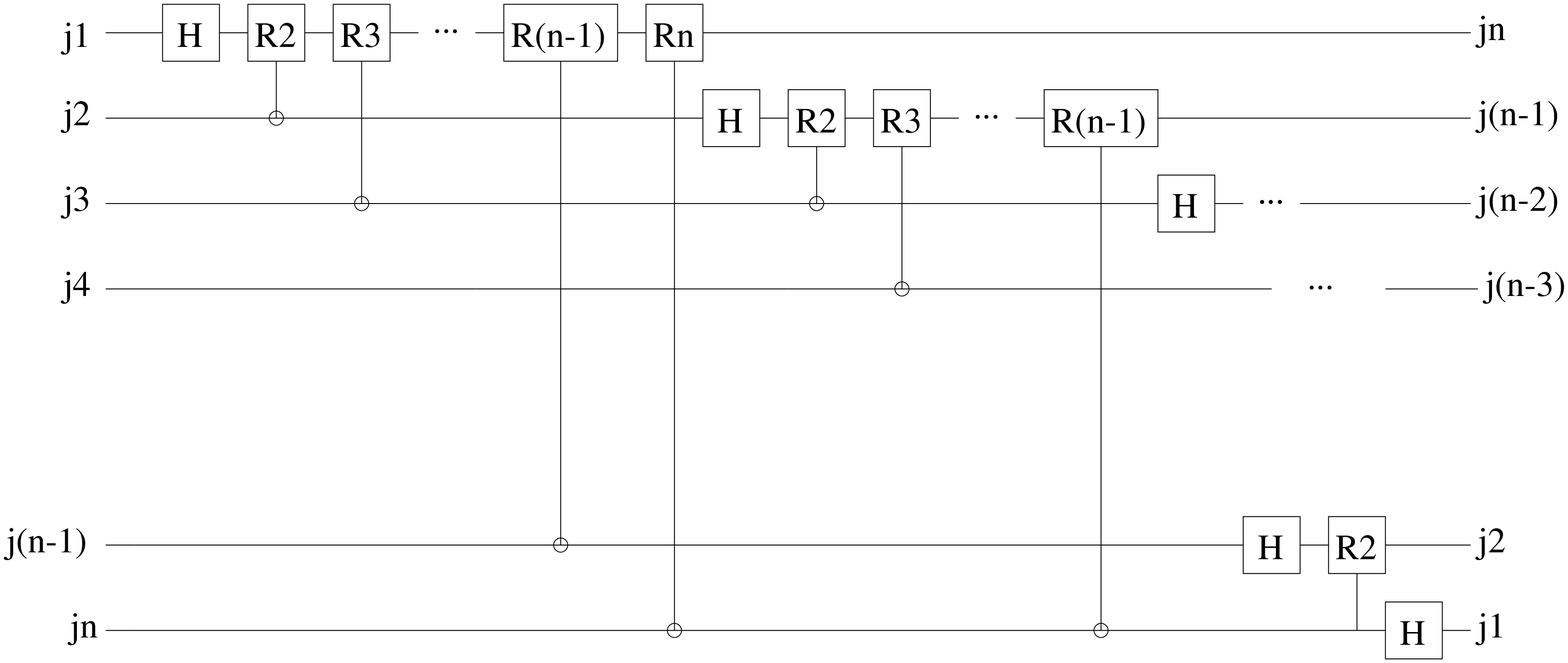,width=4in} \end{center}
  \caption{QFT over $Z_{2^n}$.}
  \label{fig:cqft}
\end{figure}
This gate array has size and depth
$\frac{n(n+1)}{2}=O(n^2)$.  The gates can easily be rearranged
so that the circuit has depth $O(n)$\cite{Moore}.  In particular,
let $G_{ij}$ for $i<j$ denote the controlled-rotation gates $R_{j-i+1}$ 
whose inputs are the $ith$ and $jth$ wires in Circuit \ref{fig:cqft}. 
Also let $G_{ii}$
be the Hadamard gate which is applied to the $ith$ bit.  It is not hard to see
the only requirement imposed by the above circuit is that
whenever $i+j<\pri+\prj$, $G_{ij}$ must precede $G_{\pri\prj}$
in the computation.  By arranging the gates in $2n-1$ stages,
where at the $k$th stage the all the gates $G_{ij}$ with $i+j=k$
are performed in parallel, the exact QFT over $2^n$
can be performed in size $\frac{n(n+1)}{2}=O(n^2)$ and depth $2n+1=O(n)$.

In practice, we are interested in merely
approximating the QFT to within an arbitrary 
inverse polynomial.  Since
most of the rotation gates in the
Circuit \ref{fig:cqft}are very small, by
just omitting the rotations in $O(\epsilon/n^2)$
a circuit of size and depth 
$O(n\log\frac{n}{\epsilon})$
which approximates the QFT over $Z_{2^n}$ to within
$\epsilon$ can be achieved \cite{Coppersmith1994}.
In particular, when 
$\epsilon$ is an inverse polynomial this
gives a gate array of
size $O\left(n\log n\right)$.  Since such approximations
suffice for any polynomial-time
computation, this is a clear
benefit of this recursive technique
over the technique of Section \ref{sect:smooth}
for which there is no similar
approximation technique.
Unfortunately, this benefit applies only to the
size of the circuits -- 
the depth of the parallel version of Circuit \ref{fig:cqft}
outlined in the previous paragraph is not further
reduced by this omission of gates.

Shor's algorithms for factoring and discrete log
can be based on either the QFT over $Z_N$ for smooth $N$
or the QFT over $Z_{2^n}$,  but the inability to transform
over an arbitrary cyclic group complicates their proof.
While there is no direct way to extend either of these
methods to encompass a larger class of cyclic groups,
the reliance of the QFT over $Z_{2^n}$ on the insight
which led to the FFT over the same domain raises a
natural question.  We have an generalized classical
FFT algorithm, i.e. an algorithm for computing
the classical DFT over $Z_N$ for arbitrary $N$ which   
has arithmetic complexity $O(N\log N)$, identical to the
standard FFT over a power of two.  Why not try to
base a QFT on these classical methods?
\section{Quantum Chirp-Z.}\label{sect:qchirp}
Since the circuit performing 
the QFT over a power of
2 is derived from the classical FFT
circuit, it is natural to try to derive a 
circuit for the QFT over a general modulus 
from the corresponding general modulus
classical method.  We first 
review this method, known as the chirp-z transform and attributed to 
Rabiner et al \cite{Rab}.  We then translate
this approach to the quantum setting 
and show that with a slight modification
we do obtain an efficient $\epsilon$-approximate QFT
which succeeds with probability $\epsilon^2$. On the one hand,
this is not strong enough to be useful in a general setting --
in particular, if an algorithm involves more than a constant number of 
QFT's replacing them all with these approximations would reduce the
success probability to below an inverse polynomial. On the other
hand, all the hidden subgroup algorithms to date use
just a constant number of QFT's in each quantum subroutine
and thus this method could be used.  It will not be as efficient
as the Eigenvalue Estimation procedure of
Section~\ref{sect:ee} and our Algorithm~\ref{alg:approxft} 
but may be of independent interest.

The classical chirp-z transform is essentially 
a method of reducing the 
transform over an arbitrary modulus 
to a combination of multiplication
and convolution.  The net result is 
that the transform over an arbitrary
modulus $N$ with $n=\lfloor\log N\rfloor$ 
can be accomplished 
via 3 FFT's over $2^{n+2}$ 
together with $O(N)$ extra arithmetic 
operations.  Thus the asymptotic
arithmetic complexity of the general 
modulus DFT is the same as that
of a power of two, namely $O(N\log N)$. 

We now describe this in some detail.  
Given $\ket{a}=\sum_{i<N}a_i\ket{i}$ we wish to compute 
the vector $\fta$ where 
$$\hat{a}_j=\ftc{j}{N}{i}{a}.$$     
We let 
$$\ket{b}=\sum_{i<N}a_i\om{N}{-i^2/2}\ket{i}
\mbox{ and }\ket{c}=\sum_{i<2^{n+1}}\om{N}{i^2/2}\ket{i}.$$
Clearly $\ket{b}$ can be generated from $\ket{a}$, and $\ket{c}$ created,
using $O(N)$
arithmetic operations.

The crucial insight is that $k$th
convolution coefficient of $\ket{b}$
and $\ket{c}$, 
$$d_k=\sum_{i<2^{n+1}}b_ic_{k-i},$$
satisfies 
$$d_k\om{N}{-k^2/2}=\ftc{k}{N}{i}{a}=\hat{a}_{k-N}$$
whenever $k\geq N$.  
Thus the convolution vector $\ket{d}$
can be used to produce the desired vector $\fta$ via $O(N)$ arithmetic
operations.  
As discussed in Section \ref{ssect:FFT}, the 
convolution vector $\ket{d}$ is obtained by computing the 
FFT mod $2^{n+1}$ of the vectors 
$\ket{b}$ and $\ket{c}$, pointwise
multiplying the two resulting vectors, and 
computing the inverse FFT
of this product.  

This method uses $O(N)$ arithmetic operations
to create the vectors $\ket{b}$ and 
$\ket{c}$, perform the pointwise
multiplications which are sandwiched 
between the FFT's, and recover
the Fourier coefficients from the 
convolution coefficients. Moreover it involves
a total of three FFT's 
over $2^{n+1}$,  leading to an
overall arithmetic complexity 
of $O\left(N\log N\right)$.

Is it possible to implement this type 
of convolution reduction
in the quantum setting? Recall that 
in this case we are given as input the
superposition 
$\keta=\sum_{i<N}\alpha_i\ket{i}$ 
and we wish to output 
the superposition $\ketha$ where 
$$\ha_i=\ftc{j}{N}{i}{\alpha}.$$
The superpositions analogous to $\ket{b}$ and $\ket{c}$
above, namely
$$\ket{\beta}=\sum_{i<N}\alpha_i\om{N}{-i^2/2}\ket{i}
\mbox{ and }\ket{\gamma}=\sum_{i<2^{n+1}}\om{N}{i^2/2}\ket{i},$$
can be created easily.  For example,
the map
$$\keta\longrightarrow\ket{\beta}$$
is achieved by computing $\frac{-i^2}{2N}$, putting this value
into the phase, and then erasing it. 

Convolution of these two superpositions poses
a problem.
We can perform the required QFT's over $2^{n+2}$
yielding the superposition
\begin{equation}\label{eq:conv}
\ket{\hat{\beta}}\ket{\hat{\gamma}}=
\sum_{i,j<2^{n+2}}\hat{\beta_i}\hat{\gamma_j}\ket{i}\ket{j}.
\end{equation}
We desire the superposition 
$\sum_{i<2^{n+2}}\hat{\beta_i}\hat{\gamma_i}\ket{i}$ corresponding
to the pointwise multiplication of
$\ket{\hat{\beta}}$ and $\ket{\hat{\gamma}}$, but the best we can do is to
subtract the first register in \ref{eq:conv} from the second and measure this
difference,
yielding
\begin{equation}
\sum_{i<2^{n+2}}\hat{\beta_i}\hat{\gamma_j}\ket{i}\ket{j-i}
\end{equation}
for each possible $(j-i)$ with equal probability.  
If $(j-i)=0$ then this is the desired superposition and
taking the inverse QFT over $2^{n+2}$
completes the convolution.  We then finish
the algorithm by collapsing the superposition 
to the interval $\{N,\dots,2N-1\}$ and shifting the phase
at $\ket{l}$ by $\om{N}{-l^2/2}$.

But the probability that $(j-i)=0$
is exponentially small.  More likely
the value in the second register will be some
non-zero $h=j-i$. Then
the output of the algorithm corresponds 
to having convolved, instead of
the desired superpositions $\ket{\beta}$ and $\ket{\gamma}$, the
superpositions $\ket{\beta^\prime}$ and $\ket{\gamma}$
where 
$$\beta^\prime_i=\alpha_i\om{2^{n+1}}{ih}\om{N}{-i^2}.$$
Thus if we collapse to the appropriate
interval $\{N,\dots,2N-1\}$ and shift phases as described above
we will have computed the transform over $N$ of the superposition
$\ket{\alpha^\prime}$ where
$$\alpha^\prime_i=\alpha_i\om{2^{n+1}}{ih}$$
instead of the desired $\ketha$.  

Now, if
\begin{equation}\label{eq:err}
\left|\om{2^{n+1}}{h}-\om{N}{k}\right|<\frac{\epsilon^2}{N}
\end{equation}
for some integer $k$ then the superposition
$\ket{\alpha^\prime}$ defined above and the superposition
$\ket{\alpha^k}$ with amplitudes
$$\alpha^\prime_i=\alpha_i\om{N}{ik}$$
have distance at most $O(\epsilon)$.  This is most easily
be seen by observing that their inner product is large.
The transform of $\ket{\alpha^k}$ over $N$ is just the 
shift$\pmod N$ by $k$ of $\ketha$, and thus 
the transform of $\ket{\alpha^\prime}$ over $N$ shifted by $k$
is $O(\epsilon)$-close to the desired $\ketha$ 
whenever Equation~\ref{eq:err} holds. 

Since the $k$ which minimizes the difference in Equation \ref{eq:err}
can be ascertained
from $h$, whenever this difference is suitably small we can
perform the required shift and achieve an $\epsilon$-approximation
to $\ketha$.
Since the condition of Equation~\ref{eq:err} holds for 
a $\epsilon^2$ fraction of the $h$ the success probability
is as claimed.
\section{Eigenvalue Estimation}\label{sect:ee}
Kitaev \cite{Kitaev1995} 
gave the first algorithm approximating the QFT over an
arbitrary cyclic group based on his method of Eigenvalue
Estimation.  These techniques were further refined in 
(\cite{CEHMM},\cite{MosEke98},\cite{CEMM1998}).
Our presentation of this QFT algorithm merges some of
these later
refinements with Kitaev's original approach.  

We first note that we can perform the map
\begin{equation}\label{eq:fbs}
\ket{i}\ket{0}\longrightarrow\ket{i}
\sum_{j<N}\om{N}{ij}{\ket{j}}=\ket{i}\ket{\hat{i}}.
\end{equation} 
Specifically, we begin by putting the second register into
an equal superposition over an appropriately large interval and
computing $\frac{ij}{N}$. This
value is then placed into the phase and the computation of $\frac{ij}{N}$
is erased.

More interestingly, it is also possible to approximate the map
\begin{equation}\label{eq:ee}
\ket{\hat{i}}\ket{0}\longrightarrow\ket{\hat{i}}\ket{i}.
\end{equation}
By combining the map \ref{eq:fbs} with \ref{eq:ee} in reverse
we achieve an approximation to the
desired transform. 

\begin{figure}[hbt]
  \begin{center} \psfig{file=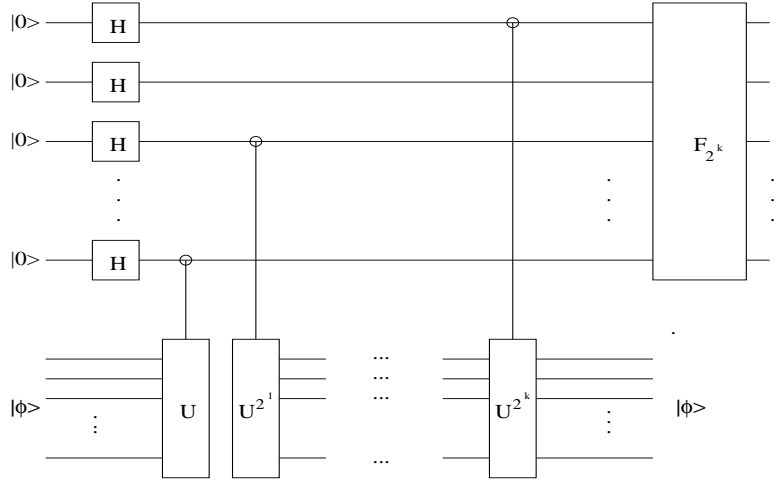,width=4in,height=2.5in} \end{center}
  \caption{Eigenvalue Estimation.}
  \label{fig:eigest}
\end{figure}

Map \ref{eq:ee} is based upon a procedure for estimating the
eigenvalues of a unitary operator.  More specifically,
suppose that we are able to perform the operations
controlled-$U$, controlled-$U^2$, ... , controlled-$U^{2^k}$ 
for some unitary operator $U$.
Assume further that we are given an eigenvector $\ket{\phi}$
of $U$ with eigenvalue $\om{}{\lambda}$.  Circuit \ref{fig:eigest}  
allows us to determine the most significant bits
of $\lambda$ with high probability.  In particular, if
$\lambda$ is exactly $k$-bits then the input to the QFT in Circuit 
\ref{fig:eigest} is
exactly $F_{2^n}^{-1}\ket{\lambda}$ and the procedure produces $\lambda$
with probability $1$.  More generally, to achieve the first 
$m$ bits of $\lambda$ with probability at least $1-\epsilon$
it suffices to choose $k=m+O\left(\log (1/\epsilon)\right)$.
How does this enable us to approximate \ref{eq:ee}?
It is easy to see that the 
the Fourier basis state $\ket{\hat{i}}$ of the QFT over $Z_N$ 
is an eigenvector with eigenvalue $\om{}{i/N}$ of the unitary operator
$U=(+1\bmod N)$.  Thus we can use the above
circuit to recover $i/N$ from $\ket{\hat{i}}$
with high probability.  Multiplying this result by $N$ allows us to
approximate the map
\begin{equation}\label{eq:junk}
\ket{\hat{i}}\ket{0}\ket{0}\longrightarrow\ket{\hat{i}}\ket{i}\ket{junk_i}
\end{equation}
where the last bits are junk deriving from the rounding off of $i/N$
to its most significant bits.

The junk produced by a map such as \ref{eq:junk} can 
always be cleaned up
using the methods outlined in the proof of Lemma \ref{lem:unreverse},
yielding an approximation to \ref{eq:ee}.
In general, if the original map is accurate to within $\epsilon$, the 
junkless version produced by this method will be accurate
to within $\sqrt{N}\epsilon$.  In this particular case, however, since a
copy of the eigenvector $\ket{\hat{i}}$ 
is maintained throughout the computation, the 
errors produced will be orthogonal and maps \ref{eq:junk} and \ref{eq:ee}
will have the same error bound.  This seems to have been overlooked
in \cite{Kitaev1995} which mentions only the more general accuracy result.

This version of Kitaev's algorithm ostensibly has 
size and depth $O(n^2)$ and $O(n)$ respectively
matching the running time of $O(n^2)$
claimed in \cite{Kitaev1995}.

\chapter{Parallel Circuits for the 
Quantum Fourier Transform over $Z_{2^n}$}
\label{chap:pcs}
The question of which quantum procedures can be performed in
parallel, i.e. by circuits of polylogarithmic depth, is of both
theoretical and practical interest.  There
are simple, natural problems, such as computing the greatest 
common divisor of two integers, which have no known classical
parallelizations.  Finding parallel quantum circuits for such a problem
would further support and elucidate the apparent power
of quantum over classical computation.  On the practical
side,  parallel computations can significantly reduce the
computational cost of fault-tolerant implementations
of quantum algorithms.  In particular, a robust
model of computational noise
must assume that an error can occur
in a qubit at a given stage in time whether or not the
qubit is undergoing a gate transformation at that particular
stage.  Under this assumption the size of the fault-tolerant implementation
of a parallel circuit -- see for example \cite{QuantumBook}, Chapter 10 --
will be smaller than the fault-tolerant
implementation of the non-parallel version by as much as a factor of $O(n)$,
where $n$ is the number of qubits in the original non-parallel circuit.

We give explicit parallel circuits for
approximating the QFT over a power of $2$ to within
an arbitrary inverse polynomial.  The existence 
of such circuits with simultaneous size and depth
$O(n\log n)$ and $O(\log n)$ respectively was proved
in \cite{cleve00fast}.  Our construction simplifes their approach
and reduces the number of qubits
required from $O(n\log n)$ to $O(n)$.
In some sense this shows that the approximate
QFT is inherently parallel, since there
is no price to be paid for parallelization --
asymptotically  
the size {\it and} width of the parallel circuits 
are the same as the apparently optimal nonparallel construction.

Our construction uses three basic maps, each of which
can be approximated by shallow depth circuits.
The first is the map 
\begin{equation}\label{eq:QFS}
\ket{j}\ket{0}\longrightarrow\ket{j}\ket{\hat{j}},
\end{equation}
which we shall refer to as the quantum Fourier state
computation, QFS for short, in keeping with \cite{cleve00fast}.
The second is the map
\begin{equation}\label{eq:COPY}
\ket{\hat{j}}\ket{0}\longrightarrow\ket{\hat{j}}\ket{\hat{j}},
\end{equation}
which copies a Fourier basis state.
Last and most interesting is the map
\begin{equation}\label{eq:FPE}
\ket{j}\ket{\hat{j}}\ket{\hat{j}}\ket{\hat{j}}=
\ket{j}\ket{\hat{j}}^3
\longrightarrow\ket{0}\ket{\hat{j}}^3
\end{equation}
which erases the identity of a Fourier basis state
from just three copies of that state.  We refer to
this as Fourier phase estimation or FPE again in keeping with 
\cite{cleve00fast}.
It is easy to compose these maps to produce a QFT in the
following manner

$$\ket{j}\ket{0}
\stackrel{\mbox{\scriptsize QFS}}{\longrightarrow}\ket{j}\ket{\hat{j}}
\stackrel{\mbox{\scriptsize QCOPYx2}}{\longrightarrow}
\ket{j}\ket{\hat{j}}^3
\stackrel{\mbox{\scriptsize FPE}}{\longrightarrow}
\ket{0}\ket{\hat{j}}^3
\stackrel{\mbox{\scriptsize reverse QCOPYx2}}{\longrightarrow}
\ket{\hat{j}}\ket{0}$$

Shallow circuits for map~\ref{eq:COPY} and an approximation to map
\ref{eq:QFS} and were exhibited in
\cite{cleve00fast}.  Their method of Fourier phase estimation, however,
uses $O(\log n)$ copies
of the Fourier basis state $\ket{\hat{j}}$
to erase its identity $\ket{j}$. This required an
ancilla of $O(n\log n)$ qubits and also 
complicated the task of copying $\ket{\hat{j}}$ --
in order to make the required $O(\log n)$ copies in
parallel classical results about prefix addition
were required. By requiring only three copies
of the Fourier basis state in our Fourier phase estimation
we are able not only to reduce the qubit requirement but also to
simplify the circuits to the point of making them
explicit, modulo our basic repetoire of arithmetic
operations (see Section~\ref{ssect:arith}).  We first turn to
this new Fourier phase estimation procedure~\ref{eq:FPE},
then give the circuits for maps~\ref{eq:QFS} and \ref{eq:COPY},
and finally  show how to combine these with a simple 
preprocessing step to
achieve an adequate approximation.

\section{Fourier Phase Estimation}\label{sect:fpe}
\begin{figure}[hbt]
  \begin{center} \psfig{file=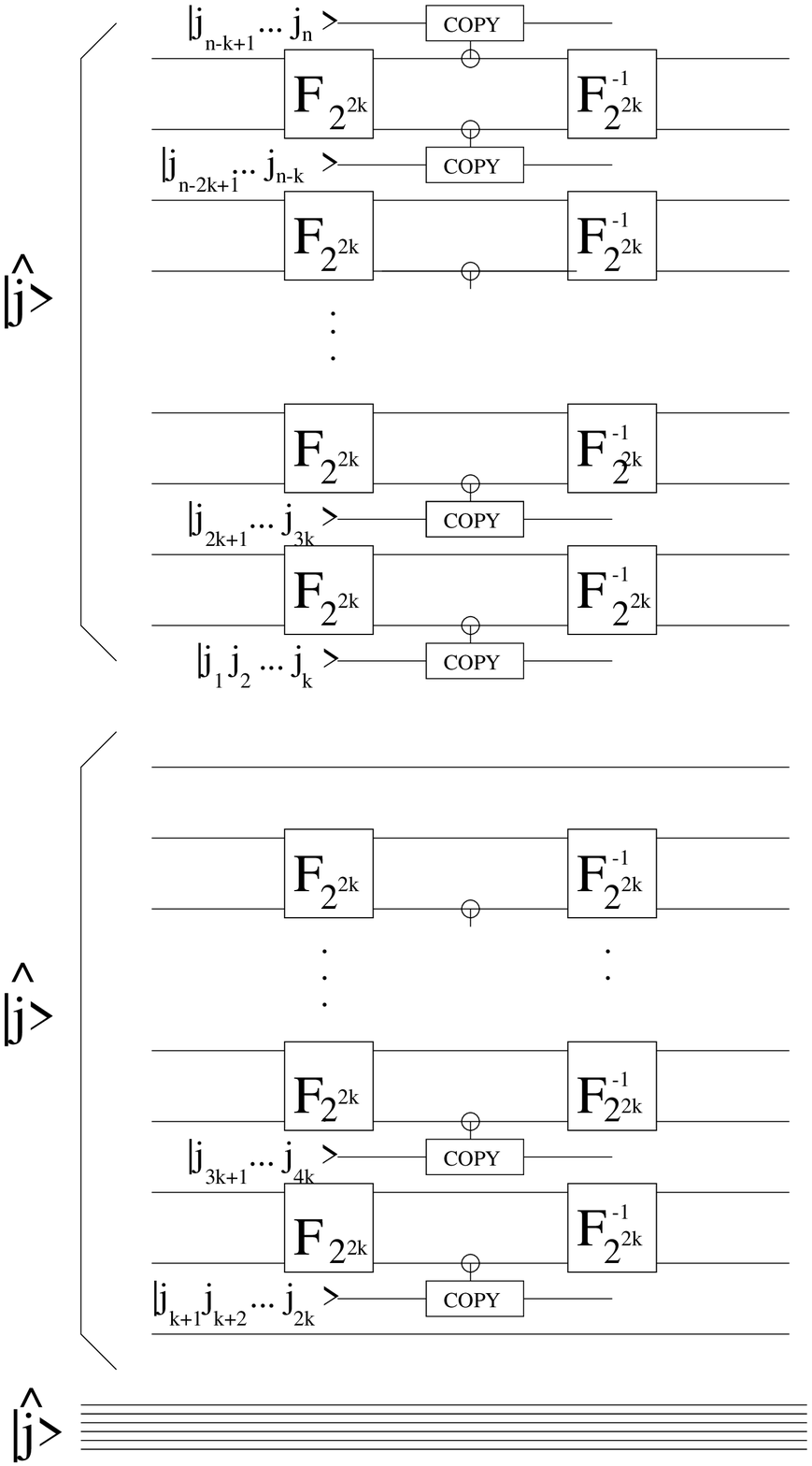,width=4in} \end{center}
  \caption{Quantum Fourier Phase Estimation (FPE): $\ket{j}\ket{\hat{j}}^3
\longrightarrow\ket{0}\ket{\hat{j}}^3$.}
  \label{fig:fpe}
\end{figure}

We now describe the circuit, pictured in Figure~\ref{fig:fpe},
which approximates the map
\begin{equation}
\ket{j}\ket{\hat{j}}^3
\longrightarrow
\ket{0}\ket{\hat{j}}^3.
\end{equation}
A collection of exact QFT's modulo
$2^{2k}$ for $k=O(\log n)$ are performed in parallel
on the bits of the first and second copies of the
Fourier basis state $\ket{\hat{j}}$.  We assume for simplicity that
$2k$ divides $n$. The first copy of the Fourier
basis state undergoes $n/2k$ QFT's modulo $2^{2k}$, 
applied in parallel to each consecutive sequence of $2k$ bits.  
The most significant $k$ bits output by each QFT
are used as an estimate for the corresponding bits
of $j$ and are thus xored into these bits to erase them.
The second copy of the Fourier basis state
undergoes $n/2k$ -$1$ QFT's modulo $2^{2k}$, 
applied in parallel to each consecutive sequence of $2k$
bits beginning with the $k+1$st bit.  
As before, the leading $k$ bits of each QFT
are xored into the corresponding bits of $j$.
The QFT computations are then reversed.
The third copy of the Fourier basis state is
left alone -- its sole purpose is to ensure 
the orthogonality of errors from distinct
basis states. 

Recall we can compute the exact QFT modulo
$2^l$ in size $l(l+1)/2$ and depth $2l-1$ as discussed in 
Section~\ref{ssect:qftton}.
Thus the above computation has depth $8k$ and size $O(kn)$.
To analyze its error we must examine the input and output
of each QFT modulo $2^{2k}$. 
Without loss of generality
we look at the topmost QFT which is applied to the 
first $2k$ bits of $\ket{\hat {j}}$, i.e the input is 
$$\frac{1}{2^k}
\left(\ket{0}+
\omega^{.j_1j_2\dotsb j_{2k}j_{2k+1}\dotsb j_n}\ket{1}\right)
\left(\ket{0}+
\omega^{.j_2\dotsb j_{2k}j_{2k+1}\dotsb j_n}\ket{1}\right)\dotsb
\left(\ket{0}+
\omega^{.j_{2k}j_{2k+1}\dotsb j_n}\ket{1}\right).
$$
The output of the QFT modulo $2^{2k}$
on this input is a smeared pointmass concentrated at integers
near the decimal $j_1j_2\dotsb j_{2k}.j_{2k+1}\dotsb j_n$.
In particular, its amplitude at $\ket{x}$ is
$$\frac{1}{2^{2k}}\sum_{l<2^{2k}}
\om{2^{2k}}{l(x-j_1j_2\dotsb j_{2k}.j_{2k+1}\dotsb j_n)}.$$
This is the sum of $2^{2k}$ equally spaced vectors which wrap around the
unit circle $$|x-j_1j_2\dotsb j_{2k}.j_{2k+1}\dotsb j_n|_{2^{2k}}$$ times 
where $\norm{\cdot}_{2^{2k}}$ is distance mod $2^{2k}$.  
The complete revolutions effectively cancel out and the only contributions
to the final amplitude come from the last fractional revolution.
There are $2^{2k}/|x-j_1j_2\dotsb j_{2k}.j_{2k+1}\dotsb j_n|_{2^{2k}}$
vectors in this fractional revolution, and each has length $\frac{1}{2^{2k}}$
leading to an amplitude which is at most
a small constant times
$$\frac{1}{|x-j_1j_2\dotsb j_{2k}.j_{2k+1}\dotsb j_n|_{2^{2k}}}.$$
See the proof of Claim~\ref{amp_knotj}, Section~\ref{sect:twocl} 
for a formal argument via geometric series of a similar bound.

It follows that the
probability, i.e. total amplitude squared, 
of being more than $t$ units away from
$j_1j_2\dotsb j_{2k}$ is
$O(1/t)$.  Since we are using the output of the
QFT modulo $2^{2k}$ to estimate just the leading bits $j_1j_2\dotsb j_k$,
we need merely ensure that with high probability
no carry into these first $k$-bits has occurred.
In other words we need to bound the probability
that the offset, $t$, combines with the bits 
$j_{k+1}j_{k+2}\dotsb j_{2k}$
to induce such a carry.  This probability is 
proportional to 
\begin{equation}
\frac{1}{\norm{j_{k+1}j_{k+2}\dotsb j_{2k}}_{2^k}}.
\end{equation}
Unfortunately, this expression is not always small.
In particular if $j_{k+1}j_{k+2}\dotsb j_{2k}$ is
very close to zero mod $2^k$ then much of the smeared pointmass
will be at points whose leading $k$ bits differ from
$j_1j_2\dotsb j_k$.  Fortunately, this will be a problem for 
only a small fraction of $j$ and we will give a simple processing 
procedure to reduce the error arising from these bad
basis states.

First we derive an expression for the total error
arising from this circuit.
Let $\bf{j_i}$ denote the $i$th sequence of $k$ bits of $j$,
that is, $j={\bf j_1j_2\dotsb j_{n/k}}$ and 
${\bf j_i}=j_{ki+1}j_{ki+2}\dotsb j_{k(i+1)}$. Then we can generalize
the above reasoning to bound the squared
error of our circuit on a fixed input $\ket{j}\ket{\hat{j}}^3$
by 
\begin{equation}
\sum_{1<i<n/k}
\frac{1}{\norm{\bf{j_i}}_{2^{k}}}.
\end{equation}
Since we have maintained a third copy of $\ket{\hat{j}}$
throughout the computation errors arising from different
$j$ are orthogonal.  Thus the total squared error of the circuit on input
$$\sum_{j}\alpha_j\ket{j}\ket{\hat{j}}^3$$
is bounded by 
$$\sum_j\norm{\alpha_j}^2\max\left(1,\norm{\sum_{1<i<n/k}
\frac{1}{\norm{\bf{j_i}}_{2^{k}}}}^2\right).$$
We define a set of bad values $j$, denoted $B$, by 
letting $j\in B$ if there exists an $i$ such that 
$\norm{\bf{j_i}}_{2^k}<2^{k/2}$.  Then the above expression
is less than
\begin{eqnarray}\label{eq:error}
\sum_{j\not\in B}\norm{\alpha_j}^2\norm{\sum_{1<i<n/k}
\frac{1}{\norm{\bf{j_i}}_{2^{k}}}}^2
+
\sum_{j\in B}\norm{\alpha_j}^2
&\leq&
\sum_{j\not\in B}\norm{\alpha_j}^2\frac{n^2}{2^k}
+\sum_{j\in B}\norm{\alpha_j}^2\\
&\leq&
\frac{n^2}{2^k}
+\sum_{j\in B}\norm{\alpha_j}^2.
\end{eqnarray}
By choosing $k\in O(\log n)$ we can make the
first of these two terms an arbitrary
inverse polynomial.  The second term
is a problem.  If the input superposition
is supported on the set $B$ then this term is one.
On the other hand, the $j\in B$ form a small fraction of the
whole -- at most an $\frac{n}{2^{k/2}}$ fraction to be precise.
Thus if the input $\balpha$ is fairly evenly distributed
this second term will also be an arbitrary inverse 
polynomial for $k\in O(\log n)$.  
We will use a simple procedure -- taking a random shift 
of our original superposition, computing the approximate QFT, and
then undoing the effect of the shift -- to mimic a uniformly distributed
input and thus ensure that the overall error is polynomially
small.  We note that 
for many important applications, such as Shor's Factoring
and Discrete Log algorithms, the input superposition is uniformly
distributed to begin with and this procedure is not required.
This will also be true when our parallel circuits for the QFT over an
arbitrary modulus invoke the parallel circuits for the
QFT over a power of $2$ as a 
subroutine.  

Finally we note that by overlapping the bit estimates
from the $2$ copies of $\ket{\hat{j}}$ and performing a $O(\log n)$-
depth carrying procedure similar to that outlined in 
\cite{cleve00fast}, one could
get rid of this problematic second term entirely.  
However, the pre- and postprocessing
procedures we have chosen are easier to express using our
set of basic arithmetic circuits and also easy to omit
when, as in the algorithms mentioned, it is unnecessary.

\section{Quantum Fourier State Computation}
\begin{figure}[hbt]
  \begin{center} \psfig{file=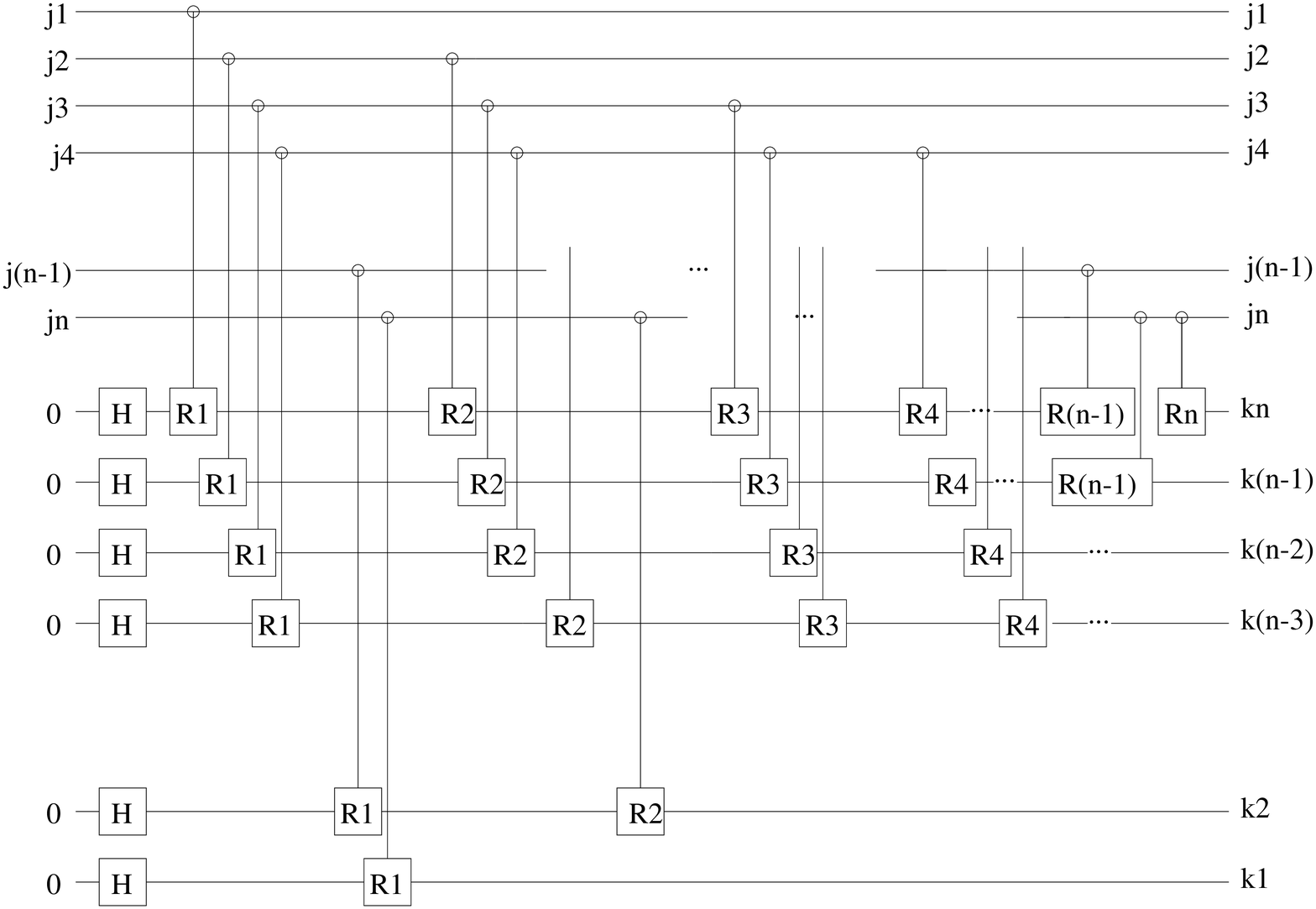,width=4in} \end{center}
  \caption{Exact 
Quantum Fourier State Computation (QFS): $\ket{j}\ket{0}\longrightarrow
            \ket{j}\ket{\hat{j}}.$  The approximate version (AQFS) just omits 
the $R_k$ for $k\in\Omega(\log n)$.}
  \label{fig:pqfs}
\end{figure}

We now turn to the task of approximating
the map
\begin{equation*}
\ket{j}\ket{0}\longrightarrow\ket{j}\ket{\hat{j}},
\end{equation*}
using parallel circuits.  The circuit pictured in
Figure~\ref{fig:pqfs}
computes this map exactly in depth $n$ and size $O(n^2)$.
By simply omitting the small rotations, i.e. the $R_k$ for
$k\in \Omega(\log n)$, in this
circuit a la Copppersmith we can approximate
this map to within an arbitrary inverse polynomial.  The resulting
circuit, which we denote AQFS, has size $O(n\log n)$ and depth $O(\log n)$.

\section{Copying a Fourier Basis State}
As was pointed out in \cite{cleve00fast}, the map
\begin{equation}
\ket{\hat{j}}\ket{0}\longrightarrow\ket{\hat{j}}\ket{\hat{j}},
\end{equation}
can easily be accomplished exactly in size and depth $O(n)$ and
$O(\log n)$.  First note that applying $n$ Hadamard gates
in parallel to the second register accomplishes the transformation
\begin{equation}
\ket{\hat{j}}\ket{0}\longrightarrow\ket{\hat{j}}\ket{\hat{0}}.
\end{equation}
Simply subtracting the second register from the first$\pmod {2^n}$ 
accomplishes the map
\begin{equation}
\ket{\hat{j}}\ket{\hat{k}}\longrightarrow\ket{\hat{j}}\ket{\widehat{j+k}}
\end{equation}
since
\begin{eqnarray}
\ket{\hat{j}}\ket{\hat{k}}
&=&\left(\sum_{i<2^n}\om{2^n}{ij}\ket{i}\right)
\left(\sum_{\pri <2^n}\om{2^n}{\pri k}\ket{\pri}\right)\\
&=&\sum_{i,\pri <2^n}\om{2^n}{ij+\pri k}\ket{i}\ket{\pri}\\
&\longrightarrow&
\sum_{i,\pri <2^n}\om{2^n}{ij+\pri k}\ket{i-\pri}\ket{\pri}\\
&=&
\sum_{i,\pri <2^n}\om{2^n}{(i-\pri)j+\pri(j+k)}\ket{i-\pri}\ket{\pri}\\
&=&\left(\sum_{i<2^n}\om{2^n}{ij}\ket{i}\right)
\left(\sum_{\pri <2^n}\om{2^n}{\pri (j+k)}\ket{\pri}\right)\\
&=&\ket{\hat{j}}\ket{\widehat{j+k}}.
\end{eqnarray}
This subtraction can be performed in size and depth $O(n)$ and
$O(\log n)$ respectively as discussed in Section~\ref{ssect:arith}.

\section{Putting it all Together}
\begin{figure}[hbt]
  \begin{center} \psfig{file=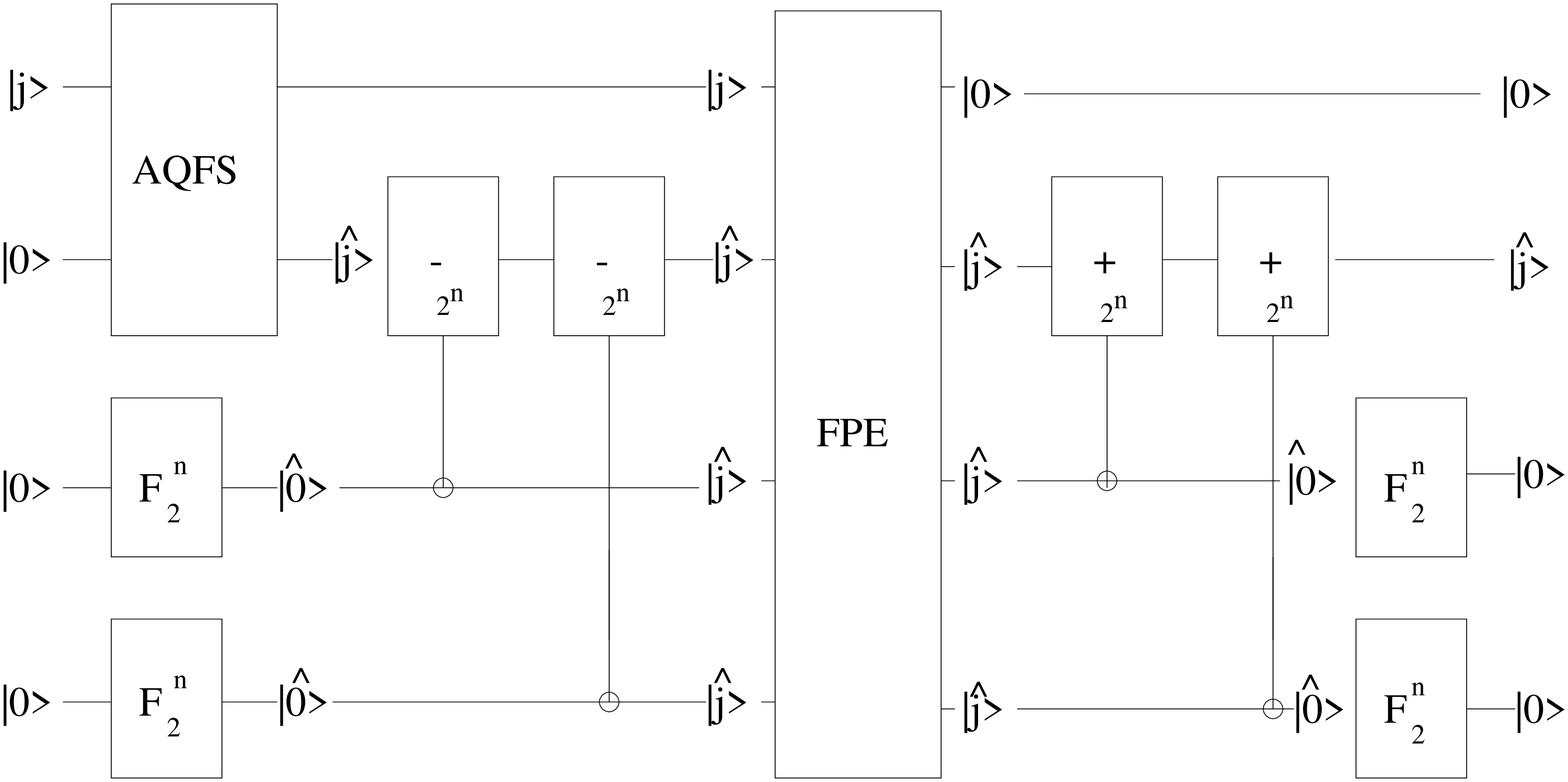,width=4in} \end{center}
  \caption{Approximate Parallel QFT for Uniform Inputs (UQFT)}
  \label{fig:uqft}
\end{figure}

\begin{figure}[hbt]
  \begin{center} \psfig{file=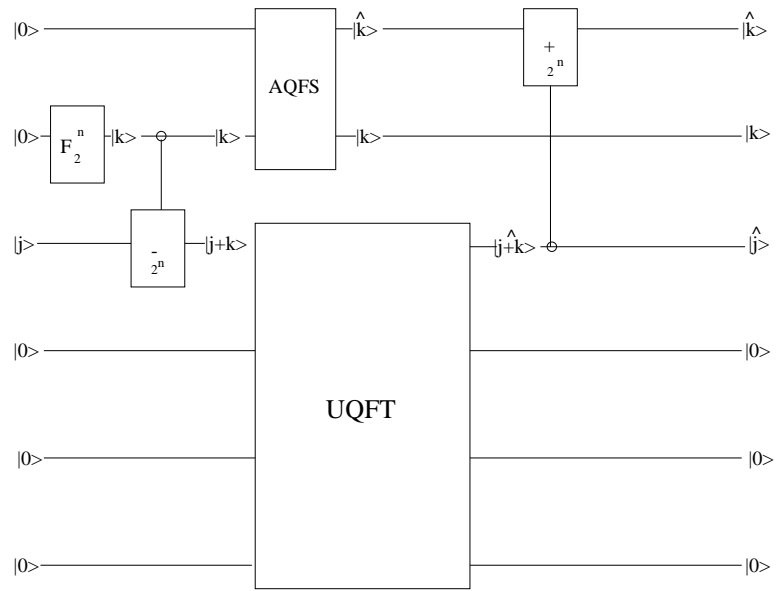,width=4in} \end{center}
  \caption{Approximate Parallel QFT}
  \label{fig:pqft}
\end{figure}

We now present two circuits.  Circuit~\ref{fig:uqft} is an approximate 
parallel QFT modulo $2^n$ which works with high
accuracy whenever
the input superposition is sufficiently uniform (in the
norms of the amplitudes) over $2^n$.  
Circuit~\ref{fig:pqft}, which calls Circuit~\ref{fig:uqft} as a 
subroutine, is an approximate parallel
QFT modulo $2^n$ which achieves arbitrary inverse polynomial precision
for all input superpositions.

Assume temporarily that all QFS maps are performed exactly.
If the FPE circuit called as a subroutine in Circuit~\ref{fig:uqft}
uses QFT's of size $2k$, and thus has depth $O(k)$, then
the size of the squared error of Circuit~\ref{fig:uqft} on input $\keta$ is
bounded by 
$$\frac{n^2}{2^k}
+\sum_{j\in B}\norm{\alpha_j}^2$$
where $B$ is the subset of indices of size $\frac{n}{2^{k/2}}$
defined in Section~\ref{sect:fpe}.  

The size of the squared error of Circuit~\ref{fig:pqft}
is then bounded by
$$\frac{1}{2^n}\sum_{k<2^n}\left(\frac{n^2}{2^k}
+\sum_{j+k\in B}\norm{\alpha_j}^2\right)=\frac{n^2}{2^k}+\frac{|B|}{2^n}=
\frac{n^2}{2^k}+\frac{n}{2^{k/2}}$$
and it suffices to choose $k\in O(\log n)$ to obtain
inverse polynomial accuracy.  

Finally, if we perform the QFS maps in depth $O(\log n)$
the resulting inverse polynomial error simply adds to the error
already analyzed and we get an overall circuit of size and depth
$O(n\log n)$ and $O(\log n)$ respectively which approximates
the QFT to within an arbitrary inverse polynomial.

\chapter{An Approximate Quantum Fourier 
Transform over an Arbitrary $Z_N$}\label{chap:fta}Let $\keta=\sum_{i<N}\alpha_i \ket{i}$ be an arbitrary
quantum superposition and let $\kfal$ denote
the quantum Fourier transform of $\keta$ over $Z_N$.  
We give quantum circuits which
approximate this QFT to within an arbitrary $\epsilon$.
When $\epsilon$ is an inverse polynomial in $n=\log N$
the circuits achieve a substantial speedup over the $O(n^2)$ method
of \cite{Kitaev1995}.  Our method continues to work for smaller $\epsilon$  
but with asymptotic size identical to earlier methods. A preliminary
version of these results can be found in \cite{FOCS::HalesH2000}.

We focus on the relevant situation of $\epsilon$ an inverse polynomial.
In particular, we show that in this case
circuits of size $O(n\log n\log\log n)$ and depth $O(\log n)$
can be achieved.

\begin{theorem}\label{fttheorems}
There are quantum circuits of size
$O\left(n\log n\log\log n\right)$
and depth $O\left(\log n\right)$
which approximate the
QFT over $Z_N$ to within
an arbitrary inverse polynomial.
\end{theorem}

More specifically, the bottleneck in our
algorithm is multiplication by our
modulus $N$ and an $n$-bit approximation to its inverse $1/N$, denoted 
$\widetilde{1/N}$.
Let $s(M)$ and $d(M)$ denote the simultaneous size
and depth of quantum circuits which multiply an arbitrary $n$-bit integer by
$M$, that is, map
$$\ket{M}\ket{j}\longrightarrow\ket{M}\ket{j}\ket{Mj}.\footnote{Notice
that both the inputs $M$ and $j$ are preserved and thus quantum
circuits for this map can be generated from classical multiplication
circuits using Lemma \ref{lem:unreverse} -- 
no division circuits are necessary.}$$
Then our Algorithm \ref{alg:approxft} yields the following:
\begin{theorem}\label{thm:fttSD}
There are quantum circuits 
of simultaneous size and depth$$O\left(s(N)+s(\widetilde{1/N})+n\log n\right)
\text{ and }O\left(d(N)+d(\widetilde{1/N})+\log n\right)$$ respectively
which approximate the
QFT over $Z_N$ to within
an arbitrary inverse polynomial.
\end{theorem}
\section{The Algorithm}\label{sect:fta}
\begin{figure}[hbt]
  \begin{center} \psfig{file=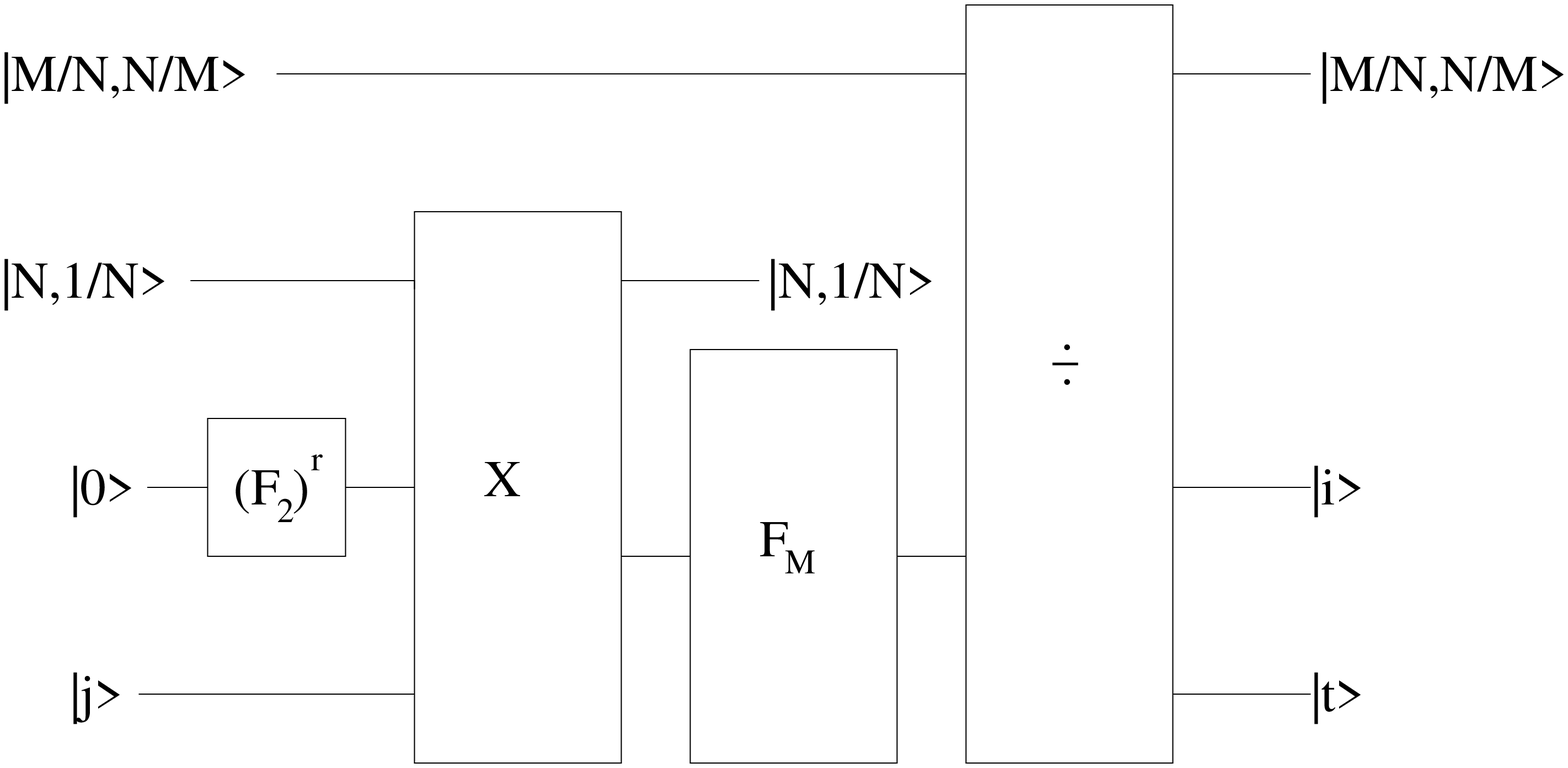,width=4in,height=2in} \end{center}
  \caption{Approximate QFT over $Z_N$.}
  \label{fig:arbqft} 
\end{figure}
We now describe the action of the circuit pictured in
Figure \ref{fig:arbqft} on input $\keta=\sum_{i<N}\alpha_i \ket{i}$
and parameters $R=2^r$ and $M\geq RN$. We will require a supply of
$\lfloor\log (M/N)\rfloor+1$ clean bits in an auxiliary register and
$O(n)$-bit approximations
to the decimals $1/N$, $N/M$, and $M/N$. 
In particular, our input will be of the form 
$$\ket{C}\ket{0}\keta$$ 
where $\ket{C}$ is a control register containing these
approximation and a copy of our modulus $N$.
Our output will be a superposition which is close
to 
$$\ket{C}\kfal\ket{\eta}$$
for some $\ket{\eta}$, where $\kfal$ denotes the QFT
over $Z_N$ of $\keta$.

\begin{algorithm}\label{alg:approxft} 
  Input: $\ket{C}\ket{0}\keta$

  \begin{enumerate}
  \item QFT over $(Z_2)^r$:
    $$\ket{0}\longrightarrow \sum_{i<R}
    \frac{1}{\sqrt{R}} \ket{i}.$$

  \item\label{tilde} Repeat $\keta$ $R$-times:

    \begin{eqnarray*}
      \ket{N,1/N}\sum_{i<R} \frac{1}{\sqrt{R}} \ket{i}\keta 
      &=&\ket{N,1/N}\sum_{i<R} \frac{1}{\sqrt{R}}
      \ket{i}\sum_{j<N} \alpha_j \ket{j}  \\
      &\longrightarrow& \ket{N,1/N}\sum_{j<N,i<R} \frac{1}{\sqrt{R}}
      \alpha_j \ket{j+iN}  \\
      &\stackrel{\mbox{\scriptsize def}}{=}&\ket{N,1/N} \kbeta .
    \end{eqnarray*}
  
  \item\label{ftm} QFT over $Z_M$,
    $$\kbeta \longrightarrow \fmkb$$

  \item\label{rounding}Division by $M/N$:
    $$\ket{j}\longrightarrow\ket{i}\ket{t}$$
    where $j=\clint{\frac{M}{N}i}+t$ 
    and $-\frac{M}{2N}\leq t<\frac{M}{2N}$.

   \end{enumerate}
\end{algorithm}
The correctness of the algorithm is a consequence of the fact that
for a typical remainder $t$, the subvector indexed
by integers of the form $\clint{\frac{M}{N}i}+t$ (and renormalized to
unit length) 
is close to the desired $\kfal$.  
It is worth noting that if $M=RN$ this is exactly true -- the only
remainder with any amplitude is zero and the subvector at 
integers $\frac{M}{N}i$ is exactly $\kfal$.  
More generally the approximation follows from Theorem \ref{thm:ftt},
which yields the following Corollary.
\begin{corollary}\label{ftalg}
Let $\ket{C}\ket{\gamma}$ be the output of the 
above algorithm.  Then there is a superposition $\ket{\eta}$ so that
$$\left\|\ket{\gamma}- \kfal\ket{\eta} \right\|< \frac{4RN}{M} + \frac{8\log N}{\sqrt{R}}.$$
\end{corollary}
In order to achieve a QFT which is accurate to within $\epsilon$,  then,
it suffices to take 
$R=\Omega\left(\frac{\log^2 N}{\epsilon^2}\right)$ and 
$M= \Omega \left(\frac{RN}{\epsilon} \right)$.

\subsection{Size and Depth Analysis}
We now return to Circuit \ref{fig:arbqft} and analyse its size and depth
requirements, restricting our analysis to the situation where $\epsilon$ is
an inverse polynomial.  First,
since we are free to choose $M$ to be a power of $2$,
the QFT at step \ref{ftm} can be implemented
by the parallel circuits of Chapter \ref{chap:pcs}.
Also note that the bounds from Corollary \ref{ftalg}
show that we can take $M$ to have only $n+O(\log n)$ bits.
The subcircuit for this transform thus has
size and depth $O(n\log n)$ and $O(\log n)$
respectively with constants approaching those
of the parallel circuits over $2^n$ itself.  

We now turn to the two multiplication procedures 
sandwiching the transform. Step \ref{tilde} involves
the multiplication of integers less than $R$
by our $n$-bit modulus $N$ and its inverse $1/N$.
But $R$ can be chosen to have only $\log n$-bits.  In this case
classical ``grade school'' multiplication 
techniques combined with carry-save adders are more
efficient than FFT related techniques and their translation to
the quantum setting yields
circuits of size and depth 
$O(n\log n)$ and $O(\log n)$
respectively.  The fact that $R$ can be taken to have
only $\log n$ bits is a consequence of the
circulant analysis of Section \ref{sect:circ}.

The bottleneck in Algorithm \ref{alg:approxft} is
the final step where we divide by the $n$-bit approximation
by $M/N$.  In other words we run Circuit \ref{fig:multmult}
on inputs $M/N,N/M$ in reverse.   
Since we choose $M$ to be a power of $2$ this
is equivalent in complexity to multiplication
by $N$ and $1/N$.
We note that if there is a special 
technique for quickly multiplying and dividing by our modulus $N$ 
the circuit size and depth can be improved.  For instance, if $N=c^m$
is a constant power then we can perform reversible multiplication
with a circuit of size $O(n\log n)$ and depth $O(\log n)$.
This gives us an algorithm which matches the asymptotic
size and depth of the QFT over a power of $2$. 
Of course, a circuit for the QFT over a modulus of this form could
also be constructed directly in analogy with the 
power of $2$ case and would achieve similar asymptotic
size and depth.  We conjecture that by just changing the 
multiplication technique used in Step \ref{rounding} to suit
the particular modulus $N$ our circuits can always be made 
asymptotically optimal.

We emphasize that the only
reversible multiplication is by the modulus and its inverse,
not between arbitrary $n$-bit integers.
The inverse can thus be prepared classically and we
can make use of Circuit \ref{fig:multmult}.  This allows us to 
avoid the problem of optimizing
the simultaneous size and depth of division circuits -- see the
discussion at the end of
Section \ref{ssect:arith}.  This is another clear benefit
of our technique over earlier approaches.  We note that there
appears to be a close relationship between the complexity of
approximating
the QFT over $Z_N$ and reversible multiplication by $N$.
Our algorithm shows that, with low ($O(n\log n)$) overhead, approximate 
circuits for reversible multiplication by $N$
lead to circuits for approximating the QFT over $Z_N$ .
On the other hand one can show that, with similar overhead, 
circuits approximating
the QFT over $Z_{N2^{n}}$ can be converted to approximate 
circuits for reversible multiplication by $N$.  Unfortunately
this relationship does not lead to a faster-than-classical
quantum multiplication algorithm, the ``tantalizing''
question posed by Shor\cite{SICOMP::Shor1997}.

\section{Fourier Sampling}\label{samsect}
In many quantum algorithms (see \cite{SICOMP::BernsteinV1997,
SICOMP::Shor1997, BonehL1995, SICOMP::Simon1997:1474}), including
the Hidden Subgroup Algorithm \ref{alg:hs}, the QFT
occurs as the final quantum step and a measurement of the
superposition immediately follows.  We refer to this procedure as
Fourier sampling \cite{SICOMP::BernsteinV1997}.  Suppose that we wish
to sample from $\distD_{\kfal}$, the distribution induced by
measuring $\kfal=F_{N}\keta$, for some given $N$ and $\keta$.  In this
situation since we need only insure that the distribution we sample from is
$\epsilon$-close to $\distD_{\kfal}$ -- we
need not worry about the phases of the amplitudes in the final
superposition.  This simplifies the computation of the previous
Section in two ways.  

First, we can reduce the size of the QFT, $F_M$,
which appears as a subroutine in our circuit.  
In particular we can choose $M$ to be any integer at least $RN$ as opposed to
requiring $M=\Omega \left(\frac{RN}{\epsilon} \right)$ 
as in the previous algorithm.  This is because we are now lumping
together the probabilities of 
all outputs of the form $j=\clint{\frac{M}{N}i}+t$ 
for $-\frac{M}{2N}\leq t<\frac{M}{2N}$ -- we are no longer concerned
with the individual superpositions corresponding to a fixed remainder $t$
or with phases of our amplitudes.

Second, and more significantly,
we can reduce the asymptotic size and depth
of the quantum circuits 
by measuring immediately after
$F_M$ and performing the final
division
classically.  
This reduces the quantum circuit size and depth to 
$O(n\log n)$ and $O(\log n)$ respectively.
\begin{figure}[hbt]
  \begin{center} \psfig{file=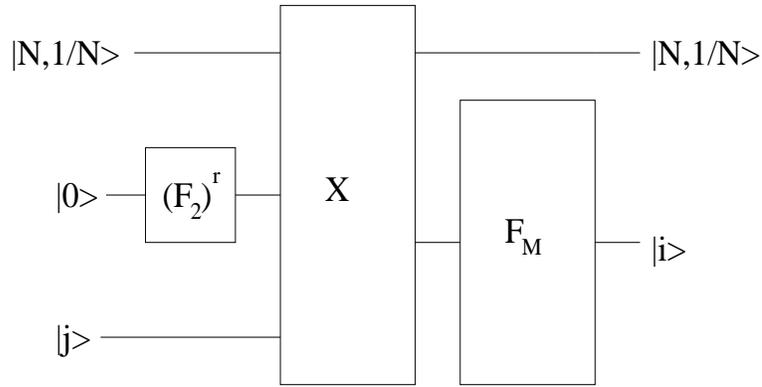,width=4in,height=2in} \end{center}
  \caption{Fourier Sampling over $Z_N$.}
  \label{fig:fszn} 
\end{figure}

\begin{algorithm}\label{samknowp} 
  Input: $\ket{N,1/N}\ket{0}\keta$

  \begin{enumerate}
  \item QFT over $(Z_2)^r$:
    $$\ket{0}\keta \longrightarrow \sum_{i<R}
    \frac{1}{\sqrt{R}} \ket{i}\keta.$$

  \item Repeat $\keta$ $R$-times:

    \begin{eqnarray*}
      \ket{N,1/N}\sum_{i<R} \frac{1}{\sqrt{R}} \ket{i}\keta 
      &=&\ket{N,1/N}\sum_{i<R} \frac{1}{\sqrt{R}}
      \ket{i}\sum_{j<N} \alpha_j \ket{j}  \\
      &\longrightarrow& \ket{N,1/N}\sum_{j<N,i<R} \frac{1}{\sqrt{R}}
      \alpha_j \ket{j+iN}  \\
      &\stackrel{\mbox{\scriptsize def}}{=}&\ket{N,1/N}\kbeta .
    \end{eqnarray*}
  
  \item QFT over $Z_M$,
    $$\kbeta \longrightarrow \fmkb$$

\item\label{it:meask}Measure
\item (Classical) Divide $\ket{j}$ by $M/N$ to output 
$i$ such that
$j=\clint{\frac{M}{N}i}+t$ 
for $-\frac{M}{2N}\leq t<\frac{M}{2N}$.
\end{enumerate}
\end{algorithm}
Let $\distD_{\kfal}$ be the distribution on $\{0,...,(N-1)\}$
induced by measuring $\kfal$ and let $\distD$ be the
distribution induced by Algorithm~\ref{samknowp}.  
Corollary \ref{ftalg} from the previous section could be used 
to prove that these distributions are close for sufficiently
large $R$ and $M>>RN$.  However, we use 
Theorem \ref{thm:ftts} to show that this is still true for any $M\geq RN$.
In particular, the following Corollary is a direct application
of this Theorem.
\begin{corollary}\label{cor:ftts}
$$\|\distD_{\kfal}-\distD\|_1<\frac{8\log N}{\sqrt{R}}.$$
\end{corollary}

\section{Fourier Sampling and The Hidden Subgroup Problem over $Z$}
\label{sect:fsp}

In the previous section we gave a procedure which, given $N$ and 
$\keta=\sum_{i<N}\alpha_i\ket{i}$ as input, sampled from
the $\distD_{\kfal}$, the distribution
induced by measuring $\kfal=F_N\keta$. 
This procedure is the basis for the finite Abelian hidden subgroup
algorithm, of which Shor's Discrete Log algorithm is a special case.  
We now give a procedure for the 
hidden subgroup problem over $Z$.  The procedure
itself is essentially identical to the quantum
portion of Shor's algorithm, but we give a more general
analysis. 

Suppose we have the ability to generate arbitrarily long repetitions
of some fixed superposition $\keta=\sum_{i<N}\alpha_i\ket{i}$, that is,
superpositions of the form $\sum_{i<M}\alpha_{(i \mod N)}\ket{i}$,
but that $\keta$ and $N$ are themselves unknown. Let $\distD_{\kfal/N}$
be the distribution on fractions with denominator $N$ and numerators
distributed according to $\distD_{\kfal}$.  
The following
algorithm allows us to sample from a distribution
which is arbitrarily close to $\distD_{\kfal/N}$.

\begin{algorithm}\label{samunknowp}
Input: $\sum_{i<M}\alpha_{(i \mod N)}\ket{i}$ 
\begin{enumerate}
\item QFT over $Z_M$
\item\label{it:measu}Measure
\item (Classical)Divide the 
result by $M$, and use the continued fractions
method to round to the nearest fraction
with denominator less than $T$, where $T$ is a known upper bound on $N$.
\end{enumerate}
\end{algorithm}

In particular, if $\distD$ is the distribution on fractions
with denominator less than $T$ output by our algorithm, then
we have the following Lemma, whose proof
appears in Section \ref{sect:pofl}.
\begin{lemma}\label{sampunp}
$$\|\distD_{\kfal/N}-\distD\|_1=O\left(\frac{T\log T}{\sqrt{M}}\right).$$
\end{lemma}
To make these distributions $\epsilon$-close, then, it
suffices to take $M=\Omega\left(\frac{T^2\log^2 T}{\epsilon^2}\right)$.
It is important to notice that even sampling {\bf exactly} 
from $\distD_{\kfal/N}$
does not immediately give us access
to the distribution $\distD_{\kfal}$ or to the value $N$
because the fractions obtained are in reduced form.

We now use this algorithm
to solve the Hidden Subgroup problem over $Z$.  
Recall that we are given
a function $f$ defined on $G$ which is both constant and distinct 
on the cosets of an unknown subgroup $H\leq G$.  The goal is to determine
$H$.  In the case of $G=Z$, $H$ must be a cyclic subgroup generated
by some element $N$ of $Z$, i.e. $H=\langle N\rangle$.  The function
$f$ can be equivalently described as a function with
period $N$ which is one-to-one within each period.  Determining the
subgroup $H$ is equivalent to determining this period.

Given an upper bound $T$ on $N$ we can easily create the superposition
$\sum_{i<M}\ket{i}\ket{f(i)}$ where
$M=\Omega\left(T^2\log^2 T\log^2\log T\right)$ is a power of $2$.  
Using this as input to the above algorithm
we can sample from a distribution very close to $\distD_{\kfal/N}$
where $\keta=\sum_{i<N}\ket{i}\ket{f(i)}$.  Now, $\distD_{\kfal}$
is uniform on $\{0,\dots,N-1\}$ -- this follows easily from the
fact that $f$ is one-to-one within its period -- and thus $\distD_{\kfal/N}$
is uniform on fractions with denominator $N$. 

We test the denominator of each
fraction output by our procedure to see if it is the period by
evaluating the function at a pair of values.  This will allow us to correctly
discard all denominators less than $N$.
We then accept the smallest value which passes our test.  This procedure
correctly recovers $N$ as long as it actually appeared as a denominator, i.e.
as long as we sampled a fraction with denominator $N$ and numerator relatively
prime to $N$.  Such numerators constitute a $c/\log n$ 
fraction of the set $\{0,\dots,N-1\}$, for some constant $c$.
By our choice of $M$ this set must constitute a similarly
sized fraction of the distribution output by our algorithm
and by sampling $O(n\log n)$ times such a numerator
will occur with exponentially high probability.. 

\subsection{Proof of Lemma \ref{sampunp}}\label{sect:pofl}
We can analyse the distribution output by Algorithm \ref{samunknowp}
on input $\sum_{i<M}\alpha_{(i\bmod N)}\ket{i}$ and parameter $T>N$
by comparison with the action of Algorithm \ref{samknowp} 
on a specially constructed input.  In particular,
let $\rketa$ be the superposition $\keta$ repeated $T$-times, i.e.
$$\rketa=\sum_{j<T}\sum_{i<N}\alpha_i\ket{i+jN}.$$
We look at the distribution output by Algorithm \ref{samknowp} 
on input $$\ket{TN}\ket{0}\rketa$$ and parameters 
$R=\lfloor\frac{M}{TN}\rfloor$ and $M$.  By Corollary \ref{cor:ftts}
this distribution is close to $\distD_{F_{TN}\rketa}$ for 
an appropriate choice of $M$.  Moreover,
since the amplitude of $F_{TN}\rketa$ at $Ti$ is identical to
the amplitude of $\kfal=F_N\keta$ at $i$, $\distD_{F_{TN}\rketa}$
is just $\distD_{\kfal}$ distributed over multiples $Ti$ of $T$.

Now, it is not hard to see that the input to the QFT over $Z_M$
when Algorithm \ref{samknowp} is run on input $\ket{TN}\ket{0}\rketa$
and the above parameters is very close to the input
to the QFT over $Z_M$ in Algorithm \ref{samunknowp}.
Thus the distribution of outputs from Step \ref{it:meask} of 
Algorithm \ref{samknowp} on $$\ket{TN}\ket{0}\rketa$$
is exponentially close to the distribution of outputs
from Step \ref{it:measu} of Algorithm \ref{samunknowp}.
By the previous paragraph
we need merely insure that an output interpreted by the former
algorithm as $Ti$ is interpreted in the latter case as $i/N$.
An output 
$k$ which the former algorithm rounded to $Ti$ must have 
satisfied
\begin{equation}\label{eq:round}
\norm{k-\frac{M}{TN}Ti}<\frac{M}{2TN}.
\end{equation}

Algorithm \ref{samunknowp} divides this same $k$ by $M$
and uses the continued fraction method to round to
the nearest fraction with denominator less than $T$.
Dividing Equation \ref{eq:round} by $M$ yields
\begin{equation}\label{eq:cf}
\norm{\frac{k}{M}-\frac{i}{N}}<\frac{1}{2TN},
\end{equation}
which implies that $\frac{i}{N}$ 
must be this nearest fraction given by the continued
fractions procedure, as desired.

This distance between $\distD$ and $\distD_{\kfal/N}$
is thus given by
$$\frac{8\log N}{\sqrt {M/TN}}+\frac{TN}{M-TN}=
O\left(\frac{T\log T}{\sqrt{M}}\right)$$
where the first term in the sum comes from the error in Algorithm 
\ref{samknowp}
and the second from the distance between the inputs to the 
QFT's in Algorithms \ref{samknowp} and \ref{samunknowp} respectively.

\chapter{A Relaxation of the Abelian Hidden Subgroup Problem}
\label{chap:nmulti}Recall the hidden subgroup problem
introduced  in Section \ref{sect:simon}.  We are given oracle
access to a function $\fh$ defined on a group $G$ and
constant on cosets of some unknown subgroup $H\leq G$.
The challenge is to find a set of generators of $H$.
The standard hidden subgroup problem assumes further that
$\fh$ is distinct on distinct cosets of $H$.  This standard
version can be solved efficiently on a quantum computer,
that is in time polynomial 
in $n=\log\left|G/H\right|$, whenever $G$
is a finitely generated Abelian group(\cite{Jozsa}). 

The related problem which relaxes
the requirement that $\fh$ be distinct on distinct cosets
of $H$ was first addressed in \cite{BonehL1995} and later in \cite{MosEke98}.
We shall refer to this as the relaxed hidden subgroup
problem.  Both \cite{BonehL1995} and \cite{MosEke98} give algorithms
which partially solve the relaxed hidden subgroup problem, with
the former addressing just the case $G=Z$ and the latter
the general problem for finitely generated Abelian $G$.
However, in both results the coset distinctness requirement is
changed only slightly.  
In particular, the function $\fh$ is allowed to map $m$ cosets
to one, but $m$ must be both polynomial in $n$ and smaller than the 
smallest prime divisor of $|G|$.  Notice that for some groups,
such as $G=\zton$, this amounts to no relaxation at all.

As noted in \cite{MosEke98}, however,
{\it some} restriction must be placed on the behavior of the function
$\fh$, since, once the distinctness requirement
is dropped, there are functions $\fh$ and $\fk$ for
$K\neq H$ which differ on an exponentially small fraction
of their inputs.  Using existing lower bound techniques
based on the unitary evolution of quantum computation
we should expect that such functions require exponentially
many queries, and thus exponential-time, to distinguish. 

We solve the relaxed hidden subgroup problem
for finitely generated Abelian groups.  In particular,
we define a stratification of the functions $\fh$ into classes,
then give a tight characterization of which classes
have polynomial-time algorithms by exhibiting both
an algorithm and a lower bound for each class.  
These results generalize and simplify our earlier work
on many-to-one periodic functions presented in \cite{FOCS::HalesH2000},
and use in a crucial way the Fourier sampling procedures
of Sections \ref{samsect} and \ref{sect:fsp}.

\section{Definitions and Main Theorems}\label{sect:multdef}
Throughout our
discussion $\fh$ will denote a function
defined on a finitely generated Abelian group
$G$ and constant on cosets of the 
subgroup $H\leq G$.  Notice that for any $K\leq H$
$\fh$ induces a well-defined function on $G/K$ which
we denote $f_{H/K}$.  
We define the input length of the hidden subgroup problem
given by the function $\fh$ on $G$ to be 
$n=\lceil\log \left|G/H\right|\rceil$.
and assume
without loss of generality that the range
of $\fh$ is contained in the set $\{0,1\}^n$.

Let $\bD$ be the generalized Hamming metric
on the functions $\fh$: 
\begin{definition}\label{def:distz}
$$\bD\left(\fh,\fk\right)$$ is the fraction of elements
in the group $G$ for which 
$f_{H}(x)\neq f_{K}(x)$. 
\end{definition}
In other words, $\fh$ and
$\fk$ are $\epsilon$-close under $\bD$
if they disagree on at most an $\epsilon$
fraction of the elements of $G$\footnote{A careful
reader should object at this point that 
this definition only really makes sense when
$G$ is a finite group.  We take the distance between $\fh$ and $\fk$
defined on an infinite $G$ to be the distance between the induced
functions $f_{H/(H\cap K)}$ and $f_{K/(H\cap K)}$
which are defined on the finite group $G/(H\cap K)$.}

Using this metric we stratify our functions into classes in the 
following manner:
\begin{definition}\label{def:codez}
  For any function $d(n)$ let
  $$C_{1/d(n)}=\lbrace \fh|\forall \fk
        \text{ with } K\not\leq H\text{, }
           \bD(\fh,\fk)>1/d(n)\rbrace.$$
\end{definition}
We think of the function $\fh$ as a codeword
for the subgroup $H$. The class $C_{1/d(n)}$ is then a code with
minimum distance $1/d(n)$.
Our results show that there is
an efficient quantum decoding procedure for $C_{1/d(n)}$
if and only if $d(n)$ 
is a polynomial.

More formally we will prove the following two theorems:
\begin{theorem}\label{Main}
  Given any polynomial $d(n)$ there is an efficient quantum algorithm
  $\boldsymbol A$\footnote{Throughout the paper we will assume that
    $\boldsymbol A$ has a blackbox subroutine for computing values of
    $f$} which, given any finitely generated
     Abelian $G$ and $\fh\in C_{1/d(n)}$,
    outputs the generators of $H$ with 
    exponentially high probability.
\end{theorem}

\begin{theorem}\label{lowbou}Let $d(n)=o(2^{n})$ be given.
  Suppose that $\boldsymbol A$ is a quantum algorithm which correctly
  computes generators of H from any $f_H\in C_{1/d(n)}$ with probability at
  least $3/4$.  Then $\boldsymbol A$ has worst-case run-time
  $\Omega(\root 4\of{d(n)})$.
\end{theorem}
Our algorithm uses the same quantum subroutine as the standard
hidden subgroup problem -- namely Fourier Sampling.  The relaxed
problem requires more repetitions of this quantum subroutine and,
in the case of $G=Z$, a more elaborate classical post-processing.

The lower bound is proved in the special case where $G=Z$
and thus the hidden subgroup 
function is periodic on $Z$ and potentially many-to-one
within each period.  The proof   
is a slight variation -- allowing for the periodic structure
of $f$ -- on the standard lower bound technique of 
\cite{BBBV}.
This is sufficient to establish the
polynomial vs. superpolynomial gap which is our primary concern.
It is likely that the more sophisticated techniques of
\cite{Ambainis00a} could be used to improve this lower bound.

\section{Finite Abelian $G$}\label{sect:fag}
We first solve the special case of the relaxed hidden subgroup 
problem where the underlying 
group $G$ is finite Abelian.  Section \ref{sect:infg}
addresses the case $G=Z$ and shows how to combine these
to give an algorithm which works for any finitely generated Abelian group.

As in our discussion of the standard hidden subgroup 
problem we assume that the group $G$ is given to us as a direct product
$\zpp$.
More concretely, the 
input to our quantum algorithm is the list 
$\left(p_1,\dotsc,p_k\right)$
and our function $\fh$ is 
defined on the {\bf set} 
$\zpp.$  As in the standard hidden subgroup problem there is a quantum procedure which
produces such a description under very general conditions \cite{Cheung}

\begin{algorithm}\label{zpp}

\begin{enumerate}

\item Prepare $$\keta=\sum_{x\in G}\ket{x}\ket{f_H(x)}.$$
\item Sample from $\dD{G}{\keta}$ where
$$F_{G}\keta=\sum_{x\in G}F_{G}\ket{x}\ket{f_H(x)}.$$
 
  \end{enumerate}

Repeat this procedure $\Omega\left(n^2d^2(n)\right)$ times
obtaining outputs $y_i$.  Solve the corresponding
system of equations $y_i\cdot_G x$  and output this solution.
\end{algorithm}
As mentioned previously, our quantum subroutine is identical
to that of the standard case but we must
increase the number of samples by a factor of $d^2(n)$.
As before,  the correctness of this algorithm
is equivalent to the condition that the samples $y_i$ 
generate the subgroup $\perH$.  We first note that,
as in the standard case, the distribution $\dD{G}{\keta}$
is supported on this subgroup.  The argument given in the
standard case (see Section \ref{sect:gsa}) 
hinges on the fact that for any $h\in H$
$$\keta=\sum_{x\in G}\ket{x}\ket{\fh(x)}$$
and
$$\ket{h*\alpha}=\ket{h}*\keta=\frac{1}{\sqrt{|G|}}
\sum_{x\in G}\ket{h+x}\ket{\fh(x)}$$
are identical, which remains true in the
relaxed problem as well.

In order to establish the correctness of the algorithm we
need further that the outputs generate $\perH$ with high probability.
Recall that in the standard case we used the fact that the
distribution was uniform on $\perH$ to argue that
$O(n^2)$ samples must generate this subgroup with high probability.
Uniformity no longer holds in the relaxed case.  Instead
we substitute the following property which limits the
probability that our samples remain trapped in some
proper subgroup of $\perH$:
\begin{lemma}\label{lem:rec}
Suppose that $\fh\in C_{1/d(n)}$.  Then for every proper subgroup 
$K<\perH$, if $y$ is chosen according to 
$\dD{G}{\keta}$
$$Pr\left(y\in K\right)< 1-1/4d^2(n).$$
 \end{lemma}
This lemma, proved in Section \ref{sect:rec}, is the main technical result
of this Chapter.  The correctness of the algorithm follows
from the Lemma by an argument similar to that of the standard case.
In particular, in order for our outputs to generate $\perH$ 
they must lie outside of any proper subgroup
$K$ of $\perH$.  There are at most $2^{n^2}$ such subgroups,
since each is determined by a set of at most $n$ generators and
$$\left|\perH\right|=\left|G/H\right|<2^n.$$
The probability that there exists
a proper subgroup of $\perH$ containing all our outputs is 
therefore upper bounded by the quantity

$$2^{n^2}\left(1-1/4d^2(n)\right)^t$$
where $t$ is the number of repetitions of the quantum subroutine.
Thus if we choose $t=\Omega\left(n^2d^2(n)\right)$
the outputs will generate $\perH$ with high probability.

\subsection{Proof of the Reconstruction Lemma}\label{sect:rec}
We prove a reformulation of Lemma \ref{lem:rec} 
which replaces the quantification over 
subgroups of $\perH$ with the quantification over subgroups of
$\perH$ which are themselves ``perps''.  Since $H=\left(\perH\right)^{\perp}$
all subgroups are of this form and the content of the lemma is 
unchanged. We first sketch how to 
establish that $H=\left(\perH\right)^{\perp}$, then proceed
to the proof of the reformulated lemma.  Notice that
it follows trivially form the definition of $\perH$ that 
$H\subseteq\left(\perH\right)^{\perp}$.
Since $H$ is finite it then suffices to show that 
$\norm{H}=\norm{\left(\perH\right)^{\perp}}$ which follows
from $\norm{G/K}=\norm{\perK}$ for all subgroups $K$.  
This last equality can be proved
by showing that the QFT over $G$ maps a subspace of dimension $\norm{G/K}$
to one of dimension $\norm{\perK}$ and using the fact that the QFT is unitary.

\begin{oldlemmalem:rec}
Suppose that $\fh\in C_{1/d(n)}$.  Then for every 
proper subgroup $\perK$ of $\perH$, if $y$ is chosen according to 
$\dD{G}{\keta}$ then 
$$Pr\left(y\in \perK\right)< 1-1/4d^2(n).$$
\end{oldlemmalem:rec}
\begin{proof}We give a proof by contradiction.
Suppose there exists a $\perK$ which violates the lemma.
We reconstruct a  function $\fk$
with $$\bD\left(\fh,\fk\right)<1/d(n),$$
This contradicts the assumption $\fh\in C_{1/d(n)}$,
since if $\perK$ is a proper subgroup of $\perH$ then
$K\not\leq H$.

We first note that for any $g\in G$
the amplitudes of $F_G\keta$ 
$F_G\ket{g*\alpha}$ at $x$ are related by the phase $\om{}{g\cdot_G x}$.
Thus for any $k\in K$ the amplitudes of $F_G\keta$ and 
$F_G\ket{k*\alpha}$
are identical at elements of $\perK$.  
Moreover by our assumption, when $y$ is
chosen according to $\dD{G}{\keta}$,
$$Pr\left(y\in\perK\right)>1-1/4d^2(n),$$ 
and thus the superpositions
$F_G\keta$ and 
$F_G\ket{k*\alpha}$ are heavily supported
on this subgroup $\perK$.  The superpositions must
therefore be close.  In particular,
$$\inprod{ F_G\keta}{
F_G\ket{k*\alpha}}>\sqrt{1-1/4d^2(n)}.$$
This implies the same lower bound for the inner product
$$\inprod{\keta}{\ket{k*\alpha}},$$
indicating that the vectors
$\keta$and $\ket{k*\alpha}$ also have
almost the same direction.  This can
only be the case if they agree 
on most of their coordinates.
In particular, if $c$ is the fraction of
$x$ for which $\fh(x)=\fh(x+_G k)$, then
$$\inprod{\keta}{\ket{k*\alpha}}
=\sqrt{c}>\sqrt{1-1/4d^2(n)}.$$ 
In other words, for every $k\in K$ 
at least a $1-1/4d^2(n)$ fraction
of the $x\in G$ satisfy
\begin{equation}\label{close}
\fh(x)=\fh(x+k).
\end{equation}

We now define our new function $\fk$ which is constant on
cosets of $K$ but still close to $\fh$.  For each
coset $xK$ we define $\fk$ to be uniformly equal to the
majority value of $\fh$ on $xK$, if one exists, and uniformly
equal to $0$ otherwise.  Clearly $\fk$ is constant on cosets
of $K$ but it remains to show that
$$\bD(\fh,\fk)<1/d(n)$$
in order to obtain a contradiction.
But by (\ref{close}) together with a standard 
averaging argument we have that for at least
a $1-1/2d(n)$ fraction
of the cosets $xK$, $\fh$ is constant on
a $1-1/2d(n)$ fraction of the 
coset.  This implies that 
$$\bD(\fh,\fk)<1/2d(n)+1/2d(n)=1/d(n),$$
as desired.
\end{proof}

\section{The Relaxed Hidden Subgroup Problem over $Z$}\label{sect:rxoverz}
We now give an algorithm for the relaxed hidden
subgroup problem over $G=Z$.  Let $\fh$ be defined
on $Z$ and constant on cosets of $H\leq Z$.
In this case $H$ must be generated
by some $N\in Z$ and we refer to $\fh$ as $\fn$. $\fn$
is equivalently a periodic function with period
$N$.  In this relaxed problem $\fn$ may not be distinct on distinct cosets,
in other words the function is potentially many-to-one within each period.
The distinctness requirement is replaced with the assumption that
$\fn\in C_{1/d(n)}$ for some polynomial $d(n)$.

Let $$\keta=\sum_{i<N}\ket{x}\ket{\fn(x)}$$
and $\ketha=F_N\keta$ be the superposition obtained
by performing the QFT over $Z_N$ of the
first register of $\keta$.
We first note that the restriction of $\fn$ to
the set $\{0,1,\dotsc,N-1\}$ is the function induced by $\fn$
on $Z/\cy{N}$.  It is easy to see that this induced function
is still in $C_{1/d(n)}$ but
now encodes the trivial subgroup $\cy{0}$.
By the results of Section \ref{sect:fag} if we sample from $\distD_{\ketha}$ 
$\Omega(n^2d^2(n))$ times we will almost surely obtain a set $\{y_i\}$ 
generating $\cy{O}^\perp=Z_N$, that is, a set $\{y_i\}$
satisfying $gcd(y_1,\dotsc,y_k,N)=1$

While we cannot create the superposition $\keta$,
we can create arbitrarily long repetitions
of $\keta$ by evaluating $\fh$ on some interval. 
The Fourier sampling procedure of Section \ref{sect:fsp}.
then allows us to sample from a distribution
exponentially close to $\distD_{\ketha/N}$,
the distribution on fractions with denominator $N$
and numerators distributed according to $\distD_{\ketha}$.
We can thus assume we are sampling exactly from the
distribution $\distD_{\ketha/N}$,
and by the above paragraph after $\Omega(n^2d^2(n))$ 
samples the numerators
of the fractions satisfy $gcd(y_1,\dotsc,y_k,N)=1$
with exponentially high probability.
By taking
the least common multiple of all denominators
we recover the
desired $N$. 

\section{Finitely Generated Abelian $G$}\label{sect:infg}
The finitely generated case can be reduced
to the finite case by restricting $f_H$ to 
each of the infinite cyclic
components of $G$ and using the algorithm of the
previous section to find the periods
of these restricted functions.  More formally, given 
a description 
$$\left(p_1,\dotsc,p_k,m\right)$$
of the group
$$G=\left(\bigoplus_{i<n} Z_{p_i}\right)+
\left(\bigoplus_{i<m}Z\right),$$
by finding the periods $N_i$ of the restriction of $f_H$
to each of the $m$ copies of $Z$ we obtain
a finite Abelian $G^\prime=\left(\bigoplus_{i<n} Z_{p_i}\right)
+\left(\bigoplus_{i<m}Z_{N_i}\right)$
so that the restriction of our function $f_H$
to $G^\prime$ now encodes a subgroup $H^\prime\leq G^\prime$
and is still in $C_{1/d(n)}$.  Moreover the generators of 
$H$ are precisely the
generators of $H^\prime$ together with the
periods $N_i$. This accomplishes the desired reduction.
\section{Proof of Lower Bound, Theorem \ref{lowbou}}
\label{sec:lowbou}

We need the following definition and theorem from
\cite{BBBV}.  Theorem \ref{lowbousea}
expresses the fact that if a quantum algorithm makes
few queries to an oracle function there must be values of
that function which have been hardly examined and thus can be 
changed without significantly changing the algorithm's behavior.
Its proof combines the unitary evolution of quantum computation
with a hybrid argument.
\begin{definition}{\cite{BBBV}}
  Let $\ket{\phi_i}$ be the superposition of $A^f$ on input $x$ at
  time $i$.  We denote by $q_y(\ket{\phi_i})$ the sum of squared
  magnitudes in $\ket{\phi_i}$ of configurations of $M$ which are
  querying the oracle on string $y$.
\end{definition}

\begin{theorem}{\cite{BBBV}}\label{lowbousea}
  Let $\ket{\phi_i}$ be the superposition of $A^f$ on input $x$ at
  time $i$.  Let $\epsilon>0$.  Let $S\subseteq [0,T-1]\times\Sigma_*$
  be a set of time-strings pairs such that $\sum_{(i,y)\in
    S}q_y(\ket{\phi_i})\leq \frac{\epsilon^2}{T}$.  Now suppose the
  answer to each query $(i,y)\in S$ is modified to some arbitrary
  fixed $a_{i,y}$ (these answers need not be consistent with an
  oracle).  Let $\ket{\phi_i^\prime}$ be the time $i$ superposition of
  $A$ on input $x$ with oracle answers modified as stated above.  Then
  $\|\ket{\phi_i}-\ket{\phi_i^\prime}\| \leq\epsilon$.
\end{theorem}

In our case we wish to use Theorem \ref{lowbousea} to show
that if a quantum algorithm computes with constant probability
the period $N$
of any $f\in C_{1/d(n)}$ defined on $G=Z$
then it 
must make at least $\Omega(\root 4\of{d(n)})$ queries to 
the function's values.  To this end we first look at the
algorithm's behavior when $f(x)=0$ for all $x$ (Note that 
the all-zeroes function is in every class $C_{1/d(n)}$).

We wish to use this behavior to generate a function $g\in C_{1/d(n)}$
which has period greater than 1 and which the algorithm cannot
distinguish from the all-zeroes function without making lots of
queries.  This is similar to earlier applications of Theorem
\ref{lowbousea} but with the added complication that $g$ must be
periodic and at least $1/d(n)$ away from any function of smaller
period.  We ensure periodicity by first deciding on the period $N$ of
$g$ and then changing the value of the function simultaneously on all
points of the form $x+kN$. We show that the latter complication can be
resolved by choosing $g$ to have prime period and to be sufficiently
different from the all-zeroes function.

\begin{proof}(Proof of Theorem \ref{lowbou})
  Given $A^f$ computing the period of any function in
  $C_{1/d(n)}$ in time $T$, we initially examine
  $A^o$ where $o$ denotes the all-zero function.  

  Fix a prime $N$ such that $\sqrt{d(n)}<N<2^n$.
  For $0\leq x<N$ let
  $$S_x=[0,T-1]\times\{y|y=x+kN\}.$$ The average value of
  $\sum_{(i,y)\in S_x}q_y(\ket{\phi_i})$ is $\frac{T}{N}$ and thus
  at least $1/2$ of the sets $S_x$ satisfy
  \begin{equation}\label{upbo}
    \sum_{(i,y)\in S_x}q_y(\ket{\phi_i})\leq 2\frac{T}{N}.
  \end{equation}
  
  Let $U$ be any set of $3N/\sqrt{d(n)}$ $x$ which satisfy (\ref{upbo}).
  We let our new function $g$ satisfy $g(x+kN)=1$ for 
  $x\in U$ and $g(x)=o(x)=0$ otherwise.
  Note that $g(x)$ has period our chosen prime $N$ and that 
  $D(o,g)\geq 3/\sqrt{d(n)}$.

  Furthermore, let
  $S_{U}=\bigcup_{x\in U}S_{x}\subseteq [0,T-1]\times\Sigma_*$.  Then
  $$\sum_{(i,y)\in S_U}q_y(\ket{\phi_i})\leq \frac{6T}{\sqrt{d(n)}},$$
  and we can take the $\epsilon$ of Theorem \ref{lowbousea} to be 
  $\frac{\sqrt{6}T}{\root 4\of{d(n)}}$.  Thus
  in order for our algorithm $A$ to distinguish between the all-zeros
  function $o$ and our new period-$N$ function $g$ with constant
  probability, $A$ must have worst-case run-time $\Omega(\root
  4\of{d(n)})$.
  
  To prove our theorem, however, we need to verify that our function $g$
  is actually in $C_{1/d(n)}$. We need the following claim whose
  simple proof is in the next section.

  \begin{claim}\label{litcl}For any periodic functions $f$ and $g$ with
    periods $N_f$ and $N_g$ respectively,
    if $D(f,g)<\epsilon^2<1/16$ then there is a
    function $h$ with period $N_h=gcd(N_f,N_g)$ and $D(h,g)<3\epsilon$.
  \end{claim}
  
  Think of the $g$ in the claim as being our $g$ constructed
  above.  We need to argue that there are no functions of smaller 
  period within $1/d(n)$ of $g$. By our claim if such a function
  $f$ existed then there would be a function $h$ with period 
  $N_h=gcd(N_f,N_g)=1$ (since the period of $g$ is a prime)
  and $D(h,g)<3/\sqrt{d(n)}$.  But $g$ and the all-zeroes function,
  which is the only plausible candidate for $h$, have distance 
  at least $3/\sqrt{d(n)}$, a contradiction.
\end{proof}
\subsection{Proof of Claim \ref{litcl}}\label{sixthapp}
\begin{proof}
Let $N_h=gcd(N_f,N_g)$.
Fix $k$ and $l$ such that $lN_f-kN_g=N_h$.
We will define a function $h$ which is constant
on flights of the form 
$\left[x+kN_h\right]=\left(x,x+N_h, x+2N_h,...\right)$ 
and within $3\epsilon$ of $g$.
Since $D(f,g)<\epsilon^2$, with probability at least 
$1-\epsilon$ when
we choose a random flight 
$\left[x+kN_h\right]$ at least a 
$1-\epsilon$ fraction
of points $y$ in that flight will satisfy $f(y)=g(y)$.
For such a ``good'' flight, choose $y$ and $z$ independently at 
random in the flight and let $j$ satisfy $jN_h=y-z$.  Then the point 
$w=y+jkN_g=z+jlN_f$ is uniformly distributed
over the flight.  Thus $f(z)=f(w)=g(w)=g(y)$ with probability at least 
$1-\epsilon$.  Putting these two facts together we get that when $y$ and
$z$ are chosen at random in a ``good'' flight, $g(y)=g(z)$ with
probability at least $1-2\epsilon$.  Using the fact that $\epsilon<1/4$,
this implies that at least
a $1-2\epsilon$ fraction of points in the flight share the same $g$ value.
We let the value of $h$ on all points in the flight be this overwhelming
$g$ value, and for ``bad'' flights we define $h$ to be uniformly 0.
Then it follows that $D(g,h)<2\epsilon+\epsilon=3\epsilon$, as claimed. 
\end{proof}

\chapter{Hidden Subgroups over the Reals}\label{chap:overr}We now expand the ideas of the previous section to show how to
find the period of certain periodic functions defined on the reals,
effectively solving the hidden cyclic subgroup problem over $<\Re,+>$.
This generalizes a recent result of \cite{hallgren2002}
which gives a quantum algorithm
finding the period of a subclass of these periodic functions
sufficient to yield
a polynomial-time quantum solution to Pell's equation.
Solving Pell's equation has been shown to be at least
as hard as factoring but no reduction in the 
opposite direction exists. 
In Chapter~\ref{chap:or} we give evidence in a
relativized setting that period-finding over the
reals is in fact harder than over the integers.
In particular, we show that the
problem over the reals lies outside
of the complexity class $MA$, a complexity
class which contains
the analogous problem over the integers.

Throughout our discussion $f$ will denote a piecewise continuous
function from $\Re\rightarrow[0,1]$ with period $p$.   
Our quantum machine is allowed
oracle access to approximate versions of $f$.  In particular,
on call $\ket{i}\ket{j}\ket{t}\ket{0}$ the oracle xors the
first $t$-bits of $f(i/j)$, denoted $f_t(i/j)$, into the last register, 
returning $\ket{i}\ket{j}\ket{t}\ket{f_t(i/j)}$.

The input
length of $f$ is $(n,k)$ if $p<2^n$
and the $n$-bit approximating 
step function $f_n$ has average step interval at least $1/2^k$,
where the average step interval is defined to be the ratio of the
period $p$ to the number of step intervals in that period.
We define a metric on these functions which is the
continuous analog of Definition~\ref{def:distz}, 
Section~\ref{sect:multdef}. 

\begin{definition}\label{def:distr}
Let $i_{f,g}(x)=1$ whenever $\norm{f(x)-g(x)}>2^{-n}$
and $0$ otherwise.  Then 
$$\bD (f,g)=\lim_{t\rightarrow\infty}\frac{1}{t}\int_{0}^{t}i_{f,g}(x)dx.$$
\end{definition}

Just as in the case of functions defined on $Z$
(Definition~\ref{def:codez}, Section~\ref{sect:multdef})
we use this metric to stratify the functions into classes.
Again, if we think of $f$
as an encoding of its period $p$ then the class $C_{1/d(n)}$ 
is a code with minimum distance $1/d(n)$.  
If $f\in C_{1/d(n)}$ then in order to reduce its encoded period
one needs to change at least a $1/d(n)$ ``fraction'' of its
values by at least $1/2^n$.  In other words, $f$
encodes its period $1/d(n)$-unambiguously and does so using just 
$n$-bits of output.
\begin{definition}\label{def:coder}
$$C_{1/d(n)}=\{f|\forall g\text{, if }p_g
<p_f \text{ then }\bD (f,g)>1/d(n)\}$$ 
\end{definition}
We can now state the main theorem of this section:
\begin{theorem}\label{thm:overr} For any polynomial $d(n)$ 
there is a quantum algorithm $A$
which generates the first $m$-bits of the period of any $f\in C_{1/d(n)}$
with exponentially high probability in time $poly(n,k,m)$.
\end{theorem}

That the condition $f\in C_{1/d(n)}$ for $d(n)$ a polynomial
is necessary for an efficient quantum algorithm to exist follows
almost immediately from the lower bound result (Theorem~\ref{lowbou})
of Chapter~\ref{chap:nmulti} -- after interpreting functions
on $Z$ with integral period as step functions on $\Re$ with
step interval $1$ in some canonical way, all that remains is to
check that the respective definitions of $C_{1/d(n)}$ do in fact coincide.

\section{Overview}\label{sect:overv}
Before we give a summary of the procedure we note that it
is sufficient to give an algorithm in the restricted case where
the given function $f$ is a step function with $n$-bit range, i.e. $f=f_n$,
and has average step interval $\geq 1$.  The first $m$ bits of the period
of an arbitrary $f\in C_{1/d(n)}$
of input length $(n,k)$ can then 
be found by running this algorithm to find the
first $m$ bits of the period of the
function $f_n(x/2^k)\in C_{1/d(n)}\subset C_{1/d(n+k)}$ which 
satisfies these restrictions and has input length $(n+k,1)$.
We will thus assume without loss of generality
that our function $f$ has $n$-bit range and
average step interval $\geq 1$.

The quantum portion of the algorithm is 
just Fourier Sampling, in this case sampling from the
distribution induced by measuring 
$F_{MN}\left(\sum_{i\in\pm\frac{MN}{2}}\ket{i}\ket{f(i/N)}\right)$
for some $M,N$.
The tricky part lies in showing that $M$ and $N$ can be chosen
simultaneously to yield the desired information about the period.
Suppose we fix $M$ and choose $N>>M$.  Then it is easy to see that
evaluating the functions $f(\frac{x}{N})$ 
and $f(\frac{px}{\clint{Np}})$ on the interval $[\pm\frac{MN}{2}]$
results in exponentially close superpositions (Lemma~\ref{lem:intcl}).
This is useful because the latter function has integral period $\clint{Np}$
(easy to see) and is in $C_{1/2d(n)}$ when regarded as a function 
on the integers (Lemma~\ref{lem:cdn}).  This allows us
to use the results of the previous chapter to analyze the
distribution output by the Fourier sampling procedure.

In particular, suppose by some fortuitous luck that $\clint{Np}$
actually divides $MN$. Then we know that the 
the Fourier sampling procedure always outputs integers
of the form $k_i=\frac{j_iMN}{\clint{Np}}$ 
and that $gcd(j_1,\dots,j_t,\clint{Np})=1$\footnote{Actually
in the end we will require, and show, that in this case the stronger
condition $gcd(j_1,\dots,j_t)=1$ is satisfied.}
is satisfied with high probability after just $O(n^2d^2(n))$ samples.
This would allow us to reconstruct $\clint{Np}$ just by taking the
$lcm$ of the denominators of the fractions $\frac{k_i}{MN}$.

Now, dropping the improbable assumption that $\clint{Np}$
divides $MN$, we can use Corollary~\ref{cor:ftts}, Section~\ref{samsect}
to conclude that if $MN>>Np\log^2(Np)$
then we will sample approximations $k_i$ to the fractions
$\frac{j_iMN}{\clint{Np}}$, where the approximations satisfy
\begin{equation}\label{eq:bnd}
\norm{k_i-\frac{j_iMN}{\clint{Np}}}<\frac{MN}{2Np}
\end{equation}
and the $j_i$ are distributed as described in the previous paragraph.
The requirement $MN>>Np\log^2(Np)$ is still
compatible with choosing $N>>M$, so if we could just 
reconstruct the fractions $\frac{j_iMN}{\clint{Np}}$
from the approximations $k_i$ we would be done.

Unfortunately, reconstructing
the $\frac{j_i}{\clint{Np}}$ from the $\frac{k_i}{MN}$
using the continued fractions method requires
a tighter bound than Equation~\ref{eq:bnd} provides -- the fractions
would need to be within $\frac{1}{2\clint{Np}^2}$ of each other.  Previous
results obtain this tighter bound by evaluating the function
past the square of its period -- see for example Algorithm \ref{samunknowp}, 
Section \ref{sect:fsp}.  This is not an option for us since it would
entail choosing $MN>>(Np)^2\log^2(Np)$, incompatible
with our initial assumption of $N>>M$.  We bypass this problem
in the following manner.  First we argue (Lemma \ref{lem:falloff})
that the approximations
$k_i$ output by our procedure
are actually very small in absolute value.  In particular, 
rather than ranging out to the maximal possible $\pm\frac{MN}{2}$,
with exponentially high probability they are within $\pm O(2^{2n}M)$,
regardless of our choice of $N$.  This implies that the
$j_i$ are within $\pm O(2^{3n})$. We can then use continued fractions
to round the ratio
$k_l/k_m$ of any pair of outputs of the Fourier sampling
procedure to the nearest fraction with denominator less than 
$2^{3n}$ and 
this modified continued fractions
procedure terminates correctly, yielding $j_l/j_m$ 
with the correct distribution, as long as
$M=\Omega(2^{11n}\log^2(MN))$.  We can thus reconstruct
$j_1$ by taking the $lcm$ of the numerators fractions $j_1/j_i$
for sufficiently many $i$.  Finally, as long as 
$M=\Omega(2^m2^{11n}\log^2(MN))$,
the leading $m$ bits
of $Mj_1/k_1$ coincide with $p$'s and we can output them
as our final answer.
\section{The Algorithm}
Choose $M=\Omega\left(2^m2^{11n}\log^2{MN}\right)$ and 
$N=\Omega\left(2^{4n}M\right)$, both powers of two.
\begin{algorithm}\label{alg:sampr}{Fourier Sampling over $\Re$}
\begin{enumerate}
\item Generate input superposition $$\sum_{i\in\pm\frac{MN}{2}}
\ket{i}\ket{f(i/N)}.$$ 
\item\label{step:sampr} Fourier Sample over $Z_{MN}$
\end{enumerate}
Repeat this quantum subroutine 
$O(n^2d^2(n))$-times.  Discard any sample
which is less than $\frac{M}{2^m2^{10n}}$ and let
$\{k_i\}$ denote the remaining valid samples.
\begin{itemize}
\item (Classical)Use the continued fractions method to 
round each fraction $k_1/k_i$ to the
closest fraction with denominator less than $2^{3n}$.
\item (Classical) Let $j_1$ be the least common multiple
of the numerators of these fractions.  Output the leading $m$ bits
of $Mj_1/k_1$.
\end{itemize}
\end{algorithm}

It suffices to show that the outputs $\{k_i\}$ from the quantum subroutine
satisfy 
\begin{equation}\label{eq:targ}
\norm{k_i-\frac{j_iMN}{\clint{Np}}}=O\left(\frac{M}{2^m2^{10n}}\right)
\end{equation}
for integers $j_i<2^{3n}$ satisfying $gcd(j_i)=1$.
This bound implies $\norm{k_1/k_i-j_1/j_i}$ is at most $1/2^{6n}$ and
thus the continued fractions procedure correctly delivers
each fraction $j_1/j_i$.  Since the $j_i$ are
relatively prime, $j_1$ will be the least common multiple of
their numerators.  And, finally, $k_1$ will be sufficiently close to
$j_1M/p$, that is, within $M/2^{m+n+1}$, 
to correctly deliver the first $m$ bits of p.

We proceed to show that the statement involving 
Equation~\ref{eq:targ} 
is true with 
exponentially high probability
via the following three lemmas, proved in Sections~\ref{sect:lemfall}
and~\ref{sect:lemseas}.
The first is used to establish that
the $j_i$ are small.  The second and third allow us to
use previous results about functions with integral period
to understand the distribution of the $j_i$ and the 
quality of the approximations $k_i$.
In each of these lemmas we assume that $f$ has
 minimal, as opposed to average, step size at least $1$.
But it is easy to show that given any $f$ with average
step size at least $1$, the function $f(x2^{-n})$
is exponentially close to a function with period
$2^np$ and {\it minimal} step-size $1$.  Thus this assumption entails
only a constant factor penalty in the run-time of the algorithm.

\begin{lemma}\label{lem:falloff}Let $f$ be an integral-valued step function
on $\Re$ with minimal step size $\geq 1$.  Let $\distD_{MN}$
be the distribution on the integers in $\pm MN/2$
induced by sampling $$F_{MN}
\left(\sum_{i\in\pm\frac{MN}{2}}\ket{i}\ket{f(i/N)}\right).$$
Then for all $k$ dividing $N$
$$Pr_{x\in\distD_{MN}}\left(\norm{x}>k^2M\right)=O(1/k).$$
\end{lemma}

\begin{lemma}\label{lem:intcl}Let $f$ be an integral-valued step function
on $\Re$ with minimal step size $\geq 1$ and period $p$.
Then for any $t\in\Re$
$$\frac{1}{MN}
\left\|\sum_{i\in\pm\frac{MN}{2}}
\ket{i}\ket{{f\left(i/N\right)}}-
\sum_{i\in\pm\frac{MN}{2}}
\ket{i}\ket{f\left(pi/(Np+t)\right)}\right\|^2
=O\left(\frac{tM}{Np}\right).$$
\end{lemma}

\begin{lemma}\label{lem:cdn}
Let $f\in C_{1/d(n)}$ be an integral-valued step function
on $\Re$ with minimal step size $\geq 1$ and period $p<2^n$.
Then any rescaling of $f$, $f(\alpha x)$ which has
integral period at least $4d(n)p$ is in $C_{1/2d(n)}$
when restricted to $Z$ (See Definition~\ref{def:codez}
Section~\ref{sect:multdef}). 
\end{lemma}

By Lemma~\ref{lem:falloff}, with exponentially high probability
our samples $k_i$ are at most $2^{2n}M$, and thus the $j_i$
are at most $2^{2n}p<2^{3n}$.  By Lemma~\ref{lem:intcl} and our choice of
$N=\Omega\left(2^{4n}M\right)$ we can assume that
we are Fourier sampling, not our given function, but instead
the function $f(pi/\clint{Np})$ which has integral period $\clint{Np}$.
By Lemma~\ref{lem:cdn} this function is in $C_{1/2d(n)}$ 
when restricted to $Z$.  Thus by the results of   
Section~\ref{sect:rxoverz}, 
after just $\Omega(n^2d^2(n))$
samples  with exponentially high probability 
we have approximations to fractions $j_iMN/\clint{Np}$ with
$gcd(j_i,\clint{Np})=1$.  In this case
we need further that $gcd(j_i)=1$.  If this was not true there
would be some common divisor $d<2^{3n}$, our bound on
the $j_i$.  Choose $r<2^{3n}$ so that $d$ divides $Np+r$.
Then by using Lemma~\ref{lem:intcl} and our choice of $N$
a second time with
the function $f(pi/Np+r)$ we get
that the $j_i$ also satisfy $gcd(j_i,Np+r)=1$ 
with exponentially high probability,
a contradiction.

Finally, we need to ensure that the $k_i$ satisfy 
$$\norm{k_i-\frac{j_iMN}{\clint{Np}}}=O\left(\frac{M}{2^m2^{10n}}\right)$$
Again we assume we are Fourier Sampling the function
$f(pi/\clint{Np})$ over $Z_{MN}$.  If $\clint{Np}$ divided
$MN$ we would be done -- the $k_i$ would exactly equal
the desired fractions.  In general we can apply the Fourier sampling
results from Section~\ref{samsect}.  By measuring 
$F_{MN}(\sum_{i\in\pm\frac{MN}{2}}\ket{i}\ket{f(i)})$
and rounding to the nearest multiple of $MN/k\clint{Np}$
we approximate the distribution gotten by Fourier sampling 
$f(pi/\clint{Np})$ over
$k\clint{Np}$, i.e. the desired distribution.  These distributions
are
exponentially close as long as the number of repetitions $MN/k\clint{Np}$
of this initial
superposition
(this ratio corresponds to the
$R$ in Corollary~\ref{cor:ftts}, Section~\ref{samsect} 
with $k\clint{Np}$ corresponding to $N$) is
$\Omega\left(2^n\log^2(k\clint{Np})\right)$.  
Since the ratio $MN/k\clint{Np}$ is also the approximation error
we must have 
$MN/k\clint{Np}=\Omega\left(2^n\log^2(k\clint{Np})\right)
=O(M/2^m2^{10n})$ which holds by
our choice of $M=\Omega\left(2^m2^{11n}\log^2{MN}\right)$.

We note that without the results of Chapter~\ref{chap:ftt},
a naive analysis -- see the
discussion in Chapter \ref{chap:ftt}, Section~\ref{sect:tb} --
would require that the number of repetitions
$MN/k\clint{Np}$ be at least $k\clint{Np}$ in order
for the distributions to be close.  This would force
$M>N$ which is incompatible with the earlier condition 
$N=\Omega\left(2^{4n}M\right)$.

\subsection{Proof of Lemma~\ref{lem:falloff}}\label{sect:lemfall}
We now prove Lemma~\ref{lem:falloff}.  Notice that this lemma
applies to any step function on $\Re$ with minimal interval 1 
-- we do not require that the
function be periodic. We are taking the Fourier transform of 
this function evaluated on
the fixed interval $\pm M/2$
and allowing the spacing of the evaluations to become finer and finer.
The resulting distributions/superpositions 
approach a fixed limit which is concentrated   
within small multiples of $\pm M/2$. Intuitively this is because
allowing the evaluations' spacing to become finer while the step function 
remains fixed does not add any large Fourier coefficients -- these 
correspond to functions which vary rapidly and our step function is appearing
increasingly smooth.
For our purposes it
suffices to prove the following Lemma about the tails of these distributions.
\begin{oldlemmalem:falloff}
Let $f$ be an integral-valued step function
on $\Re$ with minimal step size $\geq 1$.  Let $\distD_{MN}$
be the distribution on the integers in $\pm MN/2$
induced by sampling $$F_{MN}
\left(\frac{1}{\sqrt{MN}}\sum_{i\in\pm MN/2}\ket{i}\ket{f(i/N)}\right).$$
Then for all $k$ dividing $N$
$$Pr_{x\in\distD_{MN}}\left(\norm{x}>k^2M\right)=O(1/k).$$
\end{oldlemmalem:falloff}

\begin{proof}
Fix any $k$ dividing $N$.
Let $$\keta=\frac{1}{\sqrt{MN}}\sum_{i\in\pm MN/2}\ket{i}\ket{f(i/N)}$$
and $$\ket{\alpha^\prime}=\frac{1}{\sqrt{MN}}
\sum_{i\in\pm Mk/2}\sum_{j\in\pm N/2k}
\ket{\frac{N}{k}i+j}\ket{f(i/k)}.$$

We claim that $\|\keta-\ket{\alpha^\prime}\|^2=O(1/k)$.
This squared distance is just twice the fraction of 
pairs $(i,j)$ for which $f(i/k)\neq f(i/k+j/N)$.  Since
$|j|<N/2k$ this can only be true for when $i/k$ is within $1/2k$
of either end of an interval.  Since the intervals have length
at least $1$ this occurs for at most a $1/k$ fraction of the $i$
and likewise for the pairs $(i,j)$.

Now, the behavior of $F_{MN}\ket{\alpha^\prime}$ is easy to
analyze.  Its amplitude at $\ket{x}\ket{c}$ is
\begin{equation}\label{eq:ft}
\frac{1}{MN}\sum_{\scriptstyle i\in\pm Mk/2\atop\scriptstyle f(i/k)=c}
\sum_{j\in\pm N/2k}\om{MN}
{(\frac{N}{k}i+j)x}=
\left(\frac{1}{Mk}\sum_{\scriptstyle i\in\pm Mk/2\atop\scriptstyle f(i/k)=c}
\om{Mk}{ix}\right)
\left(\frac{k}{N}\sum_{j\in \pm N/2k}\om{MN}{jx}\right).
\end{equation}
The RHS of Equation~\ref{eq:ft} is easily seen to
be the product of the amplitude of
\begin{equation}\label{expone}
F_{Mk}\left(\frac{1}{\sqrt{Mk}}
    \sum_{\scriptstyle i\in\pm Mk/2\atop\scriptstyle f(i/k)=c}
\ket{i}\right)
\end{equation}
at $\ket{x\bmod Mk}$ and the amplitude of 
\begin{equation}\label{exptwo}
\sqrt{Mk}\cdot F_{MN}\left(F_{N/k}^{-1}\ket{0}\right)
\end{equation}
at $\ket{x}$.
The amplitudes in Superposition~\ref{exptwo} fall off away from zero like
$1/x$ while the amplitudes in Superposition~\ref{expone}
just keep repeating in blocks of size $Mk$.
This will allow us to show that their product also falls off quickly away from zero.  In particular, Observation~\ref{falloff}, Section~\ref{secondapp} 
gives us that the amplitude of
$F_{MN}\left(F_{N/k}^{-1}\ket{0}\right)$ at $\ket{j}$ is at most
\begin{equation}\label{eq:ob}
\sqrt{Mk}\cdot\frac{2}{|j|}.
\end{equation}
For convenience we let $\kbeta$ denote the Superposition ~\ref{expone}, 
that is
$$\kbeta=\sum_{i\in\pm Mk/2}\beta_i\ket{i}=F_{Mk}\left(\frac{1}{\sqrt{Mk}}
    \sum_{\scriptstyle i\in\pm Mk/2\atop\scriptstyle f(i/k)=c}
\ket{i}\right).$$ Then we can use (\ref{eq:ob}) to
bound the sum of the amplitudes squared
of $F_{MN}\ket{\alpha^\prime}$ 
in the $t$th block of size $Mk$ by
$$\sum_{i\in\pm Mk/2}\norm{\beta_i}^2\norm{\frac{2Mk}{tMk}}^2=
O\left(\frac{1}{t^2}\right).$$
In other words the probability
falls off as $1/t^2$ with the $t$th block of size $Mk$.  Thus the 
probability of being larger than $k^2M$ is $O(1/k)$.

Combining the closeness of $\keta$ and $\ket{\alpha^\prime}$
with this falloff of $F_{MN}\ket{\alpha^\prime}$
gives the desired result.\end{proof}

\subsection{Proofs of Lemmas~\ref{lem:intcl} and \ref{lem:cdn}}
\label{sect:lemseas}
We now prove two easy lemmas which allow us
to use results from the previous Chapters about
functions with integral period.
\begin{oldlemmalem:intcl}Let $f$ be an integral-valued step function
on $\Re$ with minimal step size $\geq 1$ and period $p\in \Re$.
Then for any $t$
$$\frac{1}{MN}\left\|\sum_{i\in\pm\frac{MN}{2}}\ket{i}{f(i/N)}-
\sum_{i\in\pm\frac{MN}{2}}\ket{i}\ket{f(pi/Np+t)}\right\|^2
=O\left(\frac{tM}{Np}\right).$$
\end{oldlemmalem:intcl}

This squared distance is just twice the probability that
$f(i/N)\neq f(pi/Np+t)$.  Since for all $i$
$$\norm{\frac{i}{N}-\frac{pi}{Np+t}}=\norm{\frac{ti}{N(Np+t)}}=
O\left(\frac{tM}{Np}\right),$$
in order for the function values to differ,
$i/N$ must be within $O(tM/Np)$ of the end of a step interval.
Since the intervals have length at least $1$ this applies to at
most a $O(tM/Np)$ fraction of the $i$.
\begin{oldlemmalem:cdn}
Let $f\in C_{1/d(n)}$
be an integral-valued step function
on $\Re$ with minimal step size $\geq 1$ and period $p<2^n$.
Then any rescaling of $f$, $f(tx)$ which has
integral period at least $4d(n)p$ is in $C_{1/2d(n)}$
when regarded as a function on $Z$ (See Definition~\ref{def:codez},
Section~\ref{sect:multdef}.) 
\end{oldlemmalem:cdn}
Let $f(tx)$ be any rescaling of $f$ with integral
period $p_t>4d(n)p$.
Suppose $f(tx)\not\in C_{1/2d(n)}$ as a function over
$Z$.  Then there exists a function $g$ with integral period $p_g<p_t$
so that $f(tx)$ and $g$ differ on less than a $1/2d(n)$ fraction
of the inputs in $[0,p_g\cdot p_t]$. We can turn $g$ into a function on $\Re$
with period $p_g$ by letting its value at a non-integral input
correspond to its value at the nearest integer.  The distance
between $f(tx)$ and $g$ when regarded as functions on $\Re$
is small.  In particular, since $f(tx)$ has been rescaled
to have step intervals of size at least $4d(n)$, they are identical
at at least a $1-1/2d(n)-2/4d(n)$ fraction of the values, leading
to a distance of at most $1/d(n)$.  Our original function
$f$ and the function $g(x/t)$ have the same distance, with
the latter function's period equal to $p_g/t<p_f$, a contradiction
to our assumption that $f\in C_{1/d(n)}$.

\chapter{Hidden Subgroups over the Reals and MA}\label{chap:or}\section{Quantum vs. Classical Complexity Classes}
A primary method for delineating the power of quantum computation
is by comparison with various classical complexity classes.  
The Arthur-Merlin
hierarchy \cite{BabaiMor} of probabilistically-checkable interactive proofs
provides a natural backdrop for measuring quantum complexity.
First, due to the inherently probabilistic nature of
quantum computation, this hierarchy is a more natural choice than
$PH$ as a basis for comparison.  In addition, problems like
Graph Isomorphism which have defied classification as 
$NP$-complete are considered the most plausible 
candidates for possessing efficient quantum algorithms
achieving exponential advantage over classical computation.
These problems also tend to have non-trivial characterizations
in the $AM$ hierarchy -- for instance, Graph Isomorphism is known to
be in $Co-AM$.

Unlike $PH$, the Arthur-Merlin hierarchy is known not to be strict.  
In particular,
$MA\subset AM$ and any constant number of rounds
of interaction can be reduced to $AM\subseteq\Pi_2$ \cite{BabaiMor}. 
However, allowing polynomially many rounds of interaction
yields all of $PSPACE$ -- this is the well-known result $IP=PSPACE$
\cite{shamir90}.
While it may be possible to show directly that $BQP$ lies inside
a particular level of the Arthur-Merlin hierarchy, 
results showing that $BQP$ lies outside a level of the hierarchy
can only be given in the relativized or oracle setting.
In particular, since it is known that 
$P\subseteq BQP\subseteq P^{\#} \subseteq PSPACE$
a direct result of this sort would prove
$P\neq PSPACE$, one of the nasty, long-standing open problems
in complexity.

There is an oracle $\mathcal O$
separating $BQP$ from $MA$, that is, for which 
$BQP^{\mathcal O}\not\subseteq MA^{\mathcal O}$.  This was first claimed
in \cite{SICOMP::BernsteinV1997} 
but the first proof was given in \cite{watrous2000} via a different
oracle.  This also implies a separation between $BQP$ and $MA\cup Co-MA$
due to the fact that $BQP$ is closed under complementation.
An open and intriguing question is whether there exists 
an oracle separating $BQP$ from $AM$.  There has been speculation
that $BQP$ is actually contained in $AM$.  
This is due to the fact that $AM$
can perform an approximate count 
of the number the accepting paths of 
an $NP$-machine.  The proof that $BQP\subseteq P^{\#}$ relies on the fact
that exact counts of this form are sufficient 
to solve any problem in $BQP$
and it has been conjectured that approximate counting might
also be sufficient.  An oracle separation of $BQP$ from $AM$ would
be an indication to the contrary. In addition it 
would show that any proof of
$BQP\subseteq AM$
must use non-relativizing techniques, in contrast
to the result $BQP\subseteq P^{\#}$. 

We exhibit two oracle promise problems which
achieve the weaker separation 
$BQP^{\mathcal O}\not\subseteq MA^{\mathcal O}$.
The first of these promise problems is just a decision
version of Simon's problem (Section~\ref{sect:simon}) and its virtue lies
in being much simpler than the oracles of \cite{SICOMP::BernsteinV1997} 
and \cite{watrous2000} --
the proof that it is outside of $MA$ is almost trivial.
We also give a simple variant of this problem which is
in $BQP$ by the results of Chapter~\ref{chap:nmulti} but which we
suspect to be outside of $AM$, in other words, a candidate for
the stronger separation result discussed above. 

The second problem which separates $BQP$ from $MA$ is the decision
version of period-finding over $\Re$, shown to have
an efficient quantum solution in
Chapter~\ref{chap:overr}.  We observe that the analogous 
problem over the integers is in $MA$, and thus demonstrate that
period-finding over the reals is more difficult than
its integral counterpart.
This may also support the current state of knowledge
about the relationship between factoring and Pell's equation.
There is a reduction from factoring, which can be reduced
to period-finding over the integers, to Pell's equation, which
can be reduced to period-finding over the reals, but no reduction
in the opposite direction exists.

\section{MA}\label{sect:ma}

We take for our definition of $MA$ a version with one-sided
error which has been shown to be equivalent (see for example \cite{zachos}) 
to the standard definition given in \cite{BabaiMor}:
\begin{definition}
A promise problem $\distP$ is in $MA$ if and only if
for all sufficiently large polynomials $q$ there is a 
polynomial $r$ and 
a predicate $R$ computable in
deterministic polynomial-time with access to $f$
such that 
$$f\in \yes_n\longrightarrow
\exists x\in \Sigma^{q(n)}
\forall y\in  \Sigma^{r(n)} 
R^f(x,y)=1$$
and
$$f\in \no_n\longrightarrow
\forall x\in \Sigma^{q(n)},
\norm{y\in \Sigma^{r(n)} 
R^f(x,y)=1}<2^{-2q(n)}2^{r(n)}.$$
\end{definition}

The following Lemma is implicit in the literature
and is useful in proving lower bounds related to $MA$.
For a given predicate $R$ and a pair of strings $x\in \Sigma^{q(n)}$
and $y\in  \Sigma^{r(n)}$, we say that two oracles $f$ and $g$ 
are equivalent under
$R(x,y)$, or $\twid{f}{g}$, if the runs of $R^f(x,y)$ and $R^g(x,y)$
produce identical oracle query/answer transcripts.
Then we have the following:
\begin{lemma}\label{lem:sma}
If a promise problem $\distP\in MA$ via $R$ then for all $n$
there exists an oracle $f\in\yes_n$ and strings and $x\in \Sigma^{q(n)}$
and $y\in  \Sigma^{r(n)}$ such that
\begin{equation}\label{eq:ratio}
\frac{Pr_{g\in \no_n}\left(\twid{g}{f}\right)}
{Pr_{g\in \yes_n}\left(\twid{g}{f}\right)}
<2^{-q(n)}.
\end{equation}
\end{lemma}
We give a proof of Lemma~\ref{lem:sma} in Section~\ref{sect:sma}.
As an easy application we give a proof that the following
decision
version of Simon's problem (Section~\ref{sect:simon}) is outside 
of $MA$.
\begin{pproblem}{$\mathcal {NBS}$(No Bit-string)}
\begin{description}
\item[$\yes_n$:]  $f:(Z_2)^n\rightarrow (Z_2)^n$ is $1-1$.

\item[$\no_n$:]  $f:(Z_2)^n\rightarrow (Z_2)^n$ is $2-1$ and there exists some $b$
such that for all $x$, $f(x)=f(x\oplus b)$.
\end{description}
\end{pproblem}
\begin{proof}[$\mathcal {NBS}\not\in MA$]
It is easy to see that $\norm{\yes_n}=2^n!$ and
$$\norm{\no_n}=2^n\frac{2^n!}{(2^n-2^{n-1})!}.$$ 
Take any $f\in\yes_n$ and strings $x,y$ and look at the
transcript of $R^f(x,y)$.  
We can assume without loss that all such transcripts
contain exactly $t$ oracle queries, where $t$ is bounded by the polynomial 
run-time of $R$.  The number of functions
in $g\in \yes_n$ such that $\twid{g}{f}$, that is, 
which induce an oracle transcript identical to $f$'s, is exactly
$(2^n-t)!$.  The number of oracles $g\in\no_n$ with $\twid{g}{f}$
is at least 
$$(2^n-t^2)\frac{(2^n-t)!}{(2^n-2^{n-1})!}$$ 
because at most $t(t+1)/2$ hidden bit strings
have been ruled out by the $t$ oracle queries.
Thus we get that
$$Pr_{g\in \no_n}\left(\twid{g}{f}\right)\geq
\frac{(2^n-t^2)\frac{(2^n-t)!}{(2^n-2^{n-1})!}}
{2^n\frac{2^n!}{(2^n-2^{n-1})!}}=
\frac{(2^n-t^2)(2^n-t)!}{2^n(2^n!)}$$
while
$$Pr_{g\in \yes_n}\left(\twid{g}{f}\right)=\frac
{(2^n-t)!}{2^n!}.$$
For all oracles $f$ and strings $x,y$
the ratio (\ref{eq:ratio}) of these two quantities is thus at least
$$\frac{2^n-t^2}{2^n}=\Omega(1)$$
since $t$ is bounded by a polynomial.
Thus we have $\mathcal{NBS}\not\in MA$.
\end{proof}

Since $\mathcal{NBS}\in BQP$ via Simon's algorithm
(modified slightly to answer the appropriate decision problem),
a routine diagonalization
procedure -- see for example \cite{watrous2000} --
gives the oracle separation result 
$BQP^{\mathcal O}\not\subseteq MA^{\mathcal O}$.

There is an easy protocol showing that $\mathcal{NBS}\in AM$ --
the verifier chooses a value $y\in(Z_2)^n$ from the possible range of $f$
and the prover provides an $x\in (Z_2)^n$ with the verifier 
accepting iff $f(x)=y$.
It is easy to see that the prover can convince the verifier with probability
$1$ if $f\in \yes$ and with probability at most $1/2$ otherwise.  
This is a simple example of an approximate counting protocol -- in this
case the size of the range of $f$ is being estimated.  Notice that the protocol
actually distinguishes between arbitrary $1-1$ and $2-1$ 
functions from $(Z_2)^n$
to $(Z_2)^n$ and has nothing to do with the hidden bit-string structure
of the $\no$ functions.


\section{Period-finding over $\Re$ is outside of $MA$}
We now prove that the period-finding problem over $R$,
for which an efficient quantum algorithm was given in 
Chapter~\ref{chap:overr},
is not in $MA$.  In particular we show this for a
decision version of the period-finding problem
which corresponds to learning the leading bit of the period.
\begin{pproblem}{$\mathcal {P}\Re$(Period-finding over $\Re$)}
\label{pp:pr}
\begin{description} 
\item[$\yes_n$:]  $f\in C_{1/3}$ 
is a step function on $\Re$ with average interval
$\geq 1$ and period $p<2^n$ satisfying $2^m\leq p<2^{m+1}$.
\item[$\no_n$:]  $f\in C_{1/3}$ 
is a step function on $\Re$ with average interval
$\geq 1$ and period $p<2^n$ which does not satisfy $2^m\leq p<2^{m+1}$.
\end{description}
\end{pproblem}
\begin{theorem}
$$\mathcal {P}\Re\not\in MA$$
\end{theorem}
The fact that this problem is outside of $MA$ supports the  
intuition that period-finding over the reals is more difficult
than over the integers.  In particular, the analogous decision
problem over $Z$,  
\begin{pproblem}{$\mathcal {PZ}$(Period-finding over $Z$)}
\begin{description} 
\item[$\yes_n$:]  $f\in C_{1/3}$ is a function on $Z$ with
integral period $p<2^n$ satisfying $2^m\leq p<2^{m+1}$.
\item[$\no_n$:]   $f\in C_{1/3}$ is a function on $Z$ with
integral period $p<2^n$ which does not satisfy $2^m\leq p<2^{m+1}$.
\end{description}
\end{pproblem}
{\it is} in $MA$.  In this case a proof that $f\in \yes_n$
could consist of the period $p$, the prime factorization of $p$,
and primality certificates for each of these prime factors.
If $f$ is one-to-one on its period
we can test this proof deterministically (and thus this 
restricted problem is in $NP$).  We would first
verify the factorization of $p$ and the validity of the
primality certificates -- see \cite{pratt75} for the proof that
$PRIMES\in NP$.  Then we check that $f(x)=f(x+p)$
for an arbitrary choice of $x$.  This insures that
the claimed $p$ is a multiple of the period.  
Finally,
for each prime $p_i$ in the factorization of $p$
we verify that for an arbitrarily chosen $x$,
$f(x)\neq f(x+p/p_i)$.
This test, which can be done efficiently since there are at most
$n$ such primes, rules out any $p$ which is a proper multiple
of the
true period.  After this verification that $p$
is in fact the period we accept iff $2^m<p<2^{m+1}$.

For a general $f\in C_{1/3}$ we need merely
randomize the function checks in the above proof, accepting
if
$f(x)=f(x+p)$ for a randomly chosen $x$
and if for each $i$ $f(x)\neq f(x+p/p_i)$ with significant 
probability.  This gives a probabilistic check of the
above proof and establishes that $\mathcal{PZ}\in MA$.

We now show that $\mathcal{P}\Re\not\in MA$.  The proof is based
on the fact that, while in the integral case there is a short proof
to rule out any multiple of the period, such a proof does not exist
when the period is allowed to be rational.  In the integral case
we can check the function at $\leq n$ pairs of points $p/p_i$ apart, 
one
for each prime $p_i$ in the factorization of $p$, and ensure that
none of the potentially exponentially many proper divisors 
of $p$ is the period.  In the real
case to ensure that the function has period $p$ we must
rule out all rationals $p/k$, $k<p$ as possible periods.  There
is no similar polynomially sized set of points 
which can accomplish this check, even probabilistically.

\begin{proof}[$\mathcal{P}\Re\not\in MA$]
We first describe the restricted distributions of $\yes$ and $\no$
oracles which we will use.  Picking the correct
restriction of the original promise problem $\mathcal {P}\Re$
is half the battle -- one must find a restriction which
is fairly structured in order to count the oracles, but too
much structure invariably reduces the problem to the integral
version which {\it does} have an $MA$ proof system.  

We first fix the parameter $m$ so that $2^m$ is superpolynomial
in $n$.  Then let 
$\{p_i, i\in I\}$ be the set of primes satisfying 
$1<p_i<2^{\frac{m}{4}}$ and note that, by the Prime Number Theorem,
$|I|$ is also superpolynomial in $n$. Finally, let $$N=k\prod_{i\in I}p_i$$
for some integer $k$.
\begin{definition}[$\yes$ and $\no$]
Our $\yes$ functions all have period $2^m$ and are specified in the
following manner.
We choose $2^m-1$ values uniformly at random in the set
$$[0,2^m]\cap\{\text{ fractions with denominator }N\}.$$
These are the endpoints of the step intervals of the function.
We then choose a value in $\{1,\dots,2^m\}$ for each of our
steps in such a way that the function is $1-1$ modulo its
step intervals.  Finally, we discard any function which has
maximal step interval at least $2^{m/3}$.  The number of
such functions is $$2^m!{N2^m\choose 2^m-1}$$
minus the functions discarded for having too long an interval.
The fraction of functions thus discarded is very small
-- the probability of having an interval of length
at least $2^{m/3}$ is less than 
$$2^{\frac{2m}{3}+1}\left(1-\frac{1}{2^{\frac{2m}{3}+1}}\right)^{2^m-1}<2^{m}e^{-2^{\frac{m}{3}}}$$
and we shall be able to ignore it in our
calculations.

We now turn to our $\no$ functions.  For each $i\in I$
$\no_i$ will be a collection of functions with period $\frac{2^m}{p_i}$.
We define the functions on their period in an manner similar to
the $\yes$ functions. We choose $\li{\frac{2^m}{p_i}}-1$ 
values uniformly at random in the set
$$\left[0,\frac{2^m}{p_i}\right]\cap\{\text{ fractions with denominator }N\}.$$
These are the endpoints of the step intervals of the function.
We then choose a value in $\{1,\dots,2^m\}$ for each of our
steps in such a way that the function is $1-1$ on its period modulo the
step intervals.  Again we discard the very small fraction of functions
which have maximal step interval at least $2^{m/3}$.
The number of such functions is 
$$\frac{2^m!}{\left(2^m-\li{\frac{2^m}{p_i}}\right)!}
{N\frac{2^m}{p_i}\choose \li{\frac{2^m}{p_i}}-1}$$
minus the small fraction of functions discarded for 
having too long an interval.  Again these form such a small
fraction of the total that we can effectively ignore them.
Finally, we shall be interested in the class $\no$ which is a weighted
union of the $\no_i,i\in I$, with each $\no_i$ reweighted to
have an equal number of functions.
\end{definition}
We note that the $\yes$ and $\no$ oracles defined above
are in fact a subclass of the original promise problem $\mathcal{P}\Re$.
Clearly they have period less than $2^n$ and average interval at least $1$. 
The fact that they are in $C_{1/3}$ follows from the cap on the
length of the maximal step interval together with the fact that
they are $1-1$ modulo their steps.  

Let $R(x,y)$ be any deterministic predicate which
runs in time $t(n)$ and purports to yield an $MA$
proof system for $\mathcal{P}\Re$ with parameters
$q(n)$ and $r(n)$. Then clearly $R(x,y)$ also
yields an $MA$ proof system for the $\yes$ and $\no$
oracles defined above with the same parameters.
We can further assume that on this restricted problem
all oracle queries are made on inputs in the interval $[0,2^m]$
by interpreting the original queries$\pmod {2^m}$.

Recall the equivalence relation $\twid{}{}$
defined in Section~\ref{sect:ma}. The following lemma
is the main technical result establishing our theorem and
is proved in Section~\ref{sect:twid}:
\begin{lemma}\label{lem:twid}
There is a constant $c>0$ such that for all sufficiently large $n$,
$f\in\yes$, $i\in I$, and strings $x\in \Sigma^{q(n)}$
and $y\in  \Sigma^{r(n)}$
\begin{equation}\label{eq:tratio}
\frac{Pr_{g\in \no_i}\left(\twid{g}{f}\right)}
{Pr_{g\in \yes}\left(\twid{g}{f}\right)}\geq c
\end{equation}
unless the transcript of $R^f(x,y)$ includes a pair of 
oracle inputs $(u,v)$ satisfying
\begin{equation}\label{eq:diff}
k\frac{2^m}{p_i}-2^{m/3}<u-v<k\frac{2^m}{p_i}+2^{m/3},
\end{equation}
for some integer $k$ satisfying $|k|<p_i$.
\end{lemma} 
Informally this says that $R$ can only distinguish between the $\yes$ oracles
which have period $2^m$ and the $\no_i$ oracles 
with period $2^m/p_i$ if it actually queries a pair of intervals
which are a multiple of $2^m/p_i$ 
apart and thus rules out the possibility
of a $\no_i$ oracle.  

We now turn to the question of distinguishing $\yes$ oracles from the
full collection of $\no$ oracles.  The idea is that a successful
$MA$ proof would have to rule out the possibility that
$f\in \no_i$ for almost all $i\in I$ and
thus examine the function on pairs $(u,v)$ of the above form  
for almost all $i\in I$, 
but this requires making exponentially many oracle queries
in polynomial-time!

We first claim that any pair of queries
$(u,v)$ to the oracle can satisfy Equation (\ref{eq:diff}) for at most
$n$ of the $i\in I$.  
Suppose $(u,v)$ satisfies Equation
(\ref{eq:diff}) for the prime $p$ and the integer $k$.
Then in order for it to also satisfy the same equation
for another prime $p_i$ we must have
$$\norm{k\frac{2^m}{p}-k_i\frac{2^m}{p_i}}<2^{m/3+1}$$
which implies
$$\norm{\frac{k}{p}-\frac{k_i}{p_i}}<\frac{1}{(2^{m/4})^2}.$$
By our choice of $p,p_i<2^{m/4}$ these fractions
must therefore be exactly equal, or $k=lpp_i$ for some integer $l$.
Since $|k|<2^n$
it can have at most $n$ distinct prime factors and thus Equation 
(\ref{eq:diff}) can be satisfied simultaneously for at most $n$ 
of the $i\in I$. 

We now show that for all $f\in \yes$ and for all $x,y$,
\begin{equation*}
\frac{Pr_{g\in \no}\left(\twid{g}{f}\right)}
{Pr_{g\in\yes}\left(\twid{g}{f}\right)}=\Omega(1)
\end{equation*}
This will establish the result since 
it is a violation of Lemma~\ref{lem:sma}.
Now,
\begin{eqnarray}
\frac{Pr_{g\in \no}\left(\twid{g}{f}\right)}
{Pr_{g\in\yes}\left(\twid{g}{f}\right)}&=&
\frac{\norm{\{g\in\no|\twid{f}{g}\}}}
{\norm{\no}}
\frac{\norm{\yes}}
{\norm{\{g\in\yes|\twid{f}{g}\}}}\\
&=&\frac{\sum_{i\in I}\norm{\{g\in\no_i|\twid{f}{g}\}}}
{\sum_{i\in I}\norm{\no_i}}
\frac{\norm{\yes}}
{\norm{\{g\in\yes|\twid{f}{g}\}}}\\
&=&\sum_{i\in I}\frac{\norm{\{g\in\no_i|\twid{f}{g}\}}}
{|I|\norm{\no_i}}
\frac{\norm{\yes}}
{\norm{\{g\in\yes|\twid{f}{g}\}}}\\
&=&\sum_{i\in I}\frac{1}{|I|}\frac{Pr_{g\in \no_i}\left(\twid{g}{f}\right)}
{Pr_{g\in \yes}\left(\twid{g}{f}\right)}\\
\end{eqnarray}
where the second to the last equation 
follows from the fact that the $\no_i$ have
been given equal weights.
By throwing out the at most $nt^2$ $i\in I$ for which 
pairs of queries satisfying Equation~(\ref{eq:diff})
have been made, and applying the bound in Lemma~\ref{lem:twid} to the rest
we have that the above quantity is
$$\Omega\left(\sum_{i<|I|-nt^2}\frac{1}{|I|}\right)=\Omega\left(\frac{|I|-nt^2}{|I|}\right)=\Omega(1)$$
where the last equality follows from the fact that $|I|$ is
superpolynomial in $n$.  This completes the proof.
\end{proof}

\subsection{Proof of Lemma~\ref{lem:twid}}\label{sect:twid}
\begin{oldlemmalem:twid}
There is a constant $c>0$ such that for all sufficiently large $n$,
$f\in\yes$, $i\in I$, and strings $x\in \Sigma^{q(n)}$
and $y\in\Sigma^{r(n)}$
\begin{equation}\label{eq:tratiotwo}
\frac{Pr_{g\in \no_i}\left(\twid{g}{f}\right)}
{Pr_{g\in \yes}\left(\twid{g}{f}\right)}\geq c
\end{equation}
unless the transcript of $R^f(x,y)$ includes a pair of 
oracle inputs $(u,v)$ satisfying
\begin{equation}\label{eq:difftwo}
k\frac{2^m}{p_i}-2^{m/3}<u-v<k\frac{2^m}{p_i}+2^{m/3},
\end{equation}
for some integer $|k|<p_i$.
\end{oldlemmalem:twid}

We first define an equivalence relation on our functions which
is a refinement of $\twid{}{}$.
We let $f{\bf\sim }g$ if both $\twid{f}{g}$ {\it and} the at most $t$
step intervals queried on a run of $R^f(x,y)$
are identical on their endpoints.  In other words, not only the values
of the steps which are queried but also the steps themselves are identical.
This is clearly a refinement of $\twid{}{}$ and thus it suffices to 
prove the above lemma with $\twid{}{}$ replaced by $\sim$.

Fix any $f\in\yes$, $i\in I$, and strings $x\in \Sigma^{q(n)}$
and $y\in\Sigma^{r(n)}$.  We can assume without loss of generality
that exactly $t$ intervals are queried on any run, where $t=t(n)$
is the run-time of $R$.  Let $T$ denote the total length of the 
step intervals which
are queried in $R^f(x,y)$ -- note that $T<t2^{m/3}$ 
since no interval is longer than $2^{m/3}$.
The number of $\yes$ functions $g$ for which $f\sim g$ is
\begin{equation*}
(2^m-t)!{N(2^m-T)\choose 2^m-2t-1}.
\end{equation*}
minus the exponentially small fraction of these functions which have 
maximal interval greater than $2^{m/3}$.

Now, if none of these $t$ intervals overlap
when they are mapped back$\pmod {2^m/p_i}$
to the interval $[0,2^m/p_i]$, and this is the case when
Equation~(\ref{eq:difftwo}) is not satisfied by any pair of queries,
then the number of $\no$ functions $g$ for which $f\sim g$ is
\begin{equation*}
\frac{\left(2^m-t\right)!}
{\left(2^m-\li{\frac{2^m}{p_i}}\right)!}
{N(\frac{2^m}{p_i}-T)\choose\clint{\frac{2^m}{p_i}}-2t-1},
\end{equation*}
minus the small fraction of these functions which have 
maximal interval greater than $2^{m/3}$. Here we also use 
the fact that $N$ is a multiple of $p_i$ for each 
$i\in I$. This ensures that the endpoints of the original intervals 
interpreted$\pmod {2^m/p_i}$ are valid choices for the $\no_i$
oracles.

By cancelling all the factorials in the ratio in question,
\begin{equation*}\label{eq:ach}
\frac{Pr_{g\in \no_i}\left(\twid{g}{f}\right)}
{Pr_{g\in \yes}\left(\twid{g}{f}\right)}\approx
\frac
{{N(\frac{2^m}{p_i}-T)\choose \clint{\frac{2^m}{p_i}}-2t-1}}
{{{N\frac{2^m}{p_i}}\choose{\clint{\frac{2^m}{p_i}}-1}}}
\frac
{{N2^m\choose 2^m-1}}
{{{N(2^m-T)}\choose{2^m-2t-1}}}.
\end{equation*}
where the approximation reflects the fact that we have thrown out
an exponentially small fraction from each class for having too large
a maximal interval.  At this point it is easy to
see that this can be ignored.  Since we are free to choose 
$N=k\prod_{i\in I}p_i$ as large as we want we use the fact that
when $N$ is sufficiently large ${N\choose l}\approx \frac{N^l}{l!}$ 
to conclude that
Equation~\ref{eq:ach} is approximately  
\begin{equation}\frac{\left(\clint{\frac{2^m}{p_i}}-1\right)!
\left(2^m-2t-1\right)!}
{\left(\clint{\frac{2^m}{p_i}}-2t-1\right)!\left(2^m-1\right)!}\cdot
\frac{\left(N\frac{2^m}{p_i}\right)^
{\clint{\frac{2^m}{p_i}}-2t-1}}
{\left(N\frac{2^m}{p_i}\right)^
{\clint{\frac{2^m}{p_i}}-1}}
\frac
{\left(N2^m\right)^
{2^m-1}}
{\left(N2^m\right)^
{2^m-2t-1}}
\cdot
\frac{\left(1-\frac{T}{2^m/p_i}\right)^{\clint{\frac{2^m}{p_i}}-2t-1}}
{\left(1-\frac{T}{2^m}\right)^{2^m-2t-1}}.
\end{equation}
The first ratio in Equation \ref{eq:ach} can be seen to approach 
$p_i^{-2t}$ as $n$ (and thus $m$) 
goes to infinity by cancelling terms in the factorials.  
In a similar manner the second ratio can be seen to approach
$p_i^{2t}$.
We now proceed to show that the third
ratio is approaches $1$ and the Lemma follows. We can rewrite this ratio
as
$$\left[\frac{\left(1-\frac{1}{2^m/Tp_i}\right)^\frac{2^m}{Tp_i}}
{\left(1-\frac{1}{2^m/T}\right)^\frac{2^m}{T}}\right]^T
\cdot
\frac{\left(1-\frac{1}{2^m/Tp_i}\right)^{\epsilon-2t-1}}
{\left(1-\frac{1}{2^m/T}\right)^{-2t-1}}$$
where $\epsilon=\clint{\frac{2^m}{p_i}}-\frac{2^m}{p_i}\leq 1/2$.
Now the second of these ratios has numerator and denominator
both close to $1$ since $t<<2^m/Tp_i$.  Thus we can
ignore this ratio and focus on the first.

We use the fact that the expression $\left(1-1/n\right)^n$
converges to $e^{-1}$ with error $O(1/n)$ to conclude that
the ratio 
$$\frac{\left(1-\frac{1}{2^m/Tp_i}\right)^\frac{2^m}{Tp_i}}
{\left(1-\frac{1}{2^m/T}\right)^\frac{2^m}{T}}$$
is within $O(Tp_i/2^m)$ of $1$.  Finally since $$T<<\frac{2^m}{Tp_i}$$
this ratio raised to the $T$th power is still
very close to $1$ and the result follows.

\subsection{Proof of Lemma~\ref{lem:sma}}\label{sect:sma}
\begin{oldlemmalem:sma}
If a promise problem $\distP\in MA$ via $R$ then for all $n$
there exists an oracle $f\in\yes_n$ and strings and $x\in \Sigma^{q(n)}$
and $y\in\Sigma^{r(n)}$ such that
\begin{equation*}\label{eq:ratiotwo}
\frac{Pr_{g\in \no_n}\left(\twid{g}{f}\right)}
{Pr_{g\in \yes_n}\left(\twid{g}{f}\right)}
<2^{-q(n)}.
\end{equation*}
\end{oldlemmalem:sma}

\begin{proof}Since there are $2^{q(n)}$ possible proof strings $x$
there exists at least one such string which serves as a valid proof
for at least a $2^{-q(n)}$ fraction of the oracles.
Fix any such proof $x$.  We have that
$$\forall y\in\Sigma^{r(n)}
\norm{\{g\in\yes|R^g(x,y)=1\}}
\geq 2^{-q(n)}|\yes|,$$
and thus
$$\sum_{y\in\Sigma^{r(n)}}
\norm{\{g\in\yes|R^g(x,y)=1\}}
\geq 2^{-q(n)}|\yes|2^{r(n)}.$$

If Equation~\ref{eq:ratiotwo} is violated for
all $f,x,$ and $y$ then by viewing each set $\{g\in\no|R^g(x,y)=1\}$
as a 
union of $\twid{}{}$ equivalence classes
we have that for all $y$
$$\norm{\{g\in\no|R^g(x,y)=1\}}\geq
2^{-q(n)}\norm{\{g\in\yes|R^g(x,y)=1\}}
\frac{|\no|}
{|\yes|}.$$
Putting these together we get that
$$\sum_{y\in\Sigma^{r(n)}}
\norm{\{g\in\no|R^g(x,y)=1\}}
=\sum_{g\in\no}
\norm{\{y\in\Sigma^{r(n)} |R^g(x,y)=1\}}
\geq 2^{-2q(n)}|\no|2^{r(n)}.$$
But this implies that there exists a $g\in\no$
such that 
$$\norm{\{y|R^g(x,y)=1\}}\geq 2^{-2q(n)}2^{r(n)},$$
contradicting the definition of $MA$.
\end{proof}

\chapter{Fourier Transform Theorems}\label{chap:ftt}
In this chapter we establish the technical results
leading to the QFT Algorithm~\ref{alg:approxft} and the Fourier Sampling
Algorithms~\ref{samknowp} and \ref{samunknowp}.
First prove a version of the Fourier Sampling Lemma~\ref{lem:fsl}
(\cite{STOC::HalesH1999},\cite{Hoyer2000})
and show how this leads to a simple algorithm
for approximating the QFT over an arbitrary cyclic group.
While this
technique, like the quantum chirp-z method of Section \ref{sect:qchirp},
can only be used to replace a finite number of QFT's in a given computation,
it may be of independent interest.  Also, the proofs of Theorems 
\ref{thm:ftt} and \ref{thm:ftts} which lead directly
to the highly efficient Algorithms~\ref{alg:approxft} and \ref{samknowp}
rely on an elaboration of the techniques used in this earlier lemma.

\section{Fourier Sampling Lemma}\label{sect:naive}
In this section we prove a relationship between
the Fourier transforms over different moduli of a fixed 
vector.  In particular, let $\ket{v}=\sum_{i<N}v_i\ket{i}$ 
be a unit vector
and let $\kfv$ and $\kfmv$ be its Fourier 
transforms mod $N$ and
$M$ respectively, where $M>N$.\footnote{We interpret $\ket{v}$ as a unit
vector of length $M$ with entries greater than $N$ uniformly equal to zero.}

We exhibit a subvector of $\kfmv$ 
whose direction is a good approximation to $\kfv$'s whenever $M$ is
sufficiently large.  In particular, let 
$\prj$ denote the integer nearest 
$\frac{\litm}{\litn}j$ with ties broken by some standard convention.  Let 
$\afv$ be the subvector of $\kfmv$ consisting of the entries
indexed by integers $\prj$
renormalized by $\sqrt{\frac{\litm}{\litn}}$.  That is,
$$\afv=\sqrt{\frac{M}{N}}\sum_{j<N}
                    \hat{v}^{\litm}_{\prj}\ket{j}.$$
Then the $L_2$ 
distance between the vector $\kfv$ and $\afv$
becomes arbitrarily small as $M$ is increased relative to $N$.
The fact that this is true for $M=\Omega(N^{3/2})$ is almost trivial,
but we show that this is already true for $M=\Omega(N\log N)$.
This exponential improvement in the ratio $\frac{\litm}{\litn}$ is crucial
for the quantum applications discussed in Section~\ref{sect:apps}.

First, it is easily seen 
that for $M=\Omega\left(\frac{N^{3/2}}{\epsilon}\right)$,
$\kfv$ and $\afv$ are 
$\epsilon$-close in $L_2$ norm. 
The square of the $L_2$ distance between the vectors $\kfv$ and $\afv$
is given by
\begin{eqnarray*}
\sum_{j<N}
    \left|v_j-
                \sqrt{\frac{M}{N}}
                    \hat{v}^{\litm}_{\prj}\right|^2
&=&
\sum_{j<N}
    \left|\frac{1}{\sqrt{N}}\sum_{i<N}v_i\om{N}{ij}-
        \frac{1}{\sqrt{N}}
            \sum_{i<N}v_i\om{M}{i\prj}\right|^2\\
&=&
\frac{1}{N}
    \sum_{j<N}
        \left|\sum_{i<N}v_i\om{N}{ij}-
            \sum_{i<N}v_i\om{N}{ij}\om{M}{i\delta_j}\right|^2\\
&=&
\frac{1}{N}
    \sum_{j<N}
            \left|\sum_{i<N}v_i\om{N}{ij}(1-\om{M}{i\delta_j})\right|^2\\
&\leq&
\frac{1}{N}
    \sum_{j<N}
        \left(\sum_{i<N}
            \left|v_i\om{N}{ij}\right|
                \left|1-\om{M}{i\delta_j}\right|\right)^2,
\end{eqnarray*}
where $\delta_j=\prj-\frac{\litm}{\litn}j<1.$

We first use the fact that since $\left|i\delta_j\right|<N$, 
$$\left|1-\om{M}{i\delta_j}\right|<\frac{N}{M},$$ 
and then apply the inequality 
$$\sum_{i<N}|u_i|<\sqrt{N}$$ 
which holds for any unit vector $\ket{u}$, to obtain
\begin{eqnarray*}
\frac{1}{N}
    \sum_{j<N}
        \left(\sum_{i<N}\left|v_i\om{N}{ij}\right|
            \left|1-\om{M}{i\delta_j}\right|\right)^2
&\leq&
\frac{1}{N}
    \sum_{j<N}
        \left(\frac{N}{M}\sum_{i<N}
            \left|v_i\om{N}{ij}\right|\right)^2\\
&\leq&
\frac{N^3}{M^2}\end{eqnarray*} from which the claim follows.

However, this relationship cannot be exploited easily 
in the quantum setting. In short, in order for
$\afv$ to be a good approximation to $\kfv$, 
$M$ must be chosen so that the ratio $\frac{\litm}{\litn}$
is exponentially large.  But then the desired
subvector of $\kfmv$ is an exponentially small fraction
of the whole of $\kfmv$ and cannot
efficiently be recovered.

But this relationship actually holds
for much smaller $M$.  In particular, we show
that that it holds for 
$M=\Omega\left(\frac{N\log N}{\epsilon}\right)$, an
exponential improvement in the ratio $\frac{\litm}{\litn}$.

\begin{theorem}\label{lem:fsl}

Given any unit vector $\ket{v}=\sum_{i<N}v_i\ket{i}$

$$\left\|\kfv-\afv\right\|=O\left(\frac{N\log N}{M}\right).$$

\end{theorem}
A version of this theorem which referred only to
the distributions induced by $\kfv$ and $\afv$
first
appeared in \cite{STOC::HalesH1999}.  
The proof was later simplified,
and the bounds improved, in \cite{Hoyer2000}.  The proof
given in Section~\ref{sect:prfsl} is based on this simplification.

\subsection{Application:  An Approximate QFT over an Arbitrary Modulus $N$}
\label{sect:apps}We give a simple algorithm for an approximate QFT over
an arbitrary modulus based on Theorem \ref{lem:fsl}.  
This algorithm suffers from the same drawbacks as the chirp-z method
discussed in Section \ref{sect:qchirp}, namely it only succeeds with
inverse polynomial probability and thus can only be used
to replace a constant number of QFT's in a given quantum procedure.
However, the number of repetitions required to achieve an $\epsilon$
approximation with high probability is now linear rather than quadratic
in $O(\frac{1}{\epsilon})$.  Furthermore the algorithm is 
extremely simple.
We note that this is particularly true in the Fourier Sampling
setting, that is, if the
transform to be approximated occurs 
as the last step in a quantum algorithm with only the
distribution induced by the final superposition being of interest.
In this case measurement can take place immediately following 
Step \ref{it:tra}
and the rounding procedure can be accomplished classically.
This gives us an very short quantum subroutine,
but one which must be repeated many times for the 
required result, a trade-off which may be very desirable
when decoherence is taken into account.

Let $N$ and $\epsilon$ be given. Choose
$M=\Omega\left(\frac{N\log N}{\epsilon}\right)$:
\begin{algorithm}
  Input: $\keta$
\begin{enumerate}
\item\label{it:tra}Transform $\keta$ over $Z_M$:
  $$\keta\longrightarrow F_M \keta$$

\item If $x=\lfloor\frac{\litm}{\litn}i\rceil$ map 

    $$\ket{x}\ket{0}\longrightarrow\ket{i}\ket{1}.$$

\item Measure the second register.

\end{enumerate}
If a $1$ is measured in the second register which occurs with
probability $\frac{\epsilon}{\log N}$, then we output the successful
approximate QFT.  
\end{algorithm}

The correctness of this procedure follows directly from
our theorem.  If a $1$ is measured in the second register
then we have collapsed to a superposition in the direction of
$$\sum_{j<N}\hat{\alpha}^{\litm}_
{\prj}\ket{j},$$
By our Theorem the vector is $\epsilon$-close to the desired 
$$\sum_{j<N}\hat{\alpha}_j\ket{j}.$$
Moreover, since 
$$\sqrt{\frac{M}{N}}\sum_{j<N}\hat{\alpha}^{\litm}_
{\prj}\ket{j},$$
is approximately a unit vector, 
$$\sum_{j<N}\left|\hat{\alpha}^{\litm}_
{\prj}\right|^2=\frac{N}{M}$$
and the success probability is also correct.

\subsection{Two Claims}\label{sect:twocl}

To prove Theorem~\ref{lem:fsl} we first examine
the special case when the initial vector $\ket{v}$ is an element
of the Fourier basis mod $N$, in other words
$$\ket{v}=\kethj=\sum_{i<N}\frac{1}{\sqrt{N}}\om{N}{-ij}\ket{i}.$$
The Fourier transform over $N$ of $\kethj$
is just the standard basis vector $\ket{j}$,
i.e. a pointmass at $j$.
We let $\jm$ denote the Fourier transform over $M$ of $\kethj$
and $\kaj$ the subvector of $\jm$ at entries
of the form $\pri=\lfloor\frac{\litm}{\litn}i\rceil$ renormalized
by $\sqrt{\frac{\litm}{\litn}}$
in keeping with our earlier notation.
The vector $\jm$
is a smeared pointmass concentrated near $\prj$
and the entries of $\kaj$ satisfy the following:

\begin{claim}
\label{amp_j}
$\left|1-\aj_j\right|\leq\pi\frac{N}{M}$
\end{claim}

\begin{claim}
\label{amp_knotj}
For $k\neq j$, 
$$\left|\aj_k\right| 
    \leq 
        \frac{1}{|k-j|_N}\frac{N}{M}$$
where $|x|_N = \left\{ \begin{array}{ll}
		x\bmod N &  \mbox {if $0\leq x\bmod N\leq \frac{N}{2}$}\\
                -x\bmod N & \mbox {otherwise}
		\end{array}
		\right.  $

\end{claim}
These claims are proved in Section~\ref{sect:clpr}.  They
yield a
version of our main theorem in the special case that $\ket{v}$ is 
a Fourier basis vector:
\begin{observation}\label{ob:fb}
If $\ket{v}=\kethj$ is an element of the Fourier basis
mod $N$ and $M=\Omega\left(\frac{N}{\epsilon}\right),$ then
$$\left\|\kfv-\afv\right\|<\epsilon.$$ 
\end{observation}
We leave the proof of Observation~\ref{ob:fb} 
to the reader.  This Observation does not lead directly to our
Theorem~\ref{lem:fsl}.  In particular
if we try to extend it linearly
to allow for an arbitrary vector $\ket{v}$
we are forced to choose $M=\Omega(\frac{N^{3/2}}{\epsilon})$
to achieve a bound of $\epsilon$ -- the argument is that of
Section~\ref{sect:naive} now expressed in the Fourier rather
than the standard basis. 
Fortunately, a more careful examiniation of 
Claim~\ref{amp_knotj} gives us crucial 
information about the structure
of the error vectors
$$\ket{j}-\kaj$$ 
which will allow us to conclude our theorem.

\subsection{Proof of Theorem~\ref{lem:fsl}}\label{sect:prfsl}

We wish to bound the quantity 

\begin{equation}
\left\|\kfv-\afv\right\|
=\left\|\sum_{j<N}
        \fv_j
            \left(\ket{j}-\kaj\right)\right\|
=\sum_{i<N}
    \left|\sum_{j<N}
        \fv_j
            \left(\ket{j}-\kaj\right)_i\right|^2.
\end{equation}

This is the squared length of the vector which results from
applying the matrix with $ij$th entry $\left(\ket{j}-\kaj\right)_i$
to the unit vector $\kfv$, in other words the
best bound on this expression is exactly the squared operator
norm of this matrix.  By Claims~\ref{amp_j} and \ref{amp_knotj}
we have 
$$\left|\left(\ket{j}-\kaj\right)_j\right|<\pi\frac{N}{M}$$
and for $i\neq j$
$$\left|\left(\ket{j}-\kaj\right)_i\right|<\frac{2}{|i-j|_N}\frac{N}{M}.$$
It suffices, then, to bound the squared operator norm of the matrix
$A$ with $$A_{ij}= \left\{ \begin{array}{ll}
		\pi\frac{N}{M}&  \mbox {if $i=j$}\\
                \frac{1}{|i-j|_N}\frac{N}{M}& \mbox {otherwise}
		\end{array}
		\right.  $$
 
  This $N\times N$ 
  matrix  has the property that each
  row is the shift by one $(\bmod N)$ of the previous row, i.e.
  $A_{i,j}=A_{i+1\bmod N,j+1\bmod N}$, and all its entries are nonnegative
  reals. Because of this shift property -- such a matrix is commonly
  referred to as circulant --  its eigenvalues
  are all of the form $\sum_{i<N} \om{N}{jk}A_{ij}$ for some integer $k$.
  Moreover since the entries $A_{ij}$ are nonnegative reals
  the maximum eigenvalue is found by setting $k=0$, corresponding to
  an eigenvector with all equal entries.
  This maximum eigenvalue is precisely the operator norm of
  $A$ and can be found by taking the sum of any row of the matrix.
  Using the fact that $\sum_{i<N}\frac{1}{i}=\log N$ we have the
  sum of a row is $O\left(\frac{N\log N}{M}\right)$ and thus
\begin{equation}
  \left\|\kfv-\afv\right\|=O
        \left(\frac{N\log N}{M}\right)
\end{equation}
which establishes our theorem.

\subsection{Proofs of Claims~\ref{amp_j} and \ref{amp_knotj}}\label{sect:clpr}
We now prove Claims~\ref{amp_j} and \ref{amp_knotj}.

\begin{proof}[Proof of Claim~\ref{amp_j}]
To establish 
$$\left|1-\aj_j\right|\leq\pi\frac{N}{M}$$
we note that
\begin{eqnarray}
\left|1-\aj_j\right|&=&\left|1-\sqrt{\frac{M}{N}}\jm_{\prj}\right|\\
&=&\left|1-\frac{1}{N}\sum_{i<N}\om{N}{-ij}\om{M}{i\prj}\right|\\
&=&\left|1-\frac{1}{N}\sum_{i<N}\om{M}{i\epsilon}\right|
\end{eqnarray}
where $\epsilon=\prj-\frac{\litm}{\litn}j\leq 1/2$
This quantity is easily seen to be less than the arclength 
$2\pi\epsilon\frac{N}{M}$ and the claim follows.
\end{proof}

\begin{proof}[Proof of Claim~\ref{amp_knotj}]
We now establish that for $k\neq j$, 
$$\left|\aj_k\right| 
    \leq 
        \frac{1}{|k-j|_N}\frac{N}{M}$$
where $|x|_N = \left\{ \begin{array}{ll}
		x\bmod N &  \mbox {if $0\leq x\bmod N\leq \frac{N}{2}$}\\
                -x\bmod N & \mbox {otherwise.}
		\end{array}
		\right.  $

\begin{eqnarray}
\left|\aj_k\right|&=&\left|\sqrt{\frac{M}{N}}\jm_{\prk}\right|\\
&=&\left|\frac{1}{N}\sum_{i<N}\om{N}{-ij}\om{M}{i\prk}\right|\\
&=&\left|\frac{1}{N}\sum_{i<N}\om{N}{i(k-j+\frac{N}{M}\epsilon)}\right|\\
&=&\frac{1}{N}\frac{\left|\om{N}{N(k-j+\frac{N}{M}\epsilon)}-1\right|}
{\left|\om{N}{k-j+\frac{N}{M}\epsilon}-1\right|},
\end{eqnarray}
where $\epsilon=\prk-\frac{\litm}{\litn}k\leq 1/2$.
The numerator $\left|\om{N}{N(k-j+\frac{N}{M}\epsilon)}-1\right|$
is at most the arclength $2\pi\frac{N}{M}\epsilon<\pi\frac{N}{M}$
and the denominator $\left|\om{N}{k-j+\frac{N}{M}\epsilon}-1\right|$
is at least $\frac{2\pi|k-j+\frac{N}{M}\epsilon|_N}{N}>\frac{\pi|k-j|_N}{N}$ 
leading directly to the claimed bound.

\end{proof}

\section{Fourier Transform Theorems}
In this section we establish the technical results 
leading to Algorithms~\ref{alg:approxft} and \ref{samknowp}.
In particular, we prove a relationship between the transform $\kfv$
over $N$ of a given vector $\ket{v}$ and the transform
over $M>N$, not of that same vector $\ket{v}$ (as in previous Section),
but of a vector consisting of many repetitions of $\ket{v}$.
By repeating the vector $\ket{v}$ many times and
transforming over a large $M$ we get a vector
with not just one length $N$ subvector whose renormalization approximates
$\kfv$ (as in the previous Section) but a vector for which most length $N$
subvectors have this property.
Analogous to the previous section, the fact that this is true
when $\ket{v}$ is repeated $\Omega(N)$ times is
easy to prove but we show, via an amplification
of the circulant argument of Section \ref{sect:prfsl}, that this holds when the
number of repetitions is only $O(\log^2 N)$.
This improvement is responsible for the improved
efficiency of Algorithms \ref{alg:approxft} and \ref{samknowp} over earlier
methods and is also used crucially in the proof of Theorem \ref{thm:overr}.

More formally, let $\ket{v}=\sum_{i<N}v_i\ket{i}$ 
be an arbitrary unit vector, and let
$\ket{w}$ be the unit vector consisting
of $R$ repetitions of $\ket{v}$, that is
$$\ket{w}=\frac{1}{\sqrt{R}}\sum_{j<R}\sum_{i<N}v_i\ket{jN+i}$$
Then we can establish a strong relationship between
the vectors $\kfv$ and $\kfmw$ for sufficiently
large $R$ and $M$.  Recall from the previous section
that $\pri$ denotes the integer nearest $i$.
\begin{theorem}\label{thm:ftt}
Let $\ket{v}$ and $\ket{w}$ be as above.  
Then for any $M>RN$
there is a vector $\ket{u}=\sum_{|t|<\frac{M}{2N}}u_t\ket{t}$
so that 
$$\left\|\kfmw-\sum_{i<N}
\fv_i\ket{u}^{\pri}\right\|<\frac{4RN}{M} + \frac{8\log N}{\sqrt{R}}$$
where $\ket{u}^{\pri}=\sum_{|t|<\frac{M}{2N}}u_t\ket{\pri + t}$ is the vector $\ket{u}$ with indices
shifted by $\pri$. 
\end{theorem}
This theorem forms the basis for the Fourier Transform algorithm
of Section~\ref{sect:fta}.  By measuring the offset
from the nearest $\pri$, the superposition $\sum_{i<N}\fv_i\ket{u}^{\pri}$
collapses exactly to the desired $\kfv$.  This property approximately
holds for the
superposition $\kfmw$ (which we can generate)
by virtue of its closeness to $\sum_{i<N}\fv_i\ket{u}^{\pri}$.

As a byproduct of the proof of Theorem~\ref{thm:ftt} we
get a related theorem which is useful in the 
Fourier Sampling setting, that is, in the case where we
are concerned with the distribution induced by the final
superposition. We let $\Dv$ be the
probability distribution on the set $\{0,...N-1\}$ induced
by measuring $\kfv$ and $\Dw$ be the distribution
on the same set induced by measuring $\kfmw$ and interpreting
integers within $M/2N$ of $\pri$ as $i$.  More formally,
$$\Dv(i)=\left|\fv_i\right|^2$$
and
$$\Dw(i)=
        \sum_{|t|<\frac{M}{2N}}
            \left|\fmw_{\pri+t}\right|^2.$$
Then we can prove the following theorem:
\begin{theorem}\label{thm:ftts}
Let $\ket{v}$ and $\ket{w}$ be as above.  
Then for any $M\geq NR$
$$\left\|\Dw-\Dv\right\|_1<O\left(\frac{\log N}{\sqrt{R}}\right).$$
\end{theorem}
Notice that in order to make these distributions close
we need only make sure that $R$ is sufficiently large
and then $M$ can be taken to be any integer greater than
$RN$.
\subsection{Proof of Theorem~\ref {thm:ftt}}\label{sectmainlem}
First note that the Fourier transform over $RN$ of 
$$\ket{w}=\frac{1}{\sqrt{R}}\sum_{j<R}\sum_{i<N}v_i\ket{jN+i}$$
is 
\begin{equation}\label{m=rn}
\kfw=\sum_{i<N}\fv_i\ket{Ri}.
\end{equation} 
Recall that we are trying to show that there exists some vector
$\ket{u}$ supported on the integers in the interval
$(-\frac{M}{2N},\frac{M}{2N})$ such that $\kfmw$ 
is close to a vector of the form 
$$\sum_{i<N}\fv_i\ket{u}^{\pri}$$
where $\ket{u}^{\pri}$ is the vector $\ket{u}$ with indices
shifted by $\pri=\clint{\frac{M}{N}i}$. 
In the case that $M=RN$ Equation (\ref{m=rn}) immediately
yields our theorem
with the vector $\ket{u}=\ket{0}$ and no error at all.
For a general $M>RN$
$$\kfmw=\sum_{i<N} \fv_iF_{M}F^{-1}_{RN}(\ket{Ri}),$$
We let $\kim=F_{M}F^{-1}_{RN}(\ket{Ri})$ 
and thus
$$\kfmw =
\sum_{i<N} \fv_i\kim.$$
The $\kim$ are neither supported on the intervals 
$(\pri-\frac{M}{2N},\pri+\frac{M}{2N})$ nor the shifts
by $\pri$ of a fixed vector, but we show that
for sufficiently large $R$ and $M$ these conditions
approximately hold.  
 
To this end we define $\bumpi$ (for ``bump'')  
to be the vector $\kim$
restricted to the integers in the open interval
$(\pri-\frac{M}{2N},\pri+\frac{M}{2N})$, an interval which we denote by 
$(\pri)$. 
We let $\taili$ (for ``tail'') be the rest of $\kim$.  Thus 
the $\taili$ are supported on the indices outside of $(\pri)$ and
we have $\taili=\kim-\bumpi$.  Note also that
$$\kfmw=\sum_{i<N} \fv_i\kim
        =\sum_{i<N} \fv_i \bumpi
            + \sum_{i<N}\fv_i\taili.$$
Finally, let
$\shibumpo$ be the vector
$\ket{b_0}$ shifted by $\pri$.
Our aim will be to show that $\ket{b_0}$
is our candidate for $\ket{u}$, in other words that
$$\kfmw =\sum_{i<N} \fv_i \bumpi
            + \sum_{i<N}\fv_i\taili
                \approx\sum_{i<N} \fv_i\shibumpo.$$
We first bound $\left\|\sum_{i<N}\fv_i\taili\right\|$, then show
that each $\bumpi$
is very close to $\shibumpo$.  Since the 
$\bumpi$'s have disjoint support the
closeness of the vectors follows.
More formally, we will prove the following two claims:
\begin{claim}\label{tailbound}
  \begin{equation}
    \left\|\kfmw -\sum_{i<N} \fv_i\bumpi\right\|
    =\left\|\sum_{i<N}\fv_i\taili\right\|
    \leq\frac{8\log N}{\sqrt{R}}
  \end{equation}
\end{claim}
Claim~\ref{tailbound} states that making $R$ large (i.e. increasing the number 
of repetitions of
$\ket{v}$) reduces the effect of these tails.
\begin{claim}\label{shiftclose}
  Let $\shibumpo$ be the superposition
  $\ket{b_0}$ shifted by $\pri$.  Then
  $$\left\|\shibumpo-\bumpi\right\|<\frac{4RN}{M}.$$
\end{claim}
From Claim~\ref{shiftclose} and the fact that the $\bumpi$
have disjoint supports,
$$\left\|\sum_{i<N}\fv_i \shibumpo-
\sum_{i<N}\fv_i\bumpi\right\|\leq\frac{4RN}{M}.$$
Combining this with Claim~\ref{tailbound} via the triangle inequality we
have,
\begin{equation}
  \left\|\kfmw -\sum_{i<N} \fv_i \shibumpo\right\|
  \leq \frac{4RN}{M} + \frac{8\log N}{\sqrt{R}},
\end{equation}
as desired.

\subsection{Proof of Theorem~\ref{thm:ftts}}

In this case we wish to show that the distribution $\Dv$
on $\{0, ..., N-1\}$ induced by sampling $\kfv$ and
the distribution $\Dw$
on $\{0, ..., N-1\}$ induced by sampling
$\kfmw$ and interpreting integers
within $M/2N$ units of $\pri$ as $i$, are close.
The closeness of these distributions turns out to
follow from Claim~\ref{tailbound} alone, allowing us to drop the
dependence of the error on the ratio $M/RN$.  In particular,
as long as $R$ is sufficiently large, any $M>RN$ will 
do.

Let  $$d(i)=\norm{\fv_i}^2\braket{b_i}{b_i}.$$ 
Then $d$ is the sub-distribution induced by measuring the (generally
sub-unit length) superposition
$\sum_{i<N}\fv_i\bumpi$ and interpreting integers
within $M/2N$ units of $\pri$ as $i$.
By Claim~\ref{tailbound} we have
$$\left\|\kfmw -\sum_{i<N}\fv_i\bumpi\right\|\leq\frac{8\log N}{\sqrt{R}}$$
from which it follows that
$$\left|\Dw-d\right|=O\left(\frac{\log N}{\sqrt{R}}\right).$$

Now, 
$$\left|d-\Dv\right|\leq\sum_{i<N}|\fv_i|^2
\left(\braket{b_i}{b_i}-1\right)$$
and $$1-\braket{b_i}{b_i}=\braket{t_i}{t_i}=
O\left(\frac{\log^2 N}{R}\right)$$
by applying Claim~\ref{tailbound} to the vector $\kfv=\ket{i}$.
The result
then follows from the triangle inequality.
\subsection{Proof of Claim~\ref{tailbound}}\label{sect:tb}
\begin{proof}
  In order to bound $\left\|\sum_{i<N}\fv_i\taili\right\|$ we will use the
  following observation which establishes that the amplitudes in
  $\kim$ fall off quickly away from $\pri$.  Recall that $\taili$
  is identical to the superposition $\kim$ except that it is missing 
  all the amplitudes at $j\in(\pri)$ where $(\pri)$ is the interval 
  $(\pri-\frac{M}{2N},\pri+\frac{M}{2N})$.  Thus this falloff applies
  to the $\taili$ as well.
  This Observation is closely related
to Claim~\ref{amp_knotj} of the previous Section and it
proof is in Section~\ref{secondapp}.

  \begin{observation} \label{falloff}
    $$\norm{\im_j} = \norm{\frac{1}{\sqrt{M}}\frac{1}{\sqrt{RN}} 
    \sum_{k<RN}
      \om{M}{k(j-\frac{M}{N}i)}} \leq
    \sqrt{\frac{M}{RN}}\frac{2}{\norm{j-\frac{M}{N}i}_{\litm}}$$
    where $|x|_{\litm}= \left\{ \begin{array}{ll}
        x\bmod M &  \mbox {if $0\leq x\bmod M\leq M/2$}\\
        -x\bmod M & \mbox {otherwise}
      \end{array}
    \right.  $

  \end{observation}
  
  We now use this to bound $\left\|\sum_{i<N}\fv_i\taili\right\|$.  
  We first note that Observation \ref{falloff} can be used 
  to show that $\left\|\taili\right\|=O\left(\frac{1}{\sqrt{R}}\right)$.  
  A naive analysis
  of the quantity $\left\|\sum_{i<N}\fv_i\taili\right\|$ -- see the
  discussion in Section~\ref{sect:naive} -- would then give a bound of 
  $O\left(\sqrt{\frac{N}{R}}\right)$.  Instead we achieve an
  improved bound by a more complex version of the circulant argument of
  Section~\ref{sect:prfsl}.  The
  first equality below is by the definition of $\taili$ and the
  second is by the above observation:

  \begin{eqnarray*}
    \left\|\sum_{i<N}\fv_i\taili\right\|^2
    &=&
    \sum_{j<M} \left|\sum_{i,j\not\in(\pri)}
    \fv_i\taili_j\right|^2  \\
    &\leq&
    \sum_{j<M} \frac{4M}{RN} \left(\sum_{i,j\not\in(\pri)}
      \frac{\norm{\fv_i}}{\norm{j-\frac{M}{N}i}_{\litm}} \right)^2.
  \end{eqnarray*}
  
  This expression is almost maximized by taking the
  $\fv_i=1/\sqrt{N}$ for all $i$.  In particular, the expression can
  be bounded by four times its value at this vector.  The proof of this 
  fact is in Section~\ref{sect:circ} and is the heart of the Theorem.
  It is proved by an
  extension of the circulant argument used in Theorem~\ref{lem:fsl}.

  \begin{equation}
  \left\|\sum_{i<N}\fv_i \taili\right\|^2 \leq
    \sum_{j<M} \frac{16M}{N^{2}R} \left( \sum_{i,j\not\in(\pri)}
      \frac{1}{\norm{j-\frac{M}{N}i}_{\litm}} \right)^2.
  \end{equation}

  Using the fact that the smallest denominator
  $\norm{j-\frac{M}{N}i}_{\litm}$ is at least $\frac{M}{2N}$ and the rest
  are spaced out by $\frac{M}{N}$ we have
  $$\sum_{i,j\not\in(\pri)} \frac{1}{\norm{j-\frac{M}{N}i}_{\litm}}
  \leq \frac{2N\log N}{M}.$$
  Therefore \begin{equation} \left\|\kfmw
    -\sum_{i<N} \fv_i \bumpi\right\| = \left\|\sum_{i<N} \fv_i
    \taili\right\| \leq \frac{8\log N}{\sqrt{R}}, \end{equation} as
  desired.
\end{proof}
\subsection{Proof of Bound in Claim~\ref{tailbound}}\label{sect:circ}

\begin{claim} \label{thm:approxcirculant}
  For any unit vector $\ket{x}$ and $M>8N$
  \begin{equation}\label{cirbound}
    \sum_{j<M}\left|
    \sum_{i<N,j\not\in(\pri)}
    \frac{x_i}{|j-\frac{M}{N}i|_M} \right|^2
  \leq\frac{4}{N}
  \sum_{j<M} \left( \sum_{i<N,j\not\in(\pri)}
\frac{1}{|j-\frac{M}{N}i|_M} \right)^2
  \end{equation}
\end{claim}

\begin{proof}The left hand side of Equation (\ref{cirbound}) 
  is at most the squared operator
  norm of the $M\times N$ matrix $A$ with entries 
  $$A_{ji}= \left\{ \begin{array}{ll}
		0&  \mbox {if $j\in(\pri)$}\\
                \frac{1}{|j-\frac{M}{N}i|_M}& \mbox {otherwise}
		\end{array}
		\right.$$  

  Note that our matrix is positive and has the property that 
each row is comprised of samples of the 
same underlying function but with the samples shifted
by $\frac{N}{M}$ from one row to the next.  
We argued in Section~\ref{sect:prfsl} that the operator
norm of any positive $N\times N$
matrix with the property that each
  row is the shift by one$\mod N$ of the previous row
  is found by applying the matrix to the
unit vector with entries uniformly equal to $\frac{1}{\sqrt{N}}$.

  Now, while our rectangular matrix is obviously not of this form,
by reindexing and
  changing the denominators of the entries only slightly -- 
so that a fixed set of integral $t$ can be used -- the
  expression $\|A(x)\|^2$ becomes
  
$$\| \hat{A}x \|^2
    = \sum_{t\in\pm\frac{M}{2N}}\sum_{k<N}\left(
      \sum_{i<N,i\neq k}
      \frac{x_i}{|\frac{M}{N}k + t -\frac{M}{N}i|_{\litm}}\right)^2.$$
  
  Notice that the matrix giving rise to each of the double sums indexed by
  $k$ and $i$ in
  this new expression is $N\times N$ and has the properties discussed
  previously -- the entries depend only on the quantity $(k-i)\mod N$.  
  Thus each individual sum, and therefore the entire
  sum is maximized by choosing the entries of $x_i$ to be equal.
  
  Finally, we can relate $\| \hat{A}x \|^2$ to our original expression as
  follows: we added at most $\pm 1$ to 
the denominators of our original matrix entries.  Since 
these denominators were all larger than $\frac{M}{2N}>4$
  this at most doubled/halved the squared sums of the entries.
  Thus we have:
  $$\frac{1}{2} \leq \frac{\|\hat{A}x\|^2}{\|A x\|^2} \leq 2.$$
  Finally, we use this to bound the squared operator norm of $A$.  Let $x_0$
  maximize $\|\hat{A}x\|^2$.  Then for any $x$ we have $\|Ax\|^2 \leq
  2 \|\hat{A}x\|^2 \leq 2 \|\hat{A}x_0\|^2 \leq 4 \|Ax_0\|^2$.  Thus
  our expression is bounded by four times it's value at the unit
  vector with entries uniformly equal to $\frac{1}{\sqrt N}$, as claimed. 
\end{proof}
  
\subsection{Proof of Observation~\ref{falloff}} \label{secondapp}
\begin{proof}
  Recall that
  $$\kim = F_M F_{RN}^{-1} \ket{Ri}=
  \frac{1}{\sqrt{M}}\frac{1}{\sqrt{RN}}\sum_{j=0}^{M-1} \sum_{k=0}^{RN-1}
  \om{}{k(\frac{j}{M}-\frac{i}{N})} \ket{j}.$$
  We have
  \begin{eqnarray*}\label{eqn:falloff}
    |\im_j| &=& \left| \frac{1}{\sqrt{M}}\frac{1}{\sqrt{RN}}
      \sum_{k=0}^{RN-1} \om{}{k(\frac{j}{M}-\frac{i}{N})}\right|
    =
    \frac{1}{\sqrt{M}}\frac{1}{\sqrt{RN}}\left|
      \frac{1-\om{}{RNj/M}}{1-\om{}{(\frac{j}{M}-\frac{i}{N})}}
    \right| \\ 
\end{eqnarray*}
  where the second equality applies
  the formula for geometric series.  Then since
  $$\left|1-\om{}{(\frac{j}{M}-\frac{i}{N})}\right|= 
    \left|1-\om{M}{|j-\frac{M}{N}i|_{\litm}}\right|
   \geq\frac{1}{M}
  \left|j-\frac{M}{N}i\right|_{\litm}$$ and $\norm{1-\om{}{RNj/M}}<2,$
     we have
    $$|\im_j|\leq\sqrt{\frac{M}{RN}}
    \frac{2}{|j-\frac{M}{N}i|_{\litm}}$$ as claimed.
\end{proof}

\subsection{Proof of Claim~\ref{shiftclose}}

\begin{proof}
  We first note that to show that the restricted ``bump'' vectors
  are close, that is,
  $$\left\|\shibumpo-\bumpi\right\|<\frac{4RN}{M}$$
   it suffices to show
  that the corresponding full vectors $\ket{0^{\litm}}$ and $\kim$
  satisfy the same bound, that is
 $$\left\|\ket{0^{\litm}}^{\pri}-\kim\right\|<  \frac{4RN}{M}.$$
  But recalling that $\kim = F_M F_{RN}^{-1} \ket{Ri}$ and using the 
fact that $F_M$ is unitary we have
  \begin{eqnarray*}
    \left\|\ket{0^{\litm}}^{\pri}-\kim\right\|^2
    &=& \left\|F_M^{-1}\left(\ket{0^{\litm}}^{\pri} - \kim\right)\right\|^2\\
    &=& \sum_{j<RN} \left| \frac{1}{\sqrt{RN}}
\om{N}{-\pri j} -
      \frac{1}{\sqrt{RN}} \om{N}{-ij} \right|^2\\
    &\leq&
    \sum_{j<RN} \left| \frac{1}{\sqrt{RN}}\om{N}{-ij}
    \left(\om{N}{\delta_ij} -1\right)\right|^2
  \end{eqnarray*}

  where $\delta_i=\pri -\frac{M}{N}i$.  But since $j$
  only goes up to $RN$,
  $$\norm{\om{M}{\delta_ij}-1}<\frac{4RN}{M}$$
  and thus
  \begin{eqnarray*}
    \sum_{j<RN} \lefteqn{\norm{\frac{1}{\sqrt{RN}} \om{N}{-ij}
      \left(\om{N}{\delta_ij}-1\right)}^2}\\
    &\leq& \sum_{j<RN} \norm{\frac{1}{\sqrt{RN}} \om{N}{-ij}}^2
    \norm{\om{N}{\delta_ij}-1}^2\\
    &\leq& \left( \frac{4RN}{M} \right)^2,
  \end{eqnarray*}
  as desired.
\end{proof}


\bibliographystyle{plain}
\bibliography{bibthesis}

\begin{thebibliography}{10}

\bibitem{Ambainis00a}
A.~Ambainis.
\newblock Quantum lower bounds by quantum arguments.
\newblock In {\em Proceedings of the 32nd Annual ACM Symposium on the Theory of
  Computing (STOC)}, 2000.

\bibitem{BabaiMor}
L.~Babai and S.~Moran.
\newblock Arthur-{M}erlin {G}ames: a randomized proof system and a hierarchy of
  complexity classes.
\newblock {\em Journal of Computer and System sciences}, 36(2):254--276, 1988.

\bibitem{Bennett}
C.~H. Bennett.
\newblock Logical reversibility of computation.
\newblock {\em IBM Journal of Research and Development}, 17:525--532, 1973.

\bibitem{Bennett-89}
C.~H. Bennett.
\newblock Time/space trade-offs for reversible computation.
\newblock {\em SIAM Journal on Computing}, 18(4):766--776, 1989.

\bibitem{BBBV}
C.~H. Bennett, E.~Bernstein, G.~Brassard, and U.~Vazirani.
\newblock Strengths and weaknesses of quantum computing.
\newblock {\em SIAM Journal on Computing}, 26(5):1510--1523, October 1997.

\bibitem{SICOMP::BernsteinV1997}
E.~Bernstein and U.~Vazirani.
\newblock Quantum complexity theory.
\newblock {\em SIAM Journal on Computing}, 26(5):1411--1473, October 1997.

\bibitem{BonehL1995}
D.~Boneh and R.~J. Lipton.
\newblock Quantum cryptanalysis of hidden linear functions (extended abstract).
\newblock In Don Coppersmith, editor, {\em Advances in
  Cryptology---CRYPTO~'95}, volume 963 of {\em Lecture Notes in Computer
  Science}, pages 424--437. Springer-Verlag, 27--31~August 1995.

\bibitem{Cheung}
K.~Cheung and M.~Mosca.
\newblock Decomposing finite {A}belian groups.
\newblock arXiv:cs.Ds/0101004.

\bibitem{Cleve94}
R.~Cleve.
\newblock A note on computing quantum {F}ourier transforms by quantum programs.
\newblock Manuscript available at
  http://www.cpsc.ucalgary.ca/~cleve/papers.html, 1994.

\bibitem{CEHMM}
R.~Cleve, A.~Ekert, L.~Henderson, C.~Macchiavello, and M.~Mosca.
\newblock On quantum algorithms.
\newblock Los Alamos Preprint Archive quantum-ph/9903061.

\bibitem{CEMM1998}
R.~Cleve, A.~Ekert, C.~Macchiavello, and M.~Mosca.
\newblock Quantum algorithms revisited.
\newblock {\em Proc. Roy. Soc. Lond. A}, 454:339--354, 1998.

\bibitem{cleve00fast}
R.~Cleve and J.~Watrous.
\newblock Fast parallel circuits for the quantum {F}ourier transform.
\newblock {\em Proceedings of the 41st Annual IEEE Symposium on Foundations of
  Computer Science (FOCS)}, 2000.

\bibitem{cooltuk}
J.~W. Cooley and J.~Tukey.
\newblock An algorithm for the machine calculation of complex {F}ourier series.
\newblock {\em Mathematics of Computation}, 19:297--301, 1965.

\bibitem{Coppersmith1994}
D.~Coppersmith.
\newblock An approximate {F}ourier transform useful in quantum factoring.
\newblock Technical Report RC19642, IBM, 1994.

\bibitem{STOC::Grover1996}
L.~K. Grover.
\newblock A fast quantum mechanical algorithm for database search.
\newblock In {\em Proceedings of the 28th Annual ACM Symposium on Theory of
  Computing (STOC)}, pages 212--219, Philadelphia, Pennsylvania, 22--24 May
  1996.

\bibitem{STOC::HalesH1999}
L.~Hales and S.~Hallgren.
\newblock Quantum {F}ourier sampling simplified.
\newblock In {\em Proceedings of the 31st Annual ACM Symposium on Theory of
  Computing (STOC)}, pages 330--338, 1999.

\bibitem{FOCS::HalesH2000}
L.~Hales and S.~Hallgren.
\newblock An improved quantum {F}ourier transform algorithm and applications.
\newblock In {\em Proceedings of the 41st Annual IEEE Symposium on Foundations
  of Computer Science (FOCS)}, 2000.

\bibitem{hallgren2002}
S.~Hallgren.
\newblock Polynomial-time quantum algorithms for {P}ell's equation and the
  principal ideal problem.
\newblock {\em Proceedings of the 34th Annual ACM Symposium on the Theory of
  Computing (STOC)}, 2002.

\bibitem{Hastad}
J.~Hastad and T.~Leighton.
\newblock Division in ${O}(\log n)$ depth using $n^{1+\epsilon}$ processors.
\newblock Unpublished manuscript.

\bibitem{Hoyer2000}
P.~H{\o}yer.
\newblock Simplified proof of the {F}ourier sampling theorem.
\newblock {\em Information Processing Letters}, 75:139--143, 2000.

\bibitem{Jozsa}
R.~Jozsa.
\newblock Quantum algorithms and the {F}ourier transform.
\newblock {\em Proceedings of the Royal Society of London A}, pages 323--337,
  January 1998.

\bibitem{Kitaev1995}
A.~Kitaev.
\newblock Quantum measurements and the {A}belian stabilizer problem.
\newblock Los Alamos Preprint Archive quantum-ph/9511026, 1995.

\bibitem{Moore}
C.~Moore and M.~Nilsson.
\newblock Parallel quantum computation and quantum codes.
\newblock Los Alamos Preprint Archive quant-ph/9808027, 1998.

\bibitem{MosEke98}
M.~Mosca and A.~Ekert.
\newblock The hidden subgroup problem and eigenvalue estimation on a quantum
  computer.
\newblock In {\em QCQS: NASA International Conference on Quantum Computing and
  Quantum Communications}, 1998.

\bibitem{QuantumBook}
M.~A. Nielsen and I.~L. Chuang.
\newblock {\em Quantum Computation and Quantum Information}.
\newblock Cambridge University Press, 2000.

\bibitem{pratt75}
V.~R. Pratt.
\newblock Every prime has a succinct certificate.
\newblock {\em SIAM Journal on Computing}, 4:214--220, 1975.

\bibitem{Rab}
L.~Rabiner, R.~Schafer, and C.~Rader.
\newblock The chirp-z transform and its applications.
\newblock {\em Bell System Technical Journal}, 48:1249--1292, 1969.

\bibitem{Reif1990}
J.~H. Reif and S.~R. Tait.
\newblock Optimal size integer division circuits.
\newblock {\em SIAM Journal on Computing}, 19(5):912--924, 1990.

\bibitem{schstr}
A.~Sch{\"{o}}nhage and V.~Strassen.
\newblock Schnelle {M}ultiplikation gro$\beta$er {Z}ahlen.
\newblock {\em Computing}, 7:281--292, 1971.

\bibitem{shamir90}
A.~Shamir.
\newblock {IP=PSPACE}.
\newblock {\em Proceedings of the 31st Annual IEEE Symposium on Foundations of
  Computer Science (FOCS)}, pages 11--15, 1990.

\bibitem{Shor94}
P.~W. Shor.
\newblock Algorithms for quantum computation: Discrete log and factoring.
\newblock In {\em Proceedings of the 35th IEEE Annual Symposium on Foundations
  of Computer Science (FOCS)}, pages 124--134, 1994.

\bibitem{SICOMP::Shor1997}
P.~W. Shor.
\newblock Polynomial-time algorithms for prime factorization and discrete
  logarithms on a quantum computer.
\newblock {\em SIAM Journal on Computing}, 26(5):1484--1509, October 1997.

\bibitem{SICOMP::Simon1997:1474}
D.~R. Simon.
\newblock On the power of quantum computation.
\newblock {\em SIAM Journal on Computing}, 26(5):1474--1483, October 1997.

\bibitem{Toffoli-80a}
T.~Toffoli.
\newblock Reversible computing.
\newblock In J.~W. de~Bakker and J.~van Leeuwen, editors, {\em Automata,
  Languages and Programming (Seventh Colloquium, Noordwijkerhout, the
  Netherlands, July 14--18, 1980)}, volume~85 of {\em Lecture Notes in Computer
  Science}, pages 632--644. Springer-Verlag, 1980.

\bibitem{Vedral-et-al-95}
V.~Vedral, A.~Barenco, and A.~Ekert.
\newblock Quantum networks for elementary arithmetic operations.
\newblock {\em Physical Review A}, November 1995.

\bibitem{Vergis-et-al-86}
A.~Vergis, K.~Steiglitz, and B.~Dickinson.
\newblock The complexity of analog computation.
\newblock {\em Mathematics and Computers in Simulation}, 28:91--113, 1986.

\bibitem{watrous2000}
J.~Watrous.
\newblock Succinct quantum proofs for properties of finite groups.
\newblock {\em Proceedings of the 41st Annual IEEE Symposium on Foundations of
  Computer Science (FOCS)}, 2000.

\bibitem{zachos}
S.~Zachos.
\newblock Probabilistic quantifiers and games.
\newblock {\em Journal of Computer and System Sciences}, 36:433--451, 1988.

\end{thebibliography}
\end{document}